\documentclass[12pt,a4paper]{iopart}
\pdfoutput=1
\usepackage{iopams,setstack,mathrsfs,amsthm,cite,enumerate,float,graphicx}
\usepackage[utf8]{inputenc}
\usepackage[T1]{fontenc}
\usepackage{lmodern}
\usepackage[colorlinks,linkcolor=blue,citecolor=blue,urlcolor=blue]{hyperref}
\newcommand{\al}{\alpha}
\newcommand{\be}{\beta}
\newcommand{\de}{\delta}
\newcommand{\ep}{\epsilon}
\newcommand{\vep}{\varepsilon}
\newcommand{\ga}{\gamma}
\newcommand{\ka}{\kappa}
\newcommand{\la}{\lambda}

\newcommand{\si}{\sigma}

\newcommand{\ze}{\zeta}
\newcommand{\De}{\Delta}

\newcommand{\La}{\Lambda}
\newcommand{\Si}{\Sigma}

\newcommand{\bde}{\boldsymbol{\delta}}

\newcommand{\bk}{\mathbf{k}}

\newcommand{\bn}{\mathbf{n}}

\newcommand{\bbs}{\mathbf{s}}

\newcommand{\bx}{\mathbf{x}}
\newcommand{\by}{\mathbf{y}}
\newcommand{\bv}{\mathbf{v}}

\newcommand{\bsi}{{\boldsymbol{\si}}}

\newcommand{\CC}{{\mathbb C}}

\newcommand{\RR}{{\mathbb R}}

\newcommand{\cE}{{\mathcal E}}

\newcommand{\cH}{{\mathcal H}}

\newcommand{\cJ}{{\mathcal J}}

\newcommand{\cN}{{\mathcal N}}
\newcommand{\cP}{{\mathcal P}}

\newcommand{\cS}{{\mathcal S}}

\newcommand{\cZ}{{\mathcal Z}}
\newcommand\Esc{E^{\mathrm{sc}}}
\newcommand{\EGS}{E_{\mathrm{GS}}}

\newcommand\Hsc{H_{\mathrm{sc}}}
\newcommand{\Egs}{E_{\mathrm{GS}}}
\newcommand\Zsc{Z_{\mathrm{sc}}}

\newcommand{\pd}{\partial}

\newcommand{\id}{1\hspace{-.25em}{\rm I}}

\newcommand{\ket}[1]{|#1\rangle}

\newcommand{\mss}{\kern 1pt}

\renewcommand{\geq}{\geqslant}
\renewcommand{\le}{\leqslant}
\renewcommand{\ge}{\geqslant}
\newcommand{\tends}[1]{\bbuildrel{\hbox to 2em{\rightarrowfill}}_{#1}^{}}

\newcommand{\operatorname}[1]{\mathop{\rm #1}\nolimits}
\newcommand{\sech}{\operatorname{sech}}

\newcommand{\sgn}{\operatorname{sgn}}
\newcommand{\Li}{\operatorname{Li}}
\let\diff\rmd
\newcommand{\oo}{\operatorname{o}}
\newcommand{\su}{\mathrm{su}}

\newcommand{\qbinom}[3]{{#1\atopwithdelims[]#2}_{\raise 3pt\hbox{$\scriptstyle #3$}}}
\renewcommand{\Im}{\operatorname{Im}}

\newcommand{\en}{\enspace}
\newcommand{\all}{\forall}
\newcommand{\pdf}[2]{\frac{\partial #1}{\partial #2}}

\newcommand{\Int}[1]{\,\mathop{\!#1}\limits^{\lower1ex\hbox{$\scriptstyle\circ$}}{}}

\theoremstyle{remark}

\let\tfrac\case
\let\eqref\eref
\newcommand{\mathclap}[1]{\hbox to0pt{\hss$\scriptstyle #1$\hss}}

\eqnobysec
\newcommand{\bN}{\mathbf N}

\newcommand{\vepmax}{\vep_{\mathrm{max}}}
\def\clap#1{\hbox to 0pt{\hss#1\hss}}

\def\mathrlap{\mathpalette\mathrlapinternal}
\def\mathclap{\mathpalette\mathclapinternal}

\def\mathrlapinternal#1#2{%
           \rlap{$\mathsurround=0pt#1{#2}$}}
\def\mathclapinternal#1#2{%
  \clap{$\mathsurround=0pt#1{#2}$}}
\newcommand{\Pij}{P_{ij}^{(m|n)}}
\begin{document}

\title[Thermodynamics and criticality of supersymmetric spin chains]{Thermodynamics and
  criticality of supersymmetric spin chains with long-range interactions}

\author{F. Finkel, A. González-López, I. León, M.A. Rodríguez}

\address{Departamento de Física Teórica, Universidad Complutense de Madrid, 28040 Madrid, Spain}

\eads{\mailto{ffinkel@ucm.es}, \mailto{artemio@ucm.es}, \mailto{ivleon@ucm.es}, \mailto{rodrigue@ucm.es}}

\date{\today}

\begin{abstract}
  We study the thermodynamics and critical behavior of $\su(m|n)$ supersymmetric spin chains of
  Haldane--Shastry type with a chemical potential term. We obtain a closed-form expression for the
  partition function and deduce a description of the spectrum in terms of the supersymmetric
  version of Haldane's motifs, which we apply to obtain an analytic expression for the free energy
  per site in the thermodynamic limit. By studying the low-temperature behavior of the free
  energy, we characterize the critical behavior of the chains with~$1\le m,n\le2$, determining the
  critical regions and the corresponding central charge. We also show that in the $\su(2|1)$,
  $\su(1|2)$ and $\su(2|2)$ chains the bosonic or fermionic densities can undergo first-order
  (discontinuous) phase transitions at~$T=0$, in contrast with the previously studied $\su(2)$
  case.
\end{abstract}


\maketitle

\section{Introduction}

Spin chains of Haldane--Shastry type have been extensively studied as the prototypical examples of
one-dimensional lattice models with long-range interactions, due to their remarkable physical and
mathematical properties. The best known of these models is the original Haldane--Shastry (HS)
chain~\cite{Ha88,Sh88}, which consists of a circular array of $N$~equispaced spins with
inverse-square two-body interactions. This chain is integrable~\cite{FM93,BGHP93} and invariant
under the quantum Yangian for arbitrary values of~$N$, which in turn makes it possible to derive a
complete description of its spectrum in terms of Haldane's motifs~\cite{HHTBP92}. From a more
applied standpoint, the HS chain has appeared in such disparate contexts as conformal field
theory~\cite{Ha91,BBS08,CS10}, fractional statistics and anyons~\cite{Ha91,HHTBP92,GS05,Gr09},
quantum chaos vs.~integrability~\cite{FG05,BFGR08epl,BFGR10,EFG09,EFG10}, quantum information
theory~\cite{GSFPA10} or quantum simulation of long-range magnetism~\cite{HGCK16}. One of the
characteristic features of the HS chain is its close connection with the (dynamical) spin
Sutherland model~\cite{HH92}, whose spin degrees of freedom are governed by the HS Hamiltonian in
the large coupling constant limit. Using this idea, known in the literature as Polychronakos's
freezing trick~\cite{Po94}, it is possible to compute in closed form the chain's partition
function~\cite{FG05}. In fact, the same approach can be applied to the long-range dynamical spin
models of Calogero~\cite{MP93} and Inozemtsev~\cite{In96}, which yield the so-called
Polychronakos--Frahm (PF)~\cite{Po93,Fr93} and Frahm--Inozemtsev (FI)~\cite{FI94} spin chains.
Although they are not translationally invariant, these chains share many fundamental properties
with the original HS chain. For this reason, we shall collectively refer in this work to the HS,
PF and FI chains as spin chains of Haldane--Shastry type.

In all the chains of HS type discussed in the previous paragraph, the term ``spin'' actually
stands for $\su(m)$ spin. In fact, Haldane himself was the first to consider an~$\su(m|n)$
supersymmetric version of the HS chain, whose sites can be occupied either by an~$\su(m)$ boson or
by an~$\su(n)$ fermion~\cite{Ha93}. An analogous supersymmetric chain of PF type was introduced
shortly afterwards~\cite{BUW99}. The partition function of both the HS and the PF supersymmetric
chains have been exactly computed using the freezing trick~\cite{BUW99,BB06,BB09}, and their
spectra have been fully described in terms of a suitable generalization of Haldane's
motifs~\cite{HB00,BBHS07,BBH10}. On the other hand, the supersymmetric version of the
Frahm--Inozemtsev chain~\cite{BBH10} has received comparatively less attention in the literature.
It should also be noted that, apart from their intrinsic interest, the supersymmetric chains of HS
type are closely related to important models in condensed matter theory describing the dynamics of
holes in a spin background. Indeed, the HS $\su(1|n)$ chain with~$n>1$ is equivalent to the
long-range $\su(n)$ $t$-$J$ model with equal exchange and transfer energies proportional
to~$(z_i-z_j)^{-2}$, originally introduced by Kuramoto and Yokoyama~\cite{KY91}.

The study of the thermodynamics of spin chains of HS type, which goes back to the early work of
Haldane~\cite{Ha91}, has received a good deal of attention. In the latter reference the spinon
description of the spectrum is used to deduce an expression for the entropy of the~$\su(2)$ HS
chain in the thermodynamic limit. A heuristic formula for the free energy of the PF chain appeared
shortly afterwards in Ref.~\cite{Fr93}. A similar result for the FI chain using the transfer
matrix method was derived by Frahm and Inozemtsev~\cite{FI94}, who also computed the magnetization
of this chain in an external constant magnetic field. More recently, a comprehensive study of the
thermodynamics of the three (spin~$1/2$) chains of HS type in a constant magnetic field was
performed in Ref.~\cite{EFG12}, using again the transfer matrix method. In the supersymmetric
case, the thermodynamic functions of the $\su(1|1)$ HS chain (with a chemical potential term) have
been exactly computed taking advantage of the equivalence of this model to a free, translationally
invariant fermion system~\cite{CFGRT16}. This approach cannot be applied to the~$\su(1|1)$ PF and
FI chains, since these models are not translationally invariant, nor in fact to any chain of HS
type with $m$ or $n$ different from $1$. To the best of our knowledge, the thermodynamics of the
supersymmetric chains of the latter type, or the non-supersymmetric ones with~$m>2$, have not been
analyzed in the literature.

The connection between the original $\su(2)$ HS chain and the level-$1$ $\su(2)$
Wess--Zumino--Novikov--Witten conformal field theory (CFT), stemming from the Yangian symmetry of
both types of models, was already numerically observed by~Haldane~\cite{Ha88} and subsequently
established by several authors in the (purely fermionic) $\su(n)$
case~\cite{Ha91,HHTBP92,BPS94,BS96}. Thus the~$\su(n)$ HS chain (with no magnetic field or
chemical potential term) is critical (gapless), with central charge~$c=n-1$. This result was later
extended to the $\su(m|n)$ PF chain (again with zero chemical potentials) in Ref.~\cite{HB00},
where it was shown that the central charge in this case is~$c=m-1+n/2$ for~$m\ge1$. The same is
true for the~$\su(m|n)$ HS chain with~$m\ge1$, by virtue of the relation between the partition
functions of the supersymmetric PF and HS chains established in Ref.~\cite{BBS08}. The criticality
of the $\su(1|1)$ HS chain with a chemical potential was also proved in Ref.~\cite{CFGRT16}, where
it was shown that the central charge is instead~$c=1$ for a certain range of nonzero values of the
chemical potential. In particular, it should be noted that in all cases reviewed above the central
charge is integer or half-integer, as in a CFT of free bosons and/or fermions.

The aim of this paper is to study the thermodynamics and the critical behavior of the three
families of $\su(m|n)$ spin chains of HS type with a general chemical potential term. To this end,
we shall first evaluate in closed form the chains' partition functions for arbitrary finite values
of the number of sites~$N$. Exploiting the connection of the latter chains with a certain
inhomogeneous vertex model, we shall achieve a concise description of the spectrum in terms of a
generalization of Haldane's motifs. This description shall then be used to compute the transfer
matrix and obtain a closed-form expression for the free energy per site in the thermodynamic
limit. With the help of this expression, we shall study the thermodynamics and criticality of the
supersymmetric chains of HS type with $1\le m,n\le2$. First of all, examining the low-temperature
behavior of the free energy per site we shall determine the values of the chemical potentials for
which these chains are critical, and compute the corresponding central charge. In particular, it
turns out that the central charge can take rational values that are not half-integers. We shall
also analyze the existence of phase transitions at zero temperature in the densities of bosons and
fermions. We shall show that these densities exhibit only second-order (continuous) transitions
for $m=n=1$, while for $m+n>2$ (and~$1\le m,n\le2$) either the bosonic or the fermionic densities
undergo a first-order (discontinuous) phase transition.

We shall end this introduction by briefly outlining the paper's organization. In
Section~\ref{sec.models} we present the three supersymmetric chains of HS type under study, and
discuss their duality under exchange of the bosonic and fermionic degrees of freedom. The
partition function of these chains is computed in closed form in Section~\ref{sec.PF} by means of
Polychronakos's freezing trick. In Section~\ref{sec.VM} we establish the equivalence of~$\su(m|n)$
supersymmetric spin chains of HS type to certain inhomogeneous vertex models, from which we deduce
a simple formula for the spectrum in terms of supersymmetric motifs. By means of this formula, in
Section~\ref{sec.TD} we evaluate the chains' free energy per site in the thermodynamic limit. We
also discuss in this section several symmetries of the free energy and the main thermodynamic
functions, with particular emphasis on the one arising from the boson-fermion duality.
Sections~\ref{sec.su11}-\ref{sec.su22} are devoted to the analysis of the critical behavior and
the existence of zero-temperature phase transitions in the spin densities for the~$\su(1|1)$,
$\su(2|1)$ and~$\su(2|2)$ chains. In Section~\ref{sec.conc} we present our conclusions and discuss
several lines for future research suggested by the present work. The paper ends with a technical
appendix providing a detailed discussion of the behavior of the free energy per site of the
$\su(1|1)$ chains for finite values of~$N$.

\section{The models}\label{sec.models}

An $\su(m|n)$ supersymmetric spin chain is a one-dimensional array of $N$ sites, each of which is
occupied by either a boson or a fermion with $m$ and $n$ degrees of freedom, respectively. Thus
the Hilbert space $\Si^{(m|n)}=\otimes_{i=1}^N\CC^{m+n}$ of the system is spanned by the basis
vectors
\begin{equation}\label{basis}
  |s_1\cdots s_N\rangle\equiv|s_1\rangle\otimes\cdots\otimes|s_N\rangle,\qquad s_i\in\{1,\dots,m+n\},
\end{equation}
where the one-particle state~$\ket{s_i}$ is regarded as bosonic if $s_i\in B\equiv\{1,\dots,m\}$
and fermionic if $s_i\in F\equiv\{m+1,\dots,m+n\}$. The $\su(m|n)$ permutation operators $\Pij$
(with $1\le i<j\le N$) are defined by
\begin{equation}
  \label{Pij}
  \Pij\ket{\cdots s_i\cdots s_j\cdots }=\ep_{ij}(s_1,\dots,s_N)\ket{\cdots s_j\cdots s_i\cdots},
\end{equation}
where the sign~$\ep_{ij}(s_1,\dots,s_N)$ is equal to $1$ if $s_i,s_j\in B$, $-1$ if
$s_i,s_j\in F$, and~$(-1)^\nu$ if $s_i$ and $s_j$ are of different type, $\nu$ being the number of
fermionic spins $s_k$ with~$k=i+1,\dots,j-1$. We shall also define the number operators~$\cN_\al$
with $\al=1,\dots,m+n$ by
\[
  \cN_\al\ket{s_1\cdots s_N}=N_\al(\bbs)\ket{s_1\cdots s_N},
\]
where
\begin{equation}
  N_\al(\bbs)\equiv\sum_{i=1}^N\de_{s_i,\al}
  \label{Nal}
\end{equation}
is the number of spins of type~$\al$ in the
state~$\ket{s_1\cdots s_N}$. The supersymmetric spin chains we shall deal with in this paper are
described by a Hamiltonian of the form
\begin{equation}
  \label{H0H1}
  \cH=\sum_{i<j}J_{ij}(1-\Pij)-\sum_{\al=1}^{m+n-1}\mu_\al\cN_\al\equiv \cH_0+\cH_1,
\end{equation}
where (as in what follows, unless otherwise stated) the sum over Latin indices ranges from $1$ to
$N$. In the last term the real constant $\mu_\al$ has a natural interpretation as the chemical
potential of the $\al$-th species, and without loss of generality (since
$\sum_{\al=1}^{m+n}\cN_\al=N$) we have chosen the normalization $\mu_{m+n}=0$. The models we shall
focus on are those for which $\cH_0$ is the Hamiltonian of the supersymmetric Haldane--Shastry,
Polychronakos--Frahm and Frahm--Inozemtsev spin chains, whose interaction strengths~$J_{ij}$ are
respectively given by
\begin{eqnarray}
  J_{ij}&=\frac{J}{2\sin^2(\xi_i-\xi_j)},\qquad &\xi_k=\frac{k\pi}N\,,\label{HS}\\
  J_{ij}&=\frac{J}{(\xi_i-\xi_j)^2}\,,\qquad &H_N(\xi_k)=0\,,\label{PF}\\
  J_{ij}&=\frac{J}{2\sinh^2(\xi_i-\xi_j)}\,,\qquad & L_N^{c-1}(\e^{2\xi_k})=0\,.\label{FI}
\end{eqnarray}
Here~$J\ne0$ is a real constant, $H_N$ denotes the Hermite polynomial of degree $N$ and
$L_N^{c-1}$ is a generalized Laguerre polynomial of degree $N$ (with $c$ a positive parameter).

In the non-supersymmetric case ($mn=0$) the model~\eqref{H0H1} with
interactions~\eqref{HS}--\eqref{FI} is the one solved in Ref.~\cite{EFG12}. Indeed, in this case
$P_{ij}=\pm S_{ij}$, where $S_{ij}$ is the operator permuting the $i$-th and $j$-th spins and the
``$+$'' (resp.~``$-$'') sign corresponds to the case $n=0$ (resp.~$m=0$). Moreover, the operators
\[
  \cJ_\al=\cN_\al-\cN_{m+n}\,,\qquad\al=1,\dots,m+n-1\,,
\]
are a basis of the standard $\su(m+n)$ Cartan subalgebra, in terms of which~$\cH_1$ can be
expressed  as
\[
  \cH_1=\sum_{\al=1}^{m+n-1}B_\al\cJ_\al+B_0
\]
for suitable real constants~$B_\al$. It also worth mentioning that the Hamiltonian~\eqref{H0H1},
for which we shall use the more explicit notation~$\cH^{(m|n)}$, is related to $\cH^{(n|m)}$ by a
duality relation that we shall now briefly explain~\cite{BBHS07,BFGR09}. To this end, let us
define the unitary operator~$U:\Si^{(m|n)}\to\Si^{(n|m)}$ by
\[
  U\ket{s_1\cdots s_N}=(-1)^{\sum_i i\pi(s_i)}\ket{s_1'\cdots s_N'}\,,
\]
where $\pi(s_i)=0$ if $s_i\in B$ (resp.~$\pi(s_i)=1$ if $s_i\in F$) and $s_i'=m+n+1-s_i$. We then
have
\[
  U^{-1}P_{ij}^{(n|m)}U=-P_{ij}^{(m|n)}\,,\qquad U^{-1}\cN_\al U=\cN_{m+n+1-\al}\,,
\]
and therefore
\begin{equation}\label{E0def}
  U^{-1}\cH^{(n|m)}U=E_0-\cH^{(m|n)}\Big|_{\mu_\al\to-\mu_{m+n+1-\al}},\qquad E_0\equiv2\sum_{i<j}J_{ij}\,.
\end{equation}
Thus the spectra of $\cH^{(n|m)}$ and $\cH^{(m|n)}$ are related by
\begin{equation}\label{Ekmnnm}
  E_k^{(n|m)}(\mu_1,\dots,\mu_{m+n})=E_0-E_k^{(m|n)}(-\mu_{m+n},\dots,-\mu_1)\,.
\end{equation}
We can therefore assume without loss of generality that $m\ge n$ in what follows.

Another basic symmetry of the spectrum of the Hamiltonian~\eqref{H0H1} is related to changes in
the labeling of the bosonic or fermionic degrees of freedom. More precisely, let
$T_{\al\be}:\Si^{(m|n)}\to\Si^{(m|n)}$ (with~$\al\ne\be\in\{1,\dots,m+n\}$) denote the linear
operator whose action on a basis element~$\ket{s_1\cdots s_N}$ consists in replacing all the
$s_k$'s equal to~$\al$ by~$\beta$, and vice versa. If~$\pi(\al)=\pi(\be)$ (i.e., if $\al$
and~$\be$ are either both bosonic or both fermionic) it is clear that~$T_{\al\be}$ commutes with
the $\su(m|n)$ permutation operators $\Pij$, and hence with $\cH_0$. It is also straightforward to
verify that
\[
  \fl
  T_{\al\be}^{-1}\,\cN_\al\,T_{\al\be}=\cN_{\be}\,,\qquad T_{\al\be}^{-1}\,\cN_\be\,
  T_{\al\be}=\cN_{\al}\,,
  \qquad T_{\al\be}^{-1}\,\cN_\ga\,T_{\al\be}=\cN_{\ga} \quad(\ga\ne\al,\be),
\]
and thus
\[
  T_{\al\be}^{-1}\,\cH\,T_{\al\be}=\cH_0-\mu_\al\cN_\be-\mu_\be\cN_\al-\sum_{\substack{\ga=1\cr\ga\ne\al,\be}}^{m+n}\mu_\ga\cN_\ga\,.
\]
It follows that
\begin{equation}
  \label{specsymm}
  \fl
  E_k^{(m|n)}(\dots,\mu_\al,\dots,\mu_{\be},\dots)=E_k^{(m|n)}(\dots,\mu_\be,\dots,
  \mu_\al,\dots)\qquad
  (\pi(\al)=\pi(\be));
\end{equation}
in other words, the spectrum of~$\cH$ is invariant under permutations of the~bosonic or fermionic
chemical potentials among themselves. Note, finally, that combining Eqs.~\eqref{Ekmnnm}
and~\eqref{specsymm} we obtain the more general relation
\[
  E_k^{(n|m)}(\mu_1,\dots,\mu_{m+n})=E_0-E_k^{(m|n)}(-\mu_{\al_1},\dots,-\mu_{\al_{m+n}})\,,
\]
where~$(\al_1,\dots,\al_{m+n})$ is a permutation of~$(1,\dots,m+n)$ such
that~$\{\al_1,\dots,\al_m\}=\{n+1,\dots,n+m\}$ (and, consequently,
$\{\al_{m+1},\dots,\al_{m+n}\}=\{1,\dots,n\})$.

\section{Partition function}\label{sec.PF}

In this section we shall compute in closed form the partition function of the
chains~\eqref{H0H1}--\eqref{FI} by exploiting their connection with the $\su(m|n)$ spin
versions~\cite{Ha93,AK96,BB06,BUW99} of the dynamical models of Sutherland, Calogero and
Inozemtsev, respectively. For definiteness, we shall present the details of the calculation only
for the PF model~\eqref{H0H1}-\eqref{PF}, which is technically the simplest.

To begin with, recall that the Hamiltonian of the $\su(m|n)$ spin Calogero model is given by
\begin{equation}\label{H0PF}
  H_0=-\sum_{i}\partial_{x_i}^2+a^2r^2+\sum_{i\ne
    j}\frac{a\big(a-{P}^{(m|n)}_{ij}\big)}{(x_i-x_j)^2},\qquad r^2\equiv\sum_ix_i^2\,,
\end{equation}
with scalar counterpart
\begin{equation}\label{HscPF}
  \Hsc=-\sum_{i}\partial_{x_i}^2+a^2r^2+\sum_{i\ne
    j}\frac{a(a-1)}{(x_i-x_j)^2}.
\end{equation}
Defining
\begin{equation}\label{Hdef}
  H=H_0+\frac{2a}J\,\cH_1
\end{equation}
we then have
\begin{equation}\label{HHsch}
  H=\Hsc+\frac{2a}J\,h(\bx)\,,
\end{equation}
with~$\bx\equiv(x_1,\dots,x_N)$ and
\[
  h(\bx)=J\sum_{i<j}\frac{1-\Pij}{(x_i-x_j)^2}-\sum_{\al=1}^{m+n-1}\mu_\al\cN_\al\,.
\]
Hence the Hamiltonian~\eqref{H0H1} with interactions~\eqref{PF} is simply~$\cH=h(\bxi)$, where
$\bxi\equiv(\xi_1,\dots,\xi_N)$ and the~$\xi_k$'s are the chain sites (i.e., the zeros of the
Hermite polynomial of degree~$N$). From Eq.~\eqref{H0PF} it follows that in the limit $a\to\infty$
the eigenfunctions of the $\su(m|n)$ spin Hamiltonian~$H$ are sharply peaked at the coordinates of
the (unique) equilibrium of the scalar potential
\begin{equation}
  \label{U}
  U=r^2+\sum_{i\ne j}\frac{1}{(x_i-x_j)^2}
\end{equation}
in the configuration space ($A_{N-1}$ Weyl chamber)
\[
  A = \{\bx\in\RR^N|x_1<\cdots<x_N\}\,,
\]
which coincide with the chain sites~$\xi_k$~\cite{ABCOP79}. By Eq.~\eqref{HHsch} and the
relation~$\cH=h(\bxi)$, it follows that for large~$a$ the eigenvalues of~$H$ are approximately
given by
\[
  E_{ij}\simeq \Esc_i+\frac{2a}J\,e_j\,,
\]
where $\Esc_i$ and~$e_j$ respectively denote two arbitrary eigenvalues of~$\Hsc$ and $\cH$. From
the latter equation it is immediate to deduce the following exact formula relating the partition
functions $Z$, $\Zsc$ and $\mathcal Z$ of the three Hamiltonians $H$, $\Hsc$ and~$\cH$:
\begin{equation}\label{freezing}
  \mathcal{Z}(T)=\lim_{a\to\infty}\frac{Z(2aT/J)}{\Zsc(2aT/J)}.
\end{equation}
The argument just outlined leading to Eq.~\eqref{freezing} is known in the literature as
Polychronakos's freezing trick~\cite{Po94}.

The partition function~$\Zsc$ of the scalar Calogero model~\eqref{HscPF}, which is well known
(see, e.g., Refs.~\cite{Po94,BFGR08epl}), is given by
\begin{equation}\label{Zsc}
  \Zsc(2aT/J)
  =q^{\frac{J\EGS}{2a}}\prod_i(1-q^{Ji})^{-1},\qquad q\equiv\e^{-1/T},
\end{equation}
where
\[
  E_{\mathrm{GS}}=aN+a^2N(N-1)
\]
is the ground-state energy of both $H_0$ and~$\Hsc$, and we have taken Boltzmann's
constant~$k_{\mathrm B}$ as $1$. We shall next outline the computation of the spectrum of the
Hamiltonian $H$ in Eq.~\eqref{Hdef}. To this end, note first of all that, although the
Hamiltonians~$H_0$ and~$H$ are naturally defined on the Hilbert space~$L^2(A)\otimes\Si^{(m|n)}$,
they are actually equivalent to any of their extensions to the
space~$L^2(\RR^N)\otimes\Si^{(m|n)}$. This is essentially due to the fact that any
point~$\bx\in\RR^N$ outside the singular hyperplanes~$x_i-x_j=0$ can be mapped in a unique way to
a point in $A$ by an appropriate permutation. For reasons that will be clear in the sequel, from
now on we shall identify $H_0$ and $H$ with their \emph{symmetric} extension, defined on the
Hilbert space~$\La(L^2(\RR^N)\otimes\Si^{(m|n)})$. Here~$\La$ is the total symmetrizer with
respect to simultaneous permutations of both the coordinates and the spin variables, determined by
the relations
\begin{equation}\label{symm}
  K_{ij}{P}^{(m|n)}_{ij}\Lambda=\Lambda K_{ij}{P}^{(m|n)}_{ij}=\Lambda,\qquad 1\le i<j\le N\,,
\end{equation}
where $K_{ij}$ denotes the operator permuting the $i$-th and $j$-th coordinates. With the latter
identification, it can be shown that~$H$ is represented by an upper triangular matrix in an
appropriate basis that we shall now describe. To this end, let
\begin{equation}\label{ketns}
  \ket{\bn,\bbs}=\La\Big(\rho(\bx)\prod_ix_i^{n_i}\cdot\ket\bbs\Big),
\end{equation}
where~$\ket\bbs\equiv\ket{s_1\cdots s_N}$ and
\begin{equation}\label{rho}
  \rho(\bx)=\e^{-ar^2/2}\prod_{i<j}|x_i-x_j|^a.
\end{equation}
The states~\eqref{ketns}, partially ordered according to the total degree $|\bn|\equiv\sum_in_i$,
are a (non-orthonormal) basis of $\La(L^2(\RR^N)\otimes\Si^{(m|n)})$ provided that (for instance)
the quantum numbers $\bn$ and $\bbs$ satisfy the following conditions
\begin{enumerate}[i)]
\item ~$n_i\ge n_{i+1}$ for all $i=1,\dots,N-1$.
\item If $n_i=n_{i+1}$ then $s_i\le s_{i+1}$ for $s_i\in B$, or $s_i<s_{i+1}$ for $s_i\in F$.
\end{enumerate}
Indeed, note first of all that if $n_i=n_j$ and $s_i=s_j\in F$ then the state~\eqref{ketns}
vanishes by antisymmetry. Otherwise, acting with the permutation operators $K_{ij}P_{ij}$ on a
state of the form~\eqref{ketns} we can always obtain a state satisfying the first condition and
differing from the original one at most by a sign. Applying to this state permutations acting on
indices~$i,j$ such that $n_i=n_j$ we arrive at a state, again differing from the initial one by at
most a sign, in which the spin variables are ordered so that the second condition is also
satisfied. Moreover, it can be shown that the states satisfying the above two conditions are
linearly independent and complete.

Proceeding as in Ref.~\cite{BFGR08epl}, it is straightforward to show that the action of the spin
Hamiltonian~$H_0$ on the basis~\eqref{ketns} is given by
\begin{equation}\label{H0ketns}
  H_0\ket{\bn,\bbs}=E_{\bn,\bbs}^0\ket{\bn,\bbs}+\sum_{|\bn'|<|\bn|,\bbs'}c_{\bn'\bbs',\bn\bbs}\ket{\bn',\bbs'}
\end{equation}
with $c_{\bn'\bbs',\bn\bbs}\in\CC$ and
\begin{equation}
  \label{E0ns}
  E_{\bn,\bbs}^0=2a|\bn|+E_{\mathrm{GS}}\,.
\end{equation}
The matrix of $H_0$ on the basis~\eqref{ketns} is upper triangular, as claimed, and its spectrum
is given by Eq.~\eqref{E0ns}. On the other hand, since~$\cH_1$ clearly commutes with the
symmetrizer $\La$ (since each~$\cN_\al$ does) and
\[
  \cH_1\ket\bbs=-\sum_{\al=1}^{m+n-1}\mu_\al N_\al(\bbs)\ket\bbs=-\Big(\sum_i\mu_{s_i}\Big)\ket\bbs
\]
(with $\mu_{m+n}=0$), we have
\begin{equation}\label{H1ns}
  \cH_1\ket{\bn,\bbs}=-\Big(\sum_i\mu_{s_i}\Big)\ket{\bn,\bbs}\,.
\end{equation}
Thus~$\cH_1$ is diagonal in the basis~\eqref{ketns}, and by Eqs.~\eqref{H0ketns}-\eqref{H1ns} the
spectrum of~$H$ is given by
\begin{equation}
  \label{Ens}
  E_{\bn,\bbs}=2a|\bn|-\frac{2a}J\,\sum_i\mu_{s_i}+E_{\mathrm{GS}}\,,
\end{equation}
where the quantum numbers $\bn$ and $\bbs$ satisfy conditions i)-ii) above.

We are now ready to evaluate in closed form the partition function of the chain~\eqref{H0H1} using
the freezing trick formula~\eqref{freezing}. In the first place, in order to compute the partition
function of the $\su(m|n)$ spin model~\eqref{Hdef} it is convenient to parametrize the
multiindex~$\bn$ satisfying condition~i) above as
\begin{equation}\label{bnks}
  \bn=(\underbrace{\nu_1,\dots,\nu_1}_{k_1},\dots,\underbrace{\nu_r,\dots,\nu_r}_{k_r}),
\end{equation}
where~$\nu_1>\cdots>\nu_r\ge0$ and $k_1+\cdots+k_r=N$ with $k_i>0$ for all $i$. In particular,
note that the vector~$\bk=(k_1,\dots,k_r)$ can be considered as an element of the set $\cP_N$ of
ordered partitions of the integer $N$. We shall also refer in what follows to each maximal group
$(\nu_i,\dots,\nu_i)$ of repeated components of the multiindex~$\bn$ as a \emph{sector}. From
Eq.~\eqref{Ens} for the spectrum of~$H$ it immediately follows that the partition function~$H$ is
given by
\begin{equation}\label{Z}
  Z(2aT/J)=q^{\frac{J\EGS}{2a}}\sum_{\mathclap{\mathbf{k}\in\cP_N}}\qquad\sum_{\mathclap{\nu_1>\cdots>\nu_r\geq 0}}q^{\sum\limits_{i=1}^{r}Jk_i\nu_i}\sum_{\mathbf{\bbs}\in\bn}q^{-\sum_j\mu_{s_j}},
\end{equation}
where the notation~$\bbs\in\bn$ stands for all possible multiindices~$\bbs\in\{1,\dots,m+n\}^N$
satisfying condition~ii) above for a given multiindex~$\bn$. Let us next evaluate the inner sum in
Eq.~\eqref{Z}
\begin{equation}
  \label{Sigmak}
  \Si(\bk)\equiv\sum_{\mathbf{\bbs}\in\bn}q^{-\sum_j\mu_{s_j}},
\end{equation}
which clearly depends on~$\bn$ only through~$\bk$. Since condition~ii) above effects only the
ordering of the spin variables $s_k$ within each sector of the multiindex~$\bn$ independently of
the others, we have
\begin{equation}
  \label{Sigmakprod}
  \Si(\bk)\equiv\prod_{i=1}^r\si(k_i)\,,
\end{equation}
where
\begin{equation}
  \si(k)=\sum_{i+j=k}\qquad
  \sum_{\mathclap{1\le s_1\le\cdots\le s_i\le m}} q^{-\sum\limits_{l=1}^{i}\mu_{s_l}}\kern.75em
  \sum_{\mathclap{1\le l_1<\cdots<l_j\le n}} q^{-\sum\limits_{p=1}^{j}\mu_{m+l_p}}
  \label{sik}
\end{equation}
is the contribution to~$\Si(\bk)$ of a sector of length~$k$ (with $i$ bosons and $j$ fermions).
The sum~$\si(k)$ is easily expressed in terms of the complete and elementary symmetric polynomials
$h_i(x_1,\dots,x_m)$ and $e_j(x_1,\dots,x_n)$ of degrees $i$ and~$j$, respectively defined by
\[
  \fl
  h_i(x_1,\dots,x_m)\equiv\sum_{p_1+\cdots+p_m=i}x_1^{p_1}\cdots x_m^{p_m}\,,\qquad
  e_j(x_1,\dots,x_n)\equiv\sum_{1\le l_1<\cdots<l_j\le n}x_{l_1}\cdots x_{l_j}\,,
\]
where it is understood that $e_j(x_1,\dots,x_n)=0$ for $j>n$. Recall that $h_i$
(respectively~$e_j$) is nothing but the Schur polynomial associated to the partition $(i)$
(resp.~$(1^j)$). We shall also need in the sequel the supersymmetric elementary polynomial of
degree~$k$ in $m$ bosonic and $n$ fermionic variables, defined by
\[
  e_k(x_1,\dots,x_m|y_1,\dots,y_n)=\sum_{i+j=k}h_i(x_1,\dots,x_m)e_j(y_1,\dots,y_n).
\]
We then have
\begin{eqnarray}
  \sum_{\mathclap{1\le s_1\le\cdots\le s_i\le m}} q^{-\sum\limits_{l=1}^{i}\mu_{s_l}}
  &=\sum_{p_1+\cdots+p_m=i}\,\prod_{l=1}^mq^{-p_l\mu_l}=h_i(q^{-\mu_1},\dots,q^{-\mu_m}),\\
  \sum_{\mathclap{1\le l_1<\cdots<l_j\le n}}
  q^{-\sum\limits_{p=1}^{j}\mu_{m+l_p}}&=e_j(q^{-\mu_{m+1}},\dots,q^{-\mu_{m+n}}),
                                          \label{sumej}
\end{eqnarray}
where (as usual) $\mu_{m+n}=0$. Thus
\begin{eqnarray*}
  \si(k)&=\sum_{i+j=k}h_i(q^{-\mu_1},\dots,q^{-\mu_m})e_j(q^{-\mu_{m+1}},\dots,q^{-\mu_{m+n}})\\
  &=e_k(q^{-\mu_1},\dots,q^{-\mu_m}|q^{-\mu_{m+1}},\dots,q^{-\mu_{m+n}}),
\end{eqnarray*}
and therefore, by Eq.~\eqref{Sigmakprod},
\begin{eqnarray}
  \Si(\bk)&=\prod_{i=1}^re_{k_i}(q^{-\mu_1},\dots,q^{-\mu_m}|q^{-\mu_{m+1}},\dots,q^{-\mu_{m+n}})\nonumber\\
          &\equiv E_{\bk}(q^{-\mu_1},\dots,q^{-\mu_m}|q^{-\mu_{m+1}},\dots,q^{-\mu_{m+n}}).
        \label{Sikfinal}
\end{eqnarray}
On the other hand, the change of variables $\nu_i-\nu_{i+1}=l_i$ (with $i=1,\dots,r$ and
$\nu_{r+1}\equiv0$) easily yields
\begin{equation}\label{Kjdef}
  \sum_{i=1}^{r}Jk_i\nu_i=J\sum_{i=1}^rk_i\sum_{j=i}^rl_j=J\sum_{j=1}^rl_jK_j,\qquad
  K_j\equiv\sum_{i=1}^jk_i\,,
\end{equation}
and hence
\begin{eqnarray*}
  \sum_{\mathclap{\nu_1>\cdots>\nu_r\geq 0}}q^{\sum\limits_{i=1}^{r}Jk_i\nu_i}
  &=
    \sum_{l_1,\dots,l_{r-1}>0,l_r\ge0}\,\prod_{j=1}^rq^{JK_jl_j}=\prod_{j=1}^{r-1}\sum_{l_j=1}^\infty
    q^{JK_jl_j}\cdot
    \sum_{l_r=0}^\infty q^{JK_rl_r}\\
  &=\prod_{j=1}^{r-1}\frac{q^{JK_j}}{1-q^{JK_j}}\cdot\frac1{1-q^{JK_r}}\,.
\end{eqnarray*}
From the latter equality and Eq.~\eqref{Z} we thus obtain
\begin{equation}\label{Zfinal}
  Z(2aT/J)=q^{\frac{J\EGS}{2a}}\sum_{\mathclap{\mathbf{k}\in\cP_N}}
  \Si(\bk)q^{\sum\limits_{i=1}^{r-1}\!JK_i}\prod_{i=1}^r(1-q^{JK_i})^{-1},
\end{equation}
with~$\Si(\bk)$ given by Eq.~\eqref{Sikfinal}. Using Eq.~\eqref{Zsc} and the freezing trick
formula~\eqref{freezing} we finally arrive at the following closed-form expression for the
partition function of the $\su(m|n)$ chain~\eqref{H0H1} with interactions~\eqref{PF}:
\begin{equation}
  \label{cZPF}
  \cZ(T)=\sum_{\mathclap{\mathbf{k}\in\cP_N}}
  \Si(\bk)q^{\sum\limits_{i=1}^{r-1}\!JK_i}\prod_{i=1}^{N-r}(1-q^{JK_i'}),
\end{equation}
where $K_i$ is given by Eq.~\eqref{Kjdef} and the integers~$K'_1<\cdots<K'_{N-r}$ are defined by
\begin{equation}
  \label{Kpi}
  \{K'_1,\dots,K'_{N-r}\}=\{1,\dots,N-1\}\setminus\{K_1,\dots,K_{r-1}\}.
\end{equation}
The above procedure can be repeated with minor modifications for the $\su(m|n)$
chains~\eqref{H0H1} with interactions~\eqref{HS} or~\eqref{FI}. It turns out that the resulting
formula for the partition function can be written in a unified way for all three
models~\eqref{HS}-\eqref{FI} as
\begin{equation}
  \label{cZall}
  \cZ(T)=\sum_{\mathclap{\mathbf{k}\in\cP_N}}
  \Si(\bk)q^{\sum\limits_{i=1}^{r-1}\!J\cE(K_i)}\prod_{i=1}^{N-r}(1-q^{J\cE(K_i')}),
\end{equation}
where the \emph{dispersion relation}~$\cE$ is given by
\begin{equation}\label{disp}
  \cE(i)=\cases{i(N-i),& for the HS chain\\
  i,&for the PF chain\\i(i+c-1),&for the FI chain.}
\end{equation}

\section{Associated vertex models}\label{sec.VM}

The Hamiltonian~$\cH_0$ in Eq.~\eqref{H0H1} is closely related to an inhomogeneous classical
vertex model first introduced in Ref.~\cite{BBH10} that we shall now briefly describe. The model
consists of a one-dimensional array of $N+1$ vertices joined by $N$ bonds $\si_i$, each of which
can be in $m+n$ states~$\{1,\dots,m\}\equiv B$ and $\{m+1,\dots,m+n\}\equiv F$. Thus a
configuration of the system is specified by a vector $\bsi=(\si_1,\dots,\si_N)$, with
$\si_i\in B\cup F$. The energy of such a configuration is then defined by
\begin{equation}\label{Emn}
  E^{(m|n)}(\bsi)=J\sum_{i=1}^{N-1}\de(\si_i,\si_{i+1})\cE(i),
\end{equation}
where
\begin{equation}
  \label{delta}
  \de(i,j)=\cases{1,&$i>j$ or\en $i=j\in F$\\
  0,&$i<j$\en or $i=j\in B$.}
\end{equation}
The authors of Ref.~\cite{BBH10} introduced the so-called generalized partition
function\footnote{From now on, with a slight abuse of notation we shall regard~$\cZ$ as a function
  of the variable~$q=\e^{-1/T}$ instead of the temperature~$T$.}
\begin{equation}\label{cZV}
  \cZ^V(q;\bx|\by)\equiv\sum_{\si_1,\dots,\si_N=1}^{m+n}\prod_{\alpha=1}^{m}x_{\al}^{N_{\al}(\bsi)}\cdot
  \prod_{\be=1}^{n}y_{\be}^{N_{m+\be}(\bsi)}\cdot q^{E^{(m|n)}(\bsi)}\,,
\end{equation}
with~$N_\al$ given by Eq.~\eqref{Nal}, in terms of which the partition function of the vertex
model with energies~\eqref{Emn} is simply
\begin{equation}
  \label{cZVxy}
  \cZ^V(q)=\cZ^V(q;1^m|1^n).
\end{equation}
As shown in Ref.~\cite{BBH10}, this generalized partition function satisfies the remarkable
identity
\begin{equation}
  \label{cZVid}
  \cZ^V(q;\bx|\by)=\sum_{\bk\in\cP_N}S_{\bk}(\bx|\by)q^{\sum\limits_{i=1}^{r-1}\!J\cE(K_i)}
\end{equation}
for all $\bx\in\RR^m$, $\by\in\RR^n$, where $S_{\bk}(\bx|\by)$ is the super Schur polynomial
associated to the border strip~$\langle k_1,\dots,k_r\rangle$ (see, e.g., Ref.~\cite{BBHS07}). In
the latter reference it is also shown that the RHS of Eq.~\eqref{cZVid} can be alternatively
expressed as
\begin{equation}
  \label{SEid}
  \sum_{\bk\in\cP_N}S_{\bk}(\bx|\by)q^{\sum\limits_{i=1}^{r-1}\!J\cE(K_i)}
  =\sum_{\bk\in\cP_N}E_{\bk}(\bx|\by)q^{\sum\limits_{i=1}^{r-1}\!J\cE(K_i)}
  \prod_{i=1}^{N-r}(1-q^{J\cE(K'_i)})\,.
\end{equation}
Combining the last two equations we obtain the identity
\begin{equation}
  \label{cZVfinal}
  \cZ^V(q;\bx|\by)=\sum_{\bk\in\cP_N}E_{\bk}(\bx|\by)q^{\sum\limits_{i=1}^{r-1}\!J\cE(K_i)}
  \prod_{i=1}^{N-r}(1-q^{J\cE(K'_i)}),
\end{equation}
valid for \emph{arbitrary} $\bx\in\RR^m$, $\by\in\RR^n$.

Equations~\eqref{cZV} and \eqref{cZVfinal} immediately yield a strikingly simple description of
the spectrum of the chain~\eqref{H0H1} with interactions~\eqref{HS}--\eqref{FI} akin to
Eq.~\eqref{Emn}. Indeed, taking into account Eq.~\eqref{Sikfinal} we immediately have
\begin{eqnarray}
  \cZ(q)&=\cZ^V(q;q^{-\mu_1},\dots,q^{-\mu_m}|q^{-\mu_{m+1}},\dots,q^{-\mu_{m+n}})\nonumber\\
        &
          =\sum_{\si_1,\dots,\si_N=1}^{m+n}q^{E^{(m|n)}(\bsi)-\sum\limits_{\al=1}^{m+n}\mu_\al N_\al(\bsi)}
          =\sum_{\si_1,\dots,\si_N=1}^{m+n}q^{E^{(m|n)}(\bsi)-\sum_{i}\mu_{\si_i}}.
          \label{ZZV}
\end{eqnarray}
Thus the spectrum of the HS-type chains~\eqref{H0H1}--\eqref{FI} is given by
\begin{equation}\label{specH}
  E(\bsi)=E^{(m|n)}(\bsi)-\sum_{i}\mu_{\si_i}=J\sum_{i=1}^{N-1}\de(\si_i,\si_{i+1})\cE(i)
  -\sum_{i}\mu_{\si_i}\,,
\end{equation}
where~$\bsi\in\{1,\dots,m+n\}^N$. In fact, the vectors $\bde(\bsi)\in\{0,1\}^{N-1}$ with
components $\de_k(\bsi)=\de(\si_k,\si_{k+1})$ are essentially the supersymmetric version of the
celebrated \emph{motifs} introduced by Haldane et al.~\cite{HHTBP92}. Equation~\eqref{specH} will
be the starting point for the evaluation of the thermodynamic functions of the chain~\eqref{H0H1}
in the next section.

\section{Thermodynamics}\label{sec.TD}

\subsection{Computation of the free energy}

The first step in the computation of the thermodynamic functions of the
chains~\eqref{H0H1}--\eqref{FI} is to suitably normalize their Hamiltonians so that the mean
energy per site tends to a finite limit as $N\to\infty$. To this end, note that
\[
  \tr P_{ij}^{(m|n)}=(m+n)^{N-2}(m-n)\,,\qquad \tr\cN_\al=N(m+n)^{N-1}\,,
\]
and hence the mean energy of the Hamiltonian~$\cH$ is given by
\[
  \mu=\frac{\tr\cH}{(m+n)^N}=
  \bigg(1-\frac{m-n}{(m+n)^2}\bigg)\sum_{i<j}J_{ij}-\frac{N}{m+n}\sum_{\al=1}^{m+n}\mu_\al\,.
\]
The sum~$\sum_{i<j}J_{ij}$ is easily evaluated by observing that it is half the maximum energy of
the Hamiltonian $\cH_0$ in the purely fermionic case~$m=0$, so that by Eq.~\eqref{specH} with
$\mu_\al=0$ for all~$\al$ we have
\begin{equation}\label{JijcE}
  \sum_{i<j}J_{ij}=\frac J2\sum_{i=1}^{N-1}\cE(i)
\end{equation}
and therefore
\begin{equation}
  \label{mufinal}
  \mu=\frac J2\,\bigg(1-\frac{m-n}{(m+n)^2}\bigg)\sum_{i=1}^{N-1}\cE(i)
  -\frac{N}{m+n}\sum_{\al=1}^{m+n}\mu_\al\,.
\end{equation}
Using Eq.~\eqref{disp} we immediately obtain
\[
  \sum_{i=1}^{N-1}\cE(i)=\cases{\frac N6(N^2-1),& for the HS chain\\
  \frac N2(N-1),&for the PF chain\\\frac N6(N-1)(2N+3c-4),&for the FI chain.}
\]
Thus the mean energy per site will tend to a constant in the thermodynamic limit $N\to\infty$
provided that the constant $J$ scales as
\begin{equation}\label{Kdef}
  J=\cases{\frac K{N^2},&for the HS and FI chains\\
    \frac KN,&for the PF chain,}
\end{equation}
where~$K$ is a real (positive or negative) constant independent of~$N$ and we have assumed that
$\lim_{N\to\infty}c/N\equiv\ga\ge0$ is finite. With this choice of $J$ we can write
\begin{equation}\label{JcEi}
  J\cE(i)=K\vep(x_i)\,,\qquad x_i\equiv\frac iN,
\end{equation}
where~$\vep(x)$ is given by
\begin{equation}\label{vepdef}
  \vep(x)=\cases{x(1-x),& for the HS chain\\ x,& for the PF chain\\ x(\ga_N+x),& for the FI chain}
\end{equation}
and we have defined~$\ga_N=(c-1)/N$. Since we shall be mainly interested in what follows in the
thermodynamic limit, from now on we shall implicitly assume that~$\ga_N$ has been replaced by its
limit~$\ga$.

Equation~\eqref{specH} for the spectrum of the chain Hamiltonian~\eqref{H0H1}-\eqref{FI}, which by
Eq.~\eqref{JcEi} can be written as
\begin{equation}
  \label{Esitransfer}
  E(\bsi)=\sum_{i=1}^{N-1}\Big[K\de(\si_i,\si_{i+1})\vep(x_i)-\frac12(\mu_{\si_i}+\mu_{\si_{i+1}})\Big]
  -\frac12(\mu_{\si_1}+\mu_{\si_{N}}),
\end{equation}
makes it possible to evaluate in closed form the free energy per site in the thermodynamic
limit by the transfer matrix method. To this end, note that from the latter equation we have
\begin{equation}\label{cZtr}
  \cZ(q)=\tr\big[A(x_0)A(x_1)\cdots A(x_{N-1})\big]\,,
\end{equation}
where $A(x)$ is the $(m+n)\times(m+n)$ matrix with entries
\begin{equation}\label{Adef}
  A_{\al\be}(x)=q^{K\vep(x)\de(\al,\be)-\frac12(\mu_\al+\mu_\be)}\,.
\end{equation}
Let~$J(x)$ denote Jordan canonical form of~$A(x)$, so that
\[
  A(x)=P(x)J(x)P(x)^{-1}
\]
for a suitable invertible matrix~$P(x)$. This matrix is of course not unique, but should be chosen
in such a way that it is a smooth function of the variable~$x\in[0,1]$. Writing, for simplicity,
\[
  A_i\equiv A(x_i)\,,\quad J_i\equiv J(x_i)\,,\quad P_i\equiv P(x_i)
\]
we then have
\[
  \cZ(q)=\tr\big[P_0J_0(P_0^{-1}P_1)J_1\cdots (P_{N-2}^{-1}P_{N-1})J_{N-1}P_{N-1}^{-1}\big]\,.
\]
On the other hand, from the smoothness of the matrix $P(x)$ it follows that
\begin{eqnarray}
  P_{i+1}&\equiv P(x_{i+1})=P(x_i)+(x_{i+1}-x_i)P'(x_i)+\oo(x_{i+1}-x_i)\\
         &\equiv
           P_i+\frac1N\,P'(x_i)+\oo(N^{-1})=P_i+\Or(N^{-1}).
\end{eqnarray}
Thus
\[
  P_i^{-1}P_{i+1}=\id+\Or(N^{-1}),
\]
and the dominant contribution to the free energy per spin~$f(T)\equiv-(T/N)\log\cZ(q)$ in the
thermodynamic limit is given by
\[
  \fl
  f(T)\simeq-\frac TN\log\tr(UJ_0\cdots J_{N-1})\,,\qquad U\equiv
  \lim_{N\to\infty}P_{N-1}^{-1}P_0=P(1)^{-1}P(0).
\]
We shall assume at this point that the matrix~$J_0\cdots J_{N-1}$ is \emph{diagonal}. In fact, it
suffices that $J_1\cdots J_{N-1}$ be diagonal, since $A_0$ is symmetric and therefore~$J_0$ is
diagonal. If this is the case, denoting by~$\la_\al(x)$ ($\al=1,\dots,m+n$) the eigenvalues
of~$A(x)$ and defining
\[
  \La_\al=\prod_{i=0}^{N-1}\la_\al(x_i)
\]
we have
\[
  \tr(UJ_0\cdots J_{N-1})=\sum_{\al=1}^{m+n}U_{\al\al}\La_\al\,.
\]
Since all the entries of~$A(x)$ are strictly positive, by the Perron--Frobenius theorem this
matrix has a positive and simple eigenvalue, that we shall take as~$\la_1(x)$, satisfying
\[
  \la_1(x)>|\la_\al(x)|\,,\qquad \all\al>1\,.
\]
(In particular, it is understood that the matrix~$P(x)$ must be chosen so that its first column is
an eigenvector corresponding to the Perron--Frobenius eigenvalue.) This is readily seen to imply
that
\[
  \lim_{N\to\infty}\frac{|\La_\al|}{\La_1}=0\,,\qquad\forall\al>1\,.
\]
Indeed, we have
\[
  \log\biggl(\frac{|\La_\al|}{\La_1}\biggr)
  =N\cdot\frac1N\sum_{i=0}^{N-1}\log\biggl(\frac{|\la_\al(x_i)|}{\la_1(x_i)}\biggr)\,,
\]
with
\[
  \frac1N\sum_{i=0}^{N-1}\log\biggl(\frac{|\la_\al(x_i)|}{\la_1(x_i)}\biggr)
  \underset{N\to\infty}{\longrightarrow}
  \int_0^1\log\biggl(\frac{|\la_\al(x)|}{\la_1(x)}\biggr)\diff x.
\]
Since the integrand is everywhere negative, the latter integral is a negative number or
$-\infty$. In either case
\[
\lim_{N\to\infty}\log\biggl(\frac{|\La_\al|}{\La_1}\biggr)=-\infty,
\]
which is equivalent to our claim. It follows that when~$N\gg1$ we have
\[
  \tr(UJ_0\cdots J_{N-1})\simeq U_{11}\La_1\equiv U_{11}\prod_{i=0}^{N-1}\la_1(x_i),
\]
provided only that $U_{11}\ne0$. If this is the case, the free energy per site in the
thermodynamic limit is given by
\begin{equation}\label{fTmu}
  f(T)=
  -T\lim_{N\to\infty}\frac1N\sum_{i=0}^{N-1}\log\la_1(x_i)=-T\int_0^1\log\la_1(x)\,\diff x\,.
\end{equation}
The latter formula, which is valid for the three types of chains~\eqref{HS}--\eqref{FI} (and
actually for any model whose energies are given by an equation of the
form~\eqref{specH}-\eqref{JcEi}), is the main result of the paper. Recall that for the validity of
Eq.~\eqref{fTmu} we have made the following two assumptions, which will be explicitly checked in
each of the examples to which we shall apply it in the next section:
\begin{enumerate}[i)]
\item The matrix~$J_1\cdots J_{N-1}$ is diagonal
\item $U_{11}\ne0$
\end{enumerate}
The last condition can in fact be somewhat simplified, as we shall next explain. Indeed, note
first of all that
\[
  A_{\al\be}(0)=v_\al v_\be\,, \qquad v_\al\equiv q^{-\mu_\al/2}.
\]
It follows that the eigenvalues of $A(0)$ are $\bv^2>0$, with corresponding eigenspace spanned by
$\bv\equiv(v_1,\dots,v_{m+n})$, and $0$, whose eigenspace is the orthogonal complement
of~$\RR\bv_1$. Hence in this case the Perron--Frobenius eigenvalue is $\la_1(0)=\bv^2$, and
$P_{\al1}(0)= \kappa v_\al\equiv\kappa q^{-\mu_\al/2}$ for some non-vanishing constant~$\ka$. We
thus have
\[
  U_{11}=\sum_{\al=1}^{m+n}\big[P(1)^{-1}\big]_{1\al}P_{\al 1}(0)=\ka
  \sum_{\al=1}^{m+n}\big[P(1)^{-1}\big]_{1\al}q^{-\frac12\mu_\al},
\]
and therefore the condition~$U_{11}\ne0$ is equivalent to
\begin{equation}
  \label{U11cond}
  \sum_{\al=1}^{m+n}\big[P(1)^{-1}\big]_{1\al}q^{-\frac12\mu_\al}\ne0.
\end{equation}
This condition can be alternatively expressed as
\begin{equation}\label{U11condalt}
  \sum_{\al=1}^{m+n}(-1)^{\al+1}M_{\al1}(1)q^{-\frac12\mu_\al}\ne0,
\end{equation}
where~$M_{\al\be}(x)$ is the cofactor of~$P_{\al\be}(x)$. It is also worth mentioning that
Eq.~\eqref{U11cond} is automatically satisfied for the $\su(m|n)$ HS
chain~\eqref{H0H1}-\eqref{HS}. Indeed, in this case~$\vep(1)=0$ implies that~$P(1)=P(0)$, and
hence
\[
  U_{11}=\sum_{\al=1}^{m+n}\big[P(0)^{-1}\big]_{1\al}P_{\al 1}(0)=1\,.
\]
Finally, it is important to note for the sequel that in the genuinely supersymmetric case $mn\ne0$
the matrix $A(x)$ always has a zero eigenvalue, since its first and last rows are proportional.
Indeed, in this case we have
\[
\de(1,\al)=0\,,\quad \de(m+n,\al)=1\,,\qquad \al=1,\dots,m+n,
\]
so that
\[
A_{m+n,\al}(x)=q^{K\vep(x)-\frac12(\mu_{m+n}+\mu_\al)}=q^{K\vep(x)-\frac12(\mu_{m+n}-\mu_1)}A_{1\al}(x)\,.
  \]
Thus, for fixed $m+n$, the  genuinely supersymmetric models are easier to treat than their
non-supersymmetric counterparts.

\begin{figure}[h]
  \includegraphics[width=.5\textwidth]{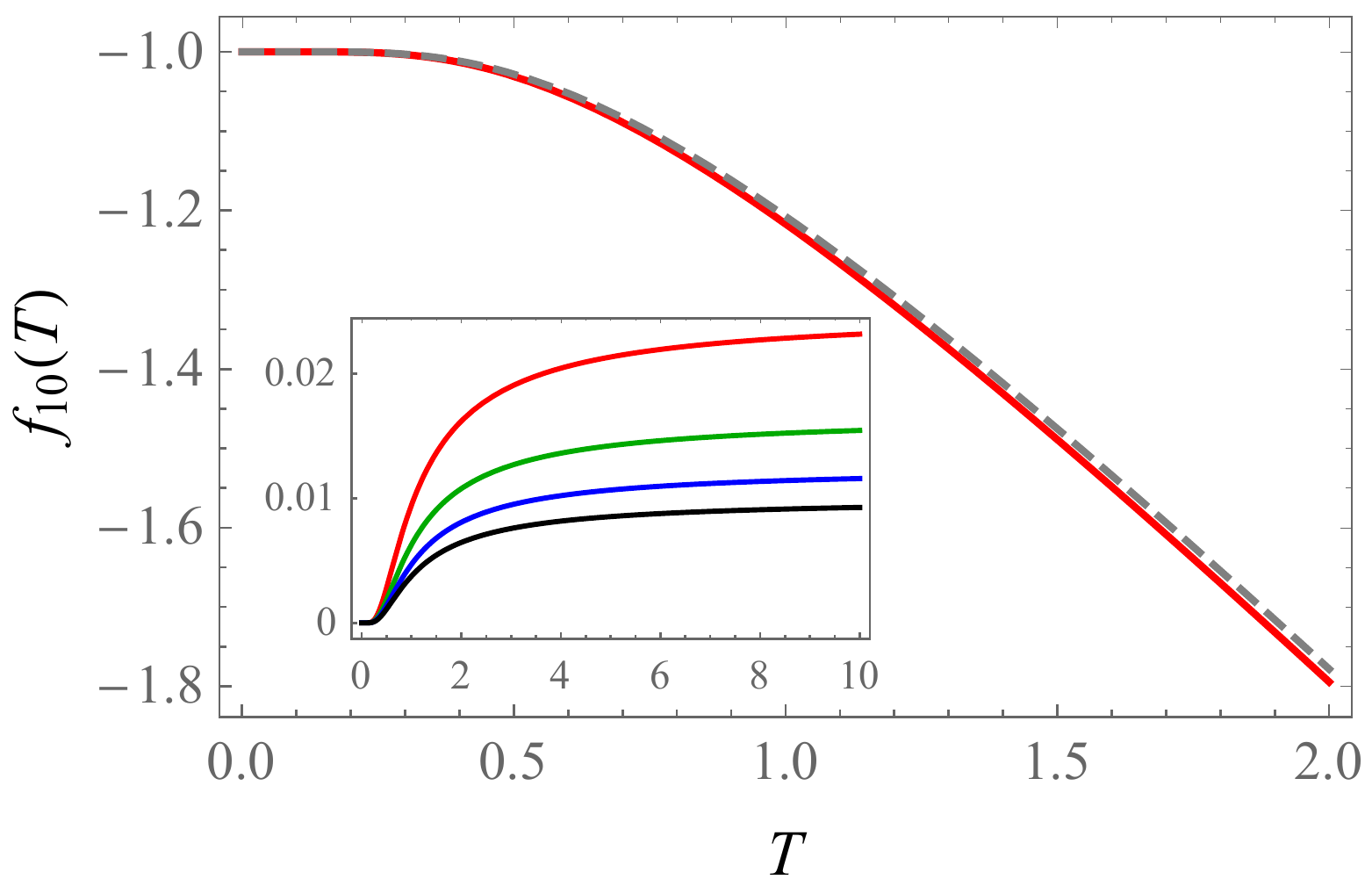}\hfill
  \includegraphics[width=.5\textwidth]{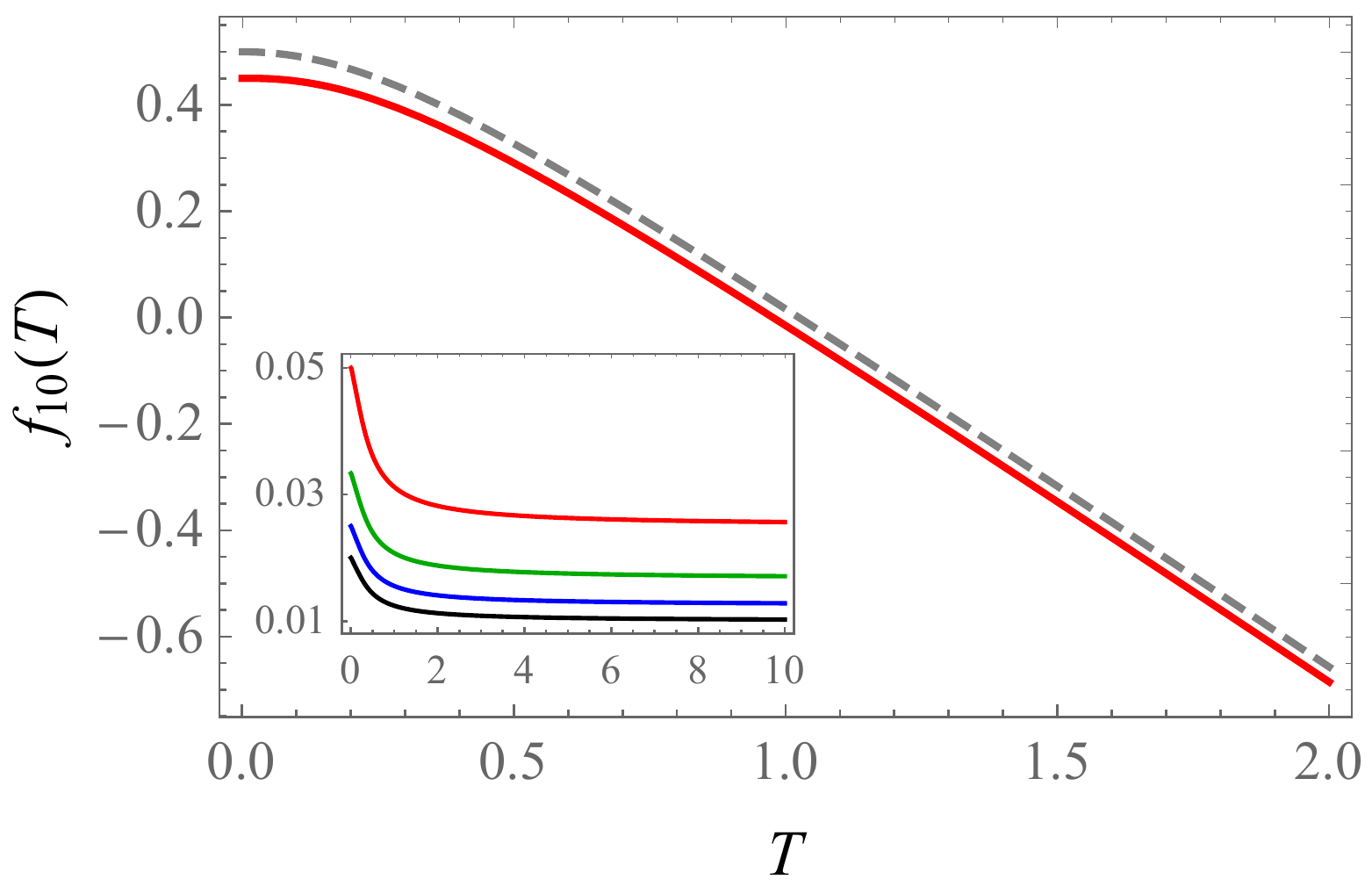}
  \caption{Left: free energy per site of the $\su(1|1)$ PF chain with~$\mu_1\equiv\mu=K>0$
    for~$N=10$ spins, $f_{10}(T)$, as a function of $T$ (solid red line) compared to its
    thermodynamic limit computed via Eq.~\eqref{fTmu} (dashed gray line). Right: same plot
    for~$\mu=-K$. Insets: difference~$f(T)-f_N(T)$ for $N=10$ (red), $15$ (green), $20$ (blue)
    and~$25$ (black) spins in the range~$0\le T\le 10$. Note: in all plots, $f_N$, $f$ and~$T$ are
    measured in units of~$K$.}
  \label{fig.fNmu}
\end{figure}

We have numerically verified that the $N=\infty$ exact equation~\eqref{fTmu} for the free energy
per site of the three $\su(m|n)$ chains of HS type provides a good approximation to its finite $N$
counterpart~$f_N(T)\equiv-(T/N)\log\cZ(q)$ for $N$ as low as $20$. For instance, in
Fig.~\ref{fig.fNmu} we have compared $f(T)$ in Eq.~\eqref{fTmu} with~$f_N(T)$ for the~$\su(1|1)$
PF chain with $N=10,15,20,25$, $K>0$ and~$\mu_1\equiv\mu=\pm K$. It is apparent that the
error~$|f(T)-f_N(T)|$ is quite small even for~$N\sim10$, and decreases steadily as $N$ increases.
It may at first seem surprising that at low temperatures this error is noticeably larger
for~$\mu=-K$ than for~$\mu=K$. A detailed explanation of this fact, which is essentially due to
the different behavior of the ground state energy for positive and negative values of~$\mu$, is
presented in the Appendix.

\subsection{Symmetries of the free energy}

We shall next deduce several symmetry properties of the free energy per site~$f$ of the
Hamiltonian~\eqref{H0H1}--\eqref{FI} stemming from Eqs.~\eqref{Ekmnnm} and~\eqref{specsymm}, which
shall be frequently applied in the following sections. To this end, in the rest of this section we
shall drop the temperature dependence but otherwise use the more descriptive
notation~$f(\mu_1,\dots,\mu_{m+n-1};K)$ for the free energy per site. Note, in particular, that
the latter notation underscores the fact that we have chosen~$\mu_{m+n}=0$ in Eq.~\eqref{H0H1}. It
is of interest for the sequel to determine how would a different ``normalization'' of the chemical
potentials like, e.g., $\mu_1=0$ affect the free energy per site. To see this, let
\[
  \hat\cH(\hat\mu_2,\dots,\hat\mu_{m+n})=\cH_0-\sum_{\al=2}^{m+n}\hat\mu_\al\cN_\al,
\]
and denote by~$\hat f(\hat\mu_2,\dots,\hat\mu_{m+n};K)$ the corresponding free energy per site (in
the thermodynamic limit). Using the identity
\[
  \cN_{m+n}=N-\sum_{\al=1}^{m+n-1}\cN_\al
\]
it is straightforward to deduce that
\[
  \hat\cH(\hat\mu_2,\dots,\hat\mu_{m+n})=\cH(\mu_1,\dots,\mu_{m+n-1})-N\hat\mu_{m+n},
\]
where $\cH(\mu_1,\dots,\mu_{m+n-1})$ is given by Eq.~\eqref{H0H1} with $\mu_{m+n}=0$ and the
chemical potentials~$\mu_\al,\hat\mu_{\al}$ are related by
\begin{equation}\label{muhatmu}
  \mu_\al=\hat\mu_\al-\hat\mu_{m+n}\quad(\hat\mu_1=0)\en\iff\en
  \hat\mu_\al=\mu_\al-\mu_1\quad(\mu_{m+n}=0).
\end{equation}
It then follows that
\begin{equation}
  \label{fmu10}
  \fl
\hat f(\hat\mu_2,\dots,\hat\mu_{m+n};K)=f(\mu_1,\dots,\mu_{m+n-1};K)-\hat\mu_{m+n}=
f(\mu_1,\dots,\mu_{m+n-1};K)+\mu_1,
\end{equation}
with~$\mu_\al$ and~$\hat\mu_\al$ related by Eq.~\eqref{muhatmu}. Consider next Eq.~\eqref{Ekmnnm}, which relates the spectra of~$\cH^{(m|n)}$ and~$\cH^{(n|m)}$.
Dividing by~$N$ and letting~$N\to\infty$ we obtain the relation
\[
  f^{(n|m)}(\mu_1,\dots,\mu_{m+n-1};K)=\lim_{N\to\infty}\frac{E_0}N+\hat
  f^{(m|n)}(\mu_{m+n-1},\dots,\mu_1;-K),
\]
where~$f^{(n|m)}$ and~$\hat f^{(m|n)}$ respectively denote the free energy per site of the
Hamiltonians~$\cH^{(n|m)}$ (with the usual choice~$\mu_{m+n}=0$) and~$\hat\cH^{(m|n)}$
(with~$\hat\mu_1=0$). From Eqs.~\eqref{E0def}, \eqref{JijcE} and~\eqref{JcEi} we have
\[
  \lim_{N\to\infty}\frac{E_0}N=K\lim_{N\to\infty}\frac1N\sum_{i=1}^{N-1}\vep(x_i)=K\int_0^1\vep(x)\diff
  x\equiv K\vep_0\,,
\]
where
\begin{equation}\label{vep0}
  \vep_0=\cases{\tfrac16\,,& for the HS chain\\
    \tfrac12\,,& for the PF chain\\
  \tfrac13+\tfrac\ga2\,,& for the FI chain.}
\end{equation}
Using Eq.~\eqref{fmu10} we finally obtain the remarkable relation
\begin{equation}
  \label{susysymm}
  \fl
  f^{(n|m)}(\mu_1,\dots,\mu_{m+n-1};K)=K\vep_0-\mu_1
  +f^{(m|n)}(-\mu_1,\mu_{m+n-1}-\mu_1,\dots,\mu_2-\mu_1;-K).
\end{equation}
Similarly, from Eq.~\eqref{specsymm} we immediately deduce that
\begin{equation}
  \label{falbe}
  \fl
  f(\dots,\mu_\al,\dots,\mu_{\al'},\dots;K)=f(\dots,\mu_{\al'},\dots,\mu_{\al},\dots;K)
  \qquad(\pi(\al)=\pi(\al')),
\end{equation}
so that~$f$ is invariant under permutations of chemical potentials of the same type (bosonic or
fermionic). Using the latter identity the relation~\eqref{susysymm} can be generalized as follows:
\begin{eqnarray}
  \fl
  f^{(n|m)}(\mu_1,\dots,\mu_{m+n-1};K)=
  &K\vep_0
    -\mu_{\al_1}\nonumber\\
  \fl
  &+f^{(m|n)}(-\mu_{\al_1},\mu_{\al_{m+n-1}}-\mu_{\al_1},\dots,\mu_{\al_2}-\mu_{\al_1};-K),
    \label{susysymmgen}
\end{eqnarray}
where~$(\al_1,\dots,\al_{m+n-1})$ is a permutation of~$(1,\dots,m+n-1)$ such
that~$\{\al_1,\dots,\al_n\}=\{1,\dots,n\}$.

\subsection{Thermodynamic functions}
Once the free energy per site is known, all the remaining thermodynamic functions can be easily
derived through standard formulas. For instance, the density of~$\su(m|n)$ spins of type $\al$
(with $\al=1,\dots,m+n-1$) is given by
\begin{equation}
  \label{ngen}
  n_{\al}=-\frac{\pd f}{\pd\mu_\al}.
\end{equation}
Indeed, note that if~$N_\al$ (with $\al=1,\dots,m+n$) are nonnegative integers such that
$N_1+\cdots+N_{m+n}=N$, the Hamiltonian $\cH_0$ leaves invariant the
subspace~$\cN(N_1,\dots,N_{m+n})\equiv\cN(\bN)$ on which $\cN_\al=N_\al$ for all~$\al$. Thus the
partition function of the chain~\eqref{H0H1} can be written as (recall that we are
setting~$\mu_{m+n}=0$)
\begin{equation}\label{ZZN}
  \cZ=\sum_{N_1+\cdots+N_{m+n}=N}\e^{\be\sum\limits_{\al=1}^{\mathclap{m+n-1}}\mu_\al
    N_\al}\cZ_\bN\,,\qquad \qquad \be\equiv1/T\,,
\end{equation}
where~$\cZ_{\bN}$ is the partition function of the restriction of~$\cH_0$ to~$\cN(\bN)$, and is
thus independent of the~$\mu_\al$. Hence the thermal average of~$\cN_\al$ is given by
\[
  \langle\cN_\al\rangle=\frac1\cZ\sum_{N_1+\cdots+N_{m+n}=N}
  N_\al\e^{\be\sum\limits_{\al=1}^{\mathclap{m+n-1}}\mu_\al
    N_\al}\cZ_{\bN}=T\,\pdf{\log\cZ}{\mu_\al}\,,
\]
which immediately yields Eq.~\eqref{ngen} for $n_\al\equiv \langle\cN_\al\rangle/N$. Note that
this result supports a natural conjecture~\cite{BaXX} according to which all the
eigenstates~$\psi_k(\bsi)$ of a supersymmetric $\su(m|n)$ chain of HS type corresponding to an
energy~$E(\bsi)$ are of the form
\[
  \psi_k(\bsi)=\sum_{P\in\cS_N}c_P(k,\bsi)P\ket{\si_1\cdots\si_N}\,,
\]
where~$\cS_N$ denotes the permutation group of $N$ elements and the coefficients~$c_P(k,\bsi)$ are
suitable complex numbers. In other words, the number of spins of each type~$\al$ in the
eigenstate~$\psi_K(\bsi)$ should coincide with the number~$N_\al(\bsi)$ of components of the
multiindex~$\bsi$ equal to~$\al$. Indeed, if the latter conjecture were true the density $n_\al$
would be given by
\[
  n_\al=\frac1{N\cZ}\sum_{\bsi}N_\al(\bsi)\e^{-\be E(\bsi)}\,,
\]
where~$E(\bsi)$, which is defined by Eq.~\eqref{specH}, can be alternatively written as
\[
  E(\bsi)=E^{(m|n)}(\bsi)-\sum_{\al'=1}^{m+n}\mu_{\al'}N_{\al'}(\bsi)\,.
\]
Since $E^{(m|n)}(\bsi)$ is independent of the $N_{\al'}$'s, from the latter equation it follows
that
\[
  n_\al=\frac{T}{N\cZ}\,\pdf{}{\mu_\al}\sum_{\bsi}\e^{-\be E(\bsi)}=\frac
  TN\pdf{\log\cZ}{\mu_\al}=-\pdf{f}{\mu_\al},
\]
which coincides with Eq.~\eqref{ngen}.

The variance (per site) of the number of spins of type~$\al$
\begin{equation}\label{nual}
  \nu_\al\equiv\frac1N\Big(\langle\cN_\al^{\,2}\rangle-\langle\cN_\al\rangle^2\Big)
\end{equation}
can be similarly computed from Eq.~\eqref{ZZN}, with the result
\begin{equation}
  \label{varalpha}
  \nu_\al=-\be^{-1}\frac{\pd^2f}{\pd\mu_\al^2}\,.
\end{equation}
The internal energy, heat capacity (at constant volume) and entropy per site are respectively
given by the usual formulas
\begin{equation}
  \label{thermo}
  u=\pdf{}{\be}(\be f),\qquad c_V=-\be^2\pdf u\be\,,\qquad s=\be^2\,\pdf f\beta=\be(u-f)\,.
\end{equation}

The symmetry properties of the free energy derived in the previous subsection yield analogous
properties of the thermodynamic functions just reviewed. For instance, it follows immediately from
Eq.~\eqref{falbe} that the thermodynamic functions~$u$, $c_V$ and~$s$ are invariant under
permutations of chemical potentials of the same type, while the particle densities~$n_\al$ (with
$\al=1,\dots,m+n-1$) behave as\footnote{In fact, from the behavior of the densities~$n_\al$
  with~$\al=1,\dots,m+n-1$ it follows that~$n_{m+n}$ is invariant under permutations of chemical
  potentials of the same type.}
\[
  \fl n_\al(\dots,\mu_{\al'},\dots,\mu_{\al''},\dots;K)
  =n_{\al}(\dots,\mu_{\al''},\dots,\mu_{\al'},\dots;K) \qquad (\pi(\al')=\pi(\al''))
\]
for $\al\ne\al',\al''$ and
\[
  \fl n_\al(\dots,\mu_\al,\dots,\mu_{\al'},\dots;K)
  =n_{\al'}(\dots,\mu_{\al'},\dots,\mu_{\al},\dots;K) \qquad(\pi(\al)=\pi(\al'))
\]
(similar relations hold for~$\nu_\al$). In view of the last equation, we can restrict ourselves
without loss of generality to studying just one bosonic and one fermionic density. Likewise,
Eq.~\eqref{susysymm} implies that
\begin{equation}\label{unmmn}
  \fl
  u^{(n|m)}(\mu_1,\dots,\mu_{m+n-1};K)=K\vep_0-\mu_1+
  u^{(m|n)}(-\mu_1,\mu_{m+n-1}-\mu_1,\dots,\mu_2-\mu_1;-K),
\end{equation}
and similar identities for~$c_V$ and~$s$. As to the boson densities,
differentiating~\eqref{susysymm} with respect to~$\mu_1$ we obtain
\begin{eqnarray}
  \fl
  n_1^{(n|m)}(\mu_1,\dots,\mu_{m+n-1};K)
  &=1
    -\sum_{\al=1}^{m+n-1}n_\al^{(m|n)}(-\mu_1,\mu_{m+n-1}-\mu_1,\dots,\mu_2-\mu_1;-K)\nonumber\\
  \fl
  &=n_{m+n}^{(m|n)}(-\mu_1,\mu_{m+n-1}-\mu_1,\dots,\mu_2-\mu_1;-K)\,.
    \label{n1mn}
\end{eqnarray}
On the other hand, differentiation of~Eq.~\eqref{susysymm} with respect to~$\mu_\al$ with
$\al=2,\dots,m+n-1$ yields
\begin{equation}\label{nalmnnm}
  \fl
  n_\al^{(n|m)}(\mu_1,\dots,\mu_{m+n-1};K)=
  n_{m+n+1-\al}^{(m|n)}(-\mu_1,\mu_{m+n-1}-\mu_1,\dots,\mu_2-\mu_1;-K)\,.
\end{equation}
Note that the latter equation is actually valid for~$\al=1,\dots,m+n$, on account of
Eq.~\eqref{n1mn} and the identity~$n_{m+n}=1-\sum_{\al=1}^{m+n-1}n_\al$. Of course, similar
relations hold for the variances per site~$\nu_\al$. In particular, when~$m=n$ Eqs.~\eqref{unmmn}
(and its analogues for $c_V$ and~$s$) and~\eqref{nalmnnm} imply that we can restrict ourselves
without loss of generality to positive values of $K$.

In Sections~\ref{sec.su11}--\ref{sec.su22} we shall apply the results of this section to study the
thermodynamics of the $\su(1|1)$, $\su(2|1)$ (or, equivalently, $\su(1|2)$) and $\su(2|2)$
supersymmetric chains~\eqref{H0H1}--\eqref{FI}.

\section{The $\su(1|1)$ chains}\label{sec.su11}

\subsection{Free energy per site}
In this case the transfer matrix~$A(x)$ is simply
\[
  A(x)=
  \left(
  \begin{array}{cc}
    q^{-\mu} & q^{-\frac{\mu}{2}} \\
    q^{K\vep(x)-\frac{\mu}{2}} & q^{K\vep(x)}
  \end{array}
  \right),\qquad \mu\equiv\mu_1\,,
\]
with eigenvalues zero and
\[
\la_1(x)=q^{K\vep(x)}+q^{-\mu}\,.
\]
In particular, the matrix~$A(x)$ is diagonalizable for all~$x\in[0,1]$, and condition~i)
in the previous section is thus trivially satisfied. Condition~ii) is also easily verified, as we
can simply take
\[
  P(x) = \left(
    \begin{array}{cc}
      q^{-(K\vep(x)+\frac{\mu}2)}& -q^{\frac{\mu}2}\\
      1& 1
    \end{array}
  \right)
\]
and therefore
\[
  \sum_{\al=1}^2(-1)^{\al+1}M_{\al1}(1)q^{-\frac12\mu_\al}=
    \left|
          \begin{array}{cc}
            q^{-\frac{\mu}2} &  -q^{\frac{\mu}{2}}\\
            1 & 1                     
    \end{array}
    \right|=q^{-\frac{\mu}2}+q^{\frac{\mu}2}>0.
\]
Thus the free energy per site is given by Eq.~\eqref{fTmu}, which in this case reads
\begin{equation}
  \label{f11}
  \fl
  f(T,\mu)=-T\int_0^1\log\bigl(q^{K\vep(x)}+q^{-\mu}\bigr)\diff x
  =-\mu-\frac1\be\int_0^1\log\Bigl(1+\e^{-\be(K\vep(x)+\mu)}\Bigr)\diff x\,.
\end{equation}
For the $\su(1|1)$ HS chain~\eqref{H0H1}-\eqref{HS}, Eq.~\eqref{f11} coincides with the formula
derived in Ref.~\cite{CFGRT16}\footnote{It should be taken into account that in
  Ref.~\cite{CFGRT16} the alternative convention~$\mu_1=0$, $\mu_2=-\la$ was used.}. The
derivation of Eq.~\eqref{f11} in the latter reference is based on the equivalence of
the~$\su(1|1)$ HS chain to a translation-invariant free fermion model, which in turn relies on the
symmetry of its dispersion relation~$\vep(x)=x(1-x)$ about~$x=1/2$. This derivation is therefore
not valid for the PF and FI chains, as their dispersion relations are monotonic. The approach
followed in this paper circumvents this problem since, as remarked in the previous section,
Eq.~\eqref{f11} is actually valid for the three chains of HS type~\eqref{HS}--\eqref{FI}.

As explained in the previous section, since~$m=n=1$ we can restrict ourselves in this case to
positive values of~$K$. This also follows directly from the alternative expression for the free
energy per site
\[
f=\frac12(K\vep_0-\mu)-\frac1\be\int_0^1\log\Big[2\cosh\Bigl(\tfrac\be2(K\vep(x)+\mu)\Bigr)\Big]\diff x\,,
\]
which implies (temporarily dropping the dependence of~$f$ on the temperature) that
\[
  f(-\mu;-K)=f(\mu;K)+\mu-K\vep_0
\]
(cf.~Eq.~\eqref{susysymm}).

\subsection{Thermodynamic functions}

\begin{figure}[t]
  \includegraphics[height=3.4cm]{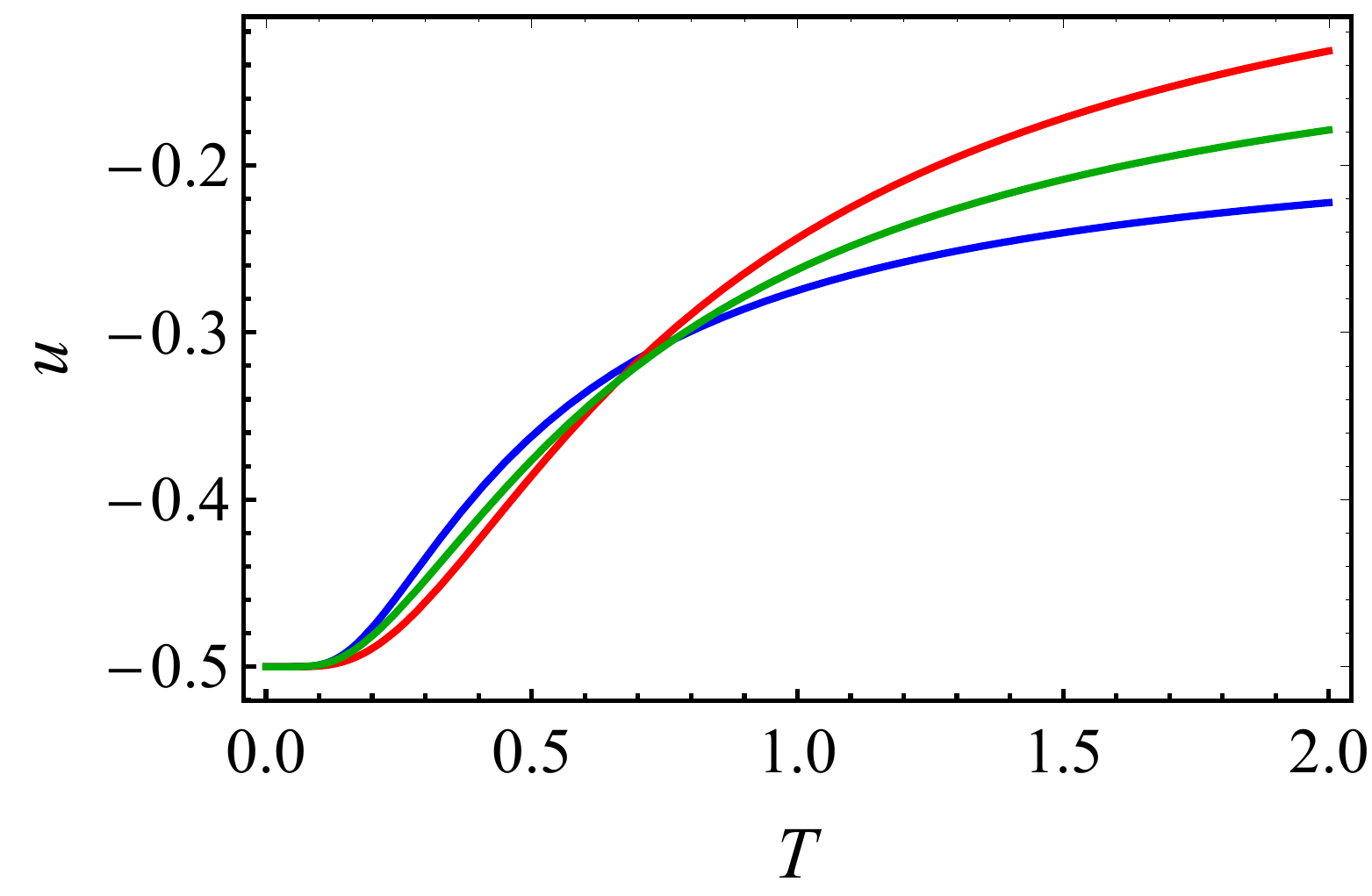}\hfill
  \includegraphics[height=3.4cm]{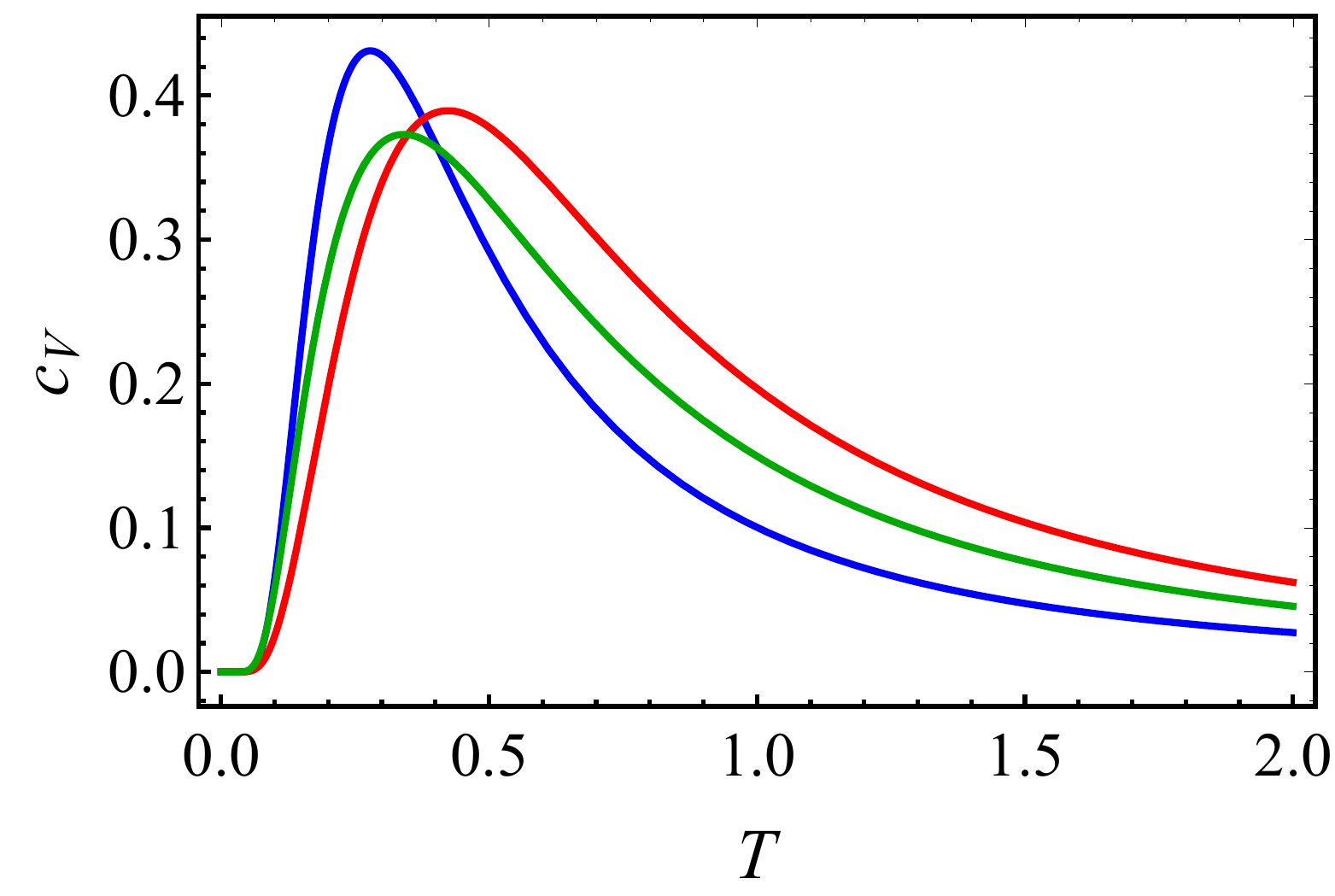}\hfill
  \includegraphics[height=3.4cm]{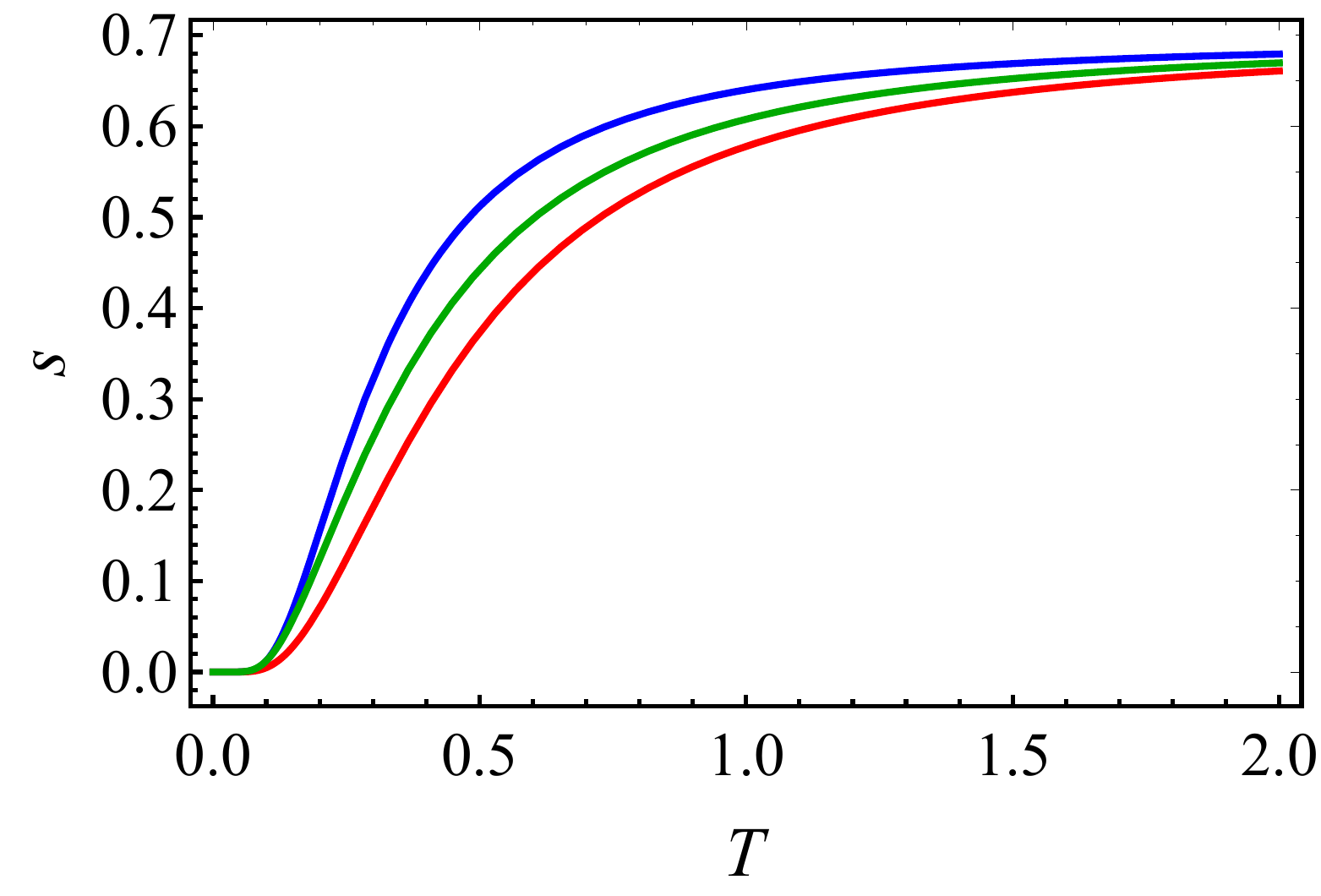}
  \caption{Internal energy (left), specific heat (center) and entropy (right) per site versus the
    temperature for the HS (blue), PF (red) and FI (with~$\ga=0$, green) $\su(1|1)$ chains with
    $\mu/K=1/2$. (The internal energy, specific heat and temperature are measured in units of
    $K$.)}
  \label{fig.ucvsplots}
\end{figure}

From Eqs.~\eqref{thermo} and~\eqref{f11} we immediately obtain the following explicit formulas for
the main thermodynamic functions of the~$\su(1|1)$ chains of HS type:
\begin{eqnarray}
  \fl
      n_1&= \int_0^1\frac{\diff x}{1+\e^{-\be(K\vep(x)+\mu)}}\,,\label{n11}\\
  \fl
     \nu_1&=\frac14\int_0^1\sech^2\bigl[\tfrac\be2(K\vep(x)+\mu)\bigr]\diff x\,,
       \label{nu11}\\
  \fl
  u&= -\mu+\int_0^1\frac{K\vep(x)+\mu}{1+\e^{\be(K\vep(x)+\mu)}}\,\diff x\,,\label{u11}\\
  \fl
     c_V&=\frac{\be^2}4\int_0^1(K\vep(x)+\mu)^2\sech^2\bigl[\tfrac\be2(K\vep(x)+\mu)\bigr]\diff x\,,
       \label{ucv11}\\
  \fl
  s&=\int_0^1\bigg\{\log\Bigl[2\cosh\bigl(\tfrac\be2(K\vep(x)+\mu)\bigr)\Bigr]
     -\frac\be2\,(K\vep(x)+\mu)\tanh\bigl(\tfrac\be2(K\vep(x)+\mu)\bigr)\bigg\}\,\diff x\,.
     &\label{s11}
\end{eqnarray}
In Fig.~\ref{fig.ucvsplots} we present a plot of the internal energy, specific heat and entropy
per site as a function of~$T$ for the three~$\su(1|1)$ chains of HS type with~$\mu=1/2$. In fact,
we have found that the latter functions have the same qualitative behavior for a wide range of
values of~$\mu$, which is essentially the same as their $\su(2|0)$ counterparts analyzed in
Ref.~\cite{EFG12}. In particular, the specific heat exhibits the so called Schottky peak,
characteristic of two-level systems like the Ising model at zero magnetic field or paramagnetic
spin $1/2$ anyons~\cite{Mu10}.

As was the case for the $\su(2|0)$ PF chain studied in Ref.~\cite{EFG12}, it turns out that the
thermodynamic functions of the $\su(1|1)$ PF chain~\eqref{H0H1}-\eqref{PF} can be expressed in
closed form in terms of elementary or well-known special functions. To this end, recall first of
all the definition of the dilogarithm function~\cite{Le81,OLBC10}
\begin{equation}
  \Li_2(z)=-\int_0^z\frac{\log(1-t)}t\,\diff t,
\end{equation}
where~$\log z$ denotes the determination of the logarithm with $\Im\log z\in(-\pi,\pi)$ and the
integral is taken along any path not intersecting the branch cut on the half-line $[1,\infty)$.
Performing the change of variables~$t=-\e^{-\be(Kx+\mu)}$ in Eq.~\eqref{f11} for the PF chain we
immediately obtain
\begin{equation}
  \label{f11PF}
  f(T,\mu)=-\mu+\frac1{K\be^2}\Big[\Li_2(-\e^{-\be\mu})-\Li_2(-\e^{-\be(K+\mu)})\Big]\,.
\end{equation}
Differentiation of this expression with respect to~$\mu$ yields a remarkable closed formula in
terms of elementary functions for the density of bosons of the $\su(1|1)$ PF chain, namely
\begin{equation}
  \label{nB11PF}
  n_1=1-\frac1{K\be}\,\log\biggl(\frac{1+\e^{-\be\mu}}{1+\e^{-\be(K+\mu)}}\biggr).
\end{equation}
The remaining thermodynamic functions admit similar closed-form expressions, namely
\begin{eqnarray}
  \label{nu}
  \nu_1&=\frac1{K\be}\,\frac{\e^{-\be\mu}(1-\e^{-\be K})}{(1+\e^{-\be\mu})(1+\e^{-\be(K+\mu)})}\,,\\
  u&=\frac\mu{K\be}\log(1+\e^{-\be\mu})-\frac{K+\mu}{K\be}\,\log(1+\e^{-\be(K+\mu)})-f-2\mu\,,\\
  c_V&=\frac{2\mu}K\log(1+\e^{-\be\mu})-\frac{2(K+\mu)}K\log(1+\e^{-\be(K+\mu)})\nonumber\\
       &\hphantom{=\frac{2\mu}K\log}{}+\frac{\be\mu^2}{K(1+\e^{\be\mu})}
         -\frac{\be(K+\mu)^2}{K(1+\e^{\be(K+\mu)})}\,-2\be(f+\mu),\\
  s&=\be(u-f)\,.
\end{eqnarray} 

\subsection{Critical behavior}\label{sub.crit11}

We shall next determine the low temperature behavior of the free energy per site~\eqref{f11} for
the three HS-types chains~\eqref{HS}--\eqref{FI}. As is well known, when $T\to0$
the free energy per unit length of a ($1+1$)-dimensional CFT (in natural
units $\hbar=k_{\mathrm B}=1$) behaves as
\begin{equation}\label{fCFT}
  f(T)\simeq f(0)-\frac{\pi c T^2}{6v}\,,
\end{equation}
where $c$ is the central charge and $v$ is the effective speed of light~\cite{BCN86,Af86}. Since
the value of~$f$ at small temperatures is determined by the low energy excitations, the validity
of Eq.~\eqref{fCFT} is generally taken as a strong indication of the conformal invariance of a
quantum system. In fact, the latter equation is one of the standard methods for identifying the
central charge of the Virasoro algebra of a quantum critical system.

Let us suppose, to begin with, that the boson chemical potential~$\mu$ is strictly positive. In
this case~$K\vep(x)+\mu>0$ for all $x\in[0,1]$ (since, as remarked above, we are taking $K>0$
throughout this section), so that~$f(0,\mu)=-\mu$ and
\[
  |f(T,\mu)-f(0,\mu)|<T\int_0^1\e^{-\be(K\vep(x)+\mu)}< T\e^{-\be\mu}\,,
\]
so that the system is not critical. A similar result holds for $\mu<-K\vepmax$, where
\[
  \vepmax=\max_{0\le x\le 1}\vep(x)=\cases{\tfrac14\,,& for the HS chain\\
  1\,& for the PF chain\\ 1+\ga,& for the FI chain.}
\]

\begin{figure}[t]
  \includegraphics[height=.32\textwidth]{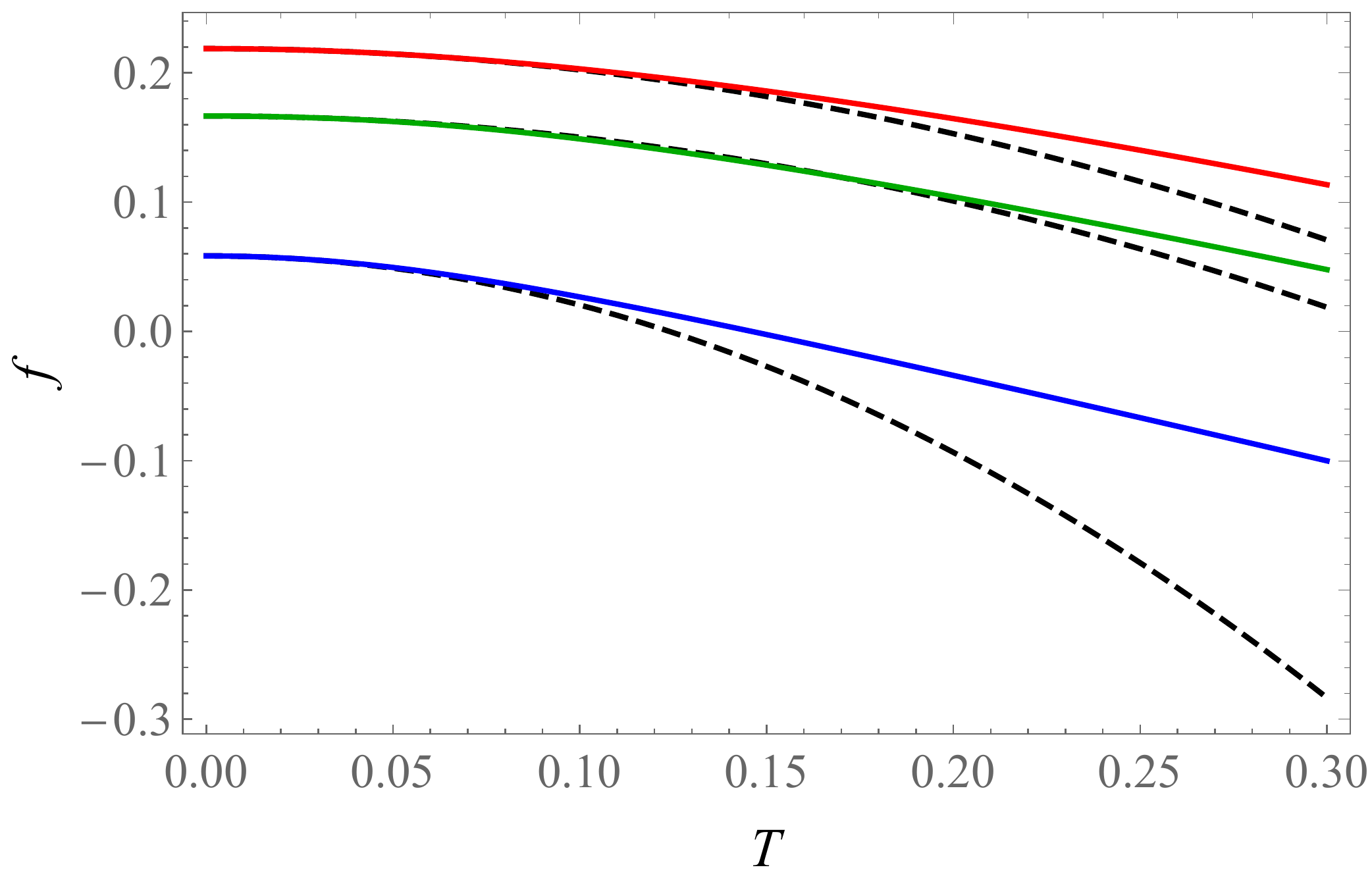}\hfill
   \includegraphics[height=.32\textwidth]{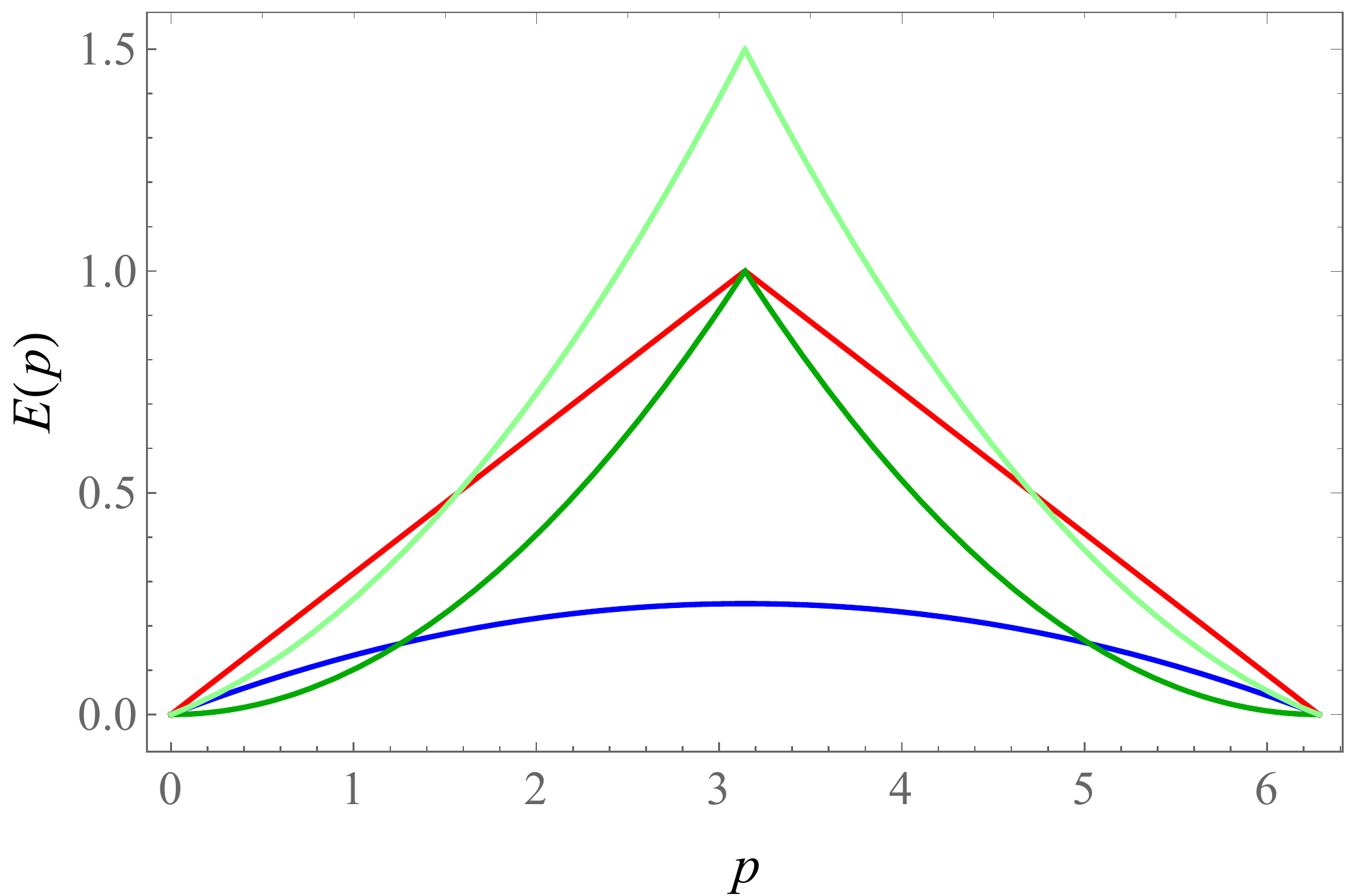}
   \caption{Left: free energy per site versus temperature (both in units of $K$) for
     the~$\su(1|1)$ HS (blue), PF (red) and FI chains (with $\ga=0$, green)
     for~$\mu/K=-\vepmax/4$. In all three cases, the dashed black line represents the
     low-temperature approximation~\eqref{fasymp}. Right: energy-momentum function~$E(p)$ for
     the~$\su(1|1)$ HS (blue), PF (red) and FI chains, with $\ga=0$ (green) and $\ga=1/2$ (light
     green) in the latter case.}
  \label{fig.dispfasym}
\end{figure}

Consider next the case~$-K\vepmax<\mu<0$. It is now convenient to rewrite Eq.~\eqref{f11} as
\begin{equation}
  \label{f11eta}
  f(T,\mu)+\mu=-\eta T\int_0^{1/\eta}\log\Bigl(1+\e^{-\be(K\vep(x)+\mu)}\Bigr)\diff x\,,
\end{equation}
where
\[
  \eta=\cases{2\,,& for the HS chain\\ 1,& for the PF and FI chains.}
\]
This is certainly possible, since the dispersion relation~$\vep(x)=x(1-x)$ is symmetric about
$x=1/2$. Let~$x_0$ denote the unique root of the equation~$K\vep(x)+\mu=0$ in the interval
$(0,1/\eta)$, namely
\begin{equation}\label{x0def}
  x_0=\cases{\tfrac12\,(1-\sqrt{1+\tfrac{4\mu}K}\,),& for the HS chain\\
  -\tfrac\mu K,& for the PF chain\\ \tfrac12\,(-\ga+\sqrt{\ga^2-\tfrac{4\mu}K}\,),& for the FI chain.}
\end{equation}
Since~$K\vep(x)+\mu$ is negative for $0\le x<x_0$ and positive for $x_0<x\le 1/\eta$, we have
\begin{equation}\label{f011}
f(0,\mu)+\mu=\eta\bigg(K\int_0^{x_0}\vep(x)\diff x+\mu x_0\bigg)
\end{equation}
and
\[
  f(T,\mu)-f(0,\mu)=
  -\eta T\int_0^{1/\eta}\log\Bigl(1+\e^{-\be|K\vep(x)+\mu|}\Bigr)\diff x.
\]
If we now fix~$\De<\min(x_0,1/\eta-x_0)$ ~independent of~$T$ and
set~$A\equiv[0,x_0-\De]\cup[x_0+\De,1/\eta]$, the latter integral can be approximated by
\[
  I(T)\equiv\int_{x_0-\De}^{x_0+\De}\log\Bigl(1+\e^{-\be|K\vep(x)+\mu|}\Bigr)\diff x
\]
with an  error
\[
  \int_A\log\Bigl(1+\e^{-\be|K\vep(x)+\mu|}\Bigr)\diff x
  <\int_A\e^{-\be|K\vep(x)+\mu|}\diff x<\e^{-\be\ka}\,,
\]
with~$\ka=\min(-\mu-\vep(x_0-\De),\mu+\vep(x_0+\De))>0$ independent of~$T$\,. Performing the
change of variables~$y=\be|K\vep(x)+\mu|$ in each of the intervals~$[x_0-\De,x_0]$
and~$[x_0,x_0+\De]$ we obtain
\begin{equation}
  \fl
  I(T)= \frac TK
 \bigg(\int_0^{\be|K\vep(x_0-\De)+\mu|}\frac{\log(1+\e^{-y})}{\vep'(x)}\,\diff y
  +\int_0^{\be|K\vep(x_0+\De)+\mu|}\frac{\log(1+\e^{-y})}{\vep'(x)}\,\diff y\bigg).
\end{equation}
Moreover, since~$\vep'$ does not vanish on~$[x_0-\De,x_0+\De]$ we have
\begin{equation}\label{vepx}
  \frac1{\vep'(x)}=\frac1{\vep'(x_0)}+\Or(x-x_0)=\frac1{\vep'(x_0)}+\Or(Ty),
\end{equation}
and therefore (taking into account that~$\int_0^\infty y\log(1+\e^{-y})\diff y$ is convergent)
\[
  \fl
  I(T)=\frac T{K\vep'(x_0)}\bigg(\int_0^{\be|K\vep(x_0-\De)+\mu|}+
  \int_0^{\be|K\vep(x_0+\De)+\mu|}\bigg)\log(1+\e^{-y})\,\diff y+\Or(T^2)\,.
\]
It can be easily checked that the error incurred by replacing the upper limits in each of the
above integrals by~$+\infty$ is $\Or(\e^{-\ka'\be})$, where again~$\ka'$ is a constant independent
of the temperature (see, e.g., Ref.~\cite{CFGR17}). Hence
\[
  I(T)=\frac{2T}{K\vep'(x_0)}\int_0^\infty\log(1+\e^{-y})\,\diff
  y+\Or(T^2)=\frac{\pi^2T}{6K\vep'(x_0)}+\Or(T^2)\,,
\]
and therefore
\[
  f(T,\mu)=f(0,\mu)-\frac{\eta\pi^2T^2}{6K\vep'(x_0)}+\Or(T^3)\,.
\]
(See Fig.~\ref{fig.dispfasym} (left) for a graphic comparison of the latter approximation to the
exact expression~\eqref{f11} for~$\mu/K=-\vepmax/4$ and~$T/K\le0.3$.)

It was shown in Ref.~\cite{CFGRT16} that the~$\su(1|1)$ HS chain~\eqref{H0H1}-\eqref{HS} can be
mapped to a translationally invariant system of free fermions with energy-momentum
relation~$E(p)=K\vep(p/(2\pi))$, where~$p\in(0,2\pi)$ is the momentum ($\bmod\,2\pi$). Moreover,
at low energies the spectrum of this chain consists of small excitations with momenta
around~$p_0\equiv 2\pi x_0$ (or $2\pi-p_0$), so that the effective speed of light is given by
\[
  v=\left.\frac{\diff E}{\diff p}\right|_{\mathrlap{p=2\pi x_0}}\en=\frac{K\vep'(x_0)}{2\pi}\qquad
  (\hbox{\rm for the HS chain})\,.
\]
Of course, the situation is quite different for the PF and FI chains, since these systems are not
translationally invariant and, in particular, their dispersion relation~$\vep(x)$ is not symmetric
around $x=1/2$. In this case we must therefore take as energy-momentum relation the symmetric
extension of~$K\vep(p/\pi)$ around~$\pi$, i.e.,
\begin{equation}\label{EpPFFI}
E(p)=K\vep(1-|1-p/\pi|)\qquad (\hbox{\rm for the PF and FI chains)}
\end{equation}
(cf.~Fig.~\ref{fig.dispfasym}, right), so that now $p=\pi x$ and the effective speed of light is
given by
\[
  v=\left.\frac{\diff E}{\diff p}\right|_{\mathrlap{p=\pi x_0}}\en=\frac{K\vep'(x_0)}\pi\qquad
  (\hbox{\rm for the PF and FI chains)}\,.
\]
Note that this implies that in the thermodynamic limit (though not for any \emph{finite} $N$) the
$\su(1|1)$ PF and FI chains are equivalent to a translation-invariant free fermion model with
energy-momentum relation~\eqref{EpPFFI}, since under the change of variables~$x=p/\pi$
Eq.~\eqref{f11} becomes
\[
  f(T,\mu)=-\mu-\frac T\pi\int_0^\pi\log(1+\e^{-\be(E(p)+\mu)})\,\diff p.
\]
Thus for all three $\su(1|1)$ chains of HS type we can write
\begin{equation}\label{vdefgen}
  v=\frac{K\vep'(x_0)}{\eta\pi}\,,
\end{equation}
and we can therefore express the asymptotic equation for the free energy per site in the unified
way
\begin{equation}
  \label{fasymp}
  f(T,\mu)=f(0,\mu)-\frac{\pi T^2}{6v}+\Or(T^3)\,,
\end{equation}
where
\begin{equation}\label{vdef}
  v=\cases{\tfrac K{2\pi}\,\sqrt{1+\tfrac{4\mu}K}\,,& for the HS chain\\
    \tfrac K{\pi}\,,& for the PF chain\\
    \tfrac K{\pi}\,\sqrt{\ga^2-\tfrac{4\mu}K}\,,& for the FI chain.
  }
\end{equation}
Comparing with Eq.~\eqref{fCFT} we conclude that for~$-K \vepmax<\mu<0$ all of these chains are
critical, with~$c=1$. In other words, the free energy per site of the three $\su(1|1)$ chains of
HS type behaves as that of a CFT with central charge $c=1$ (for instance, a free CFT with one
bosonic field).

\begin{figure}[h]
  \centering
  \includegraphics[width=7.5cm]{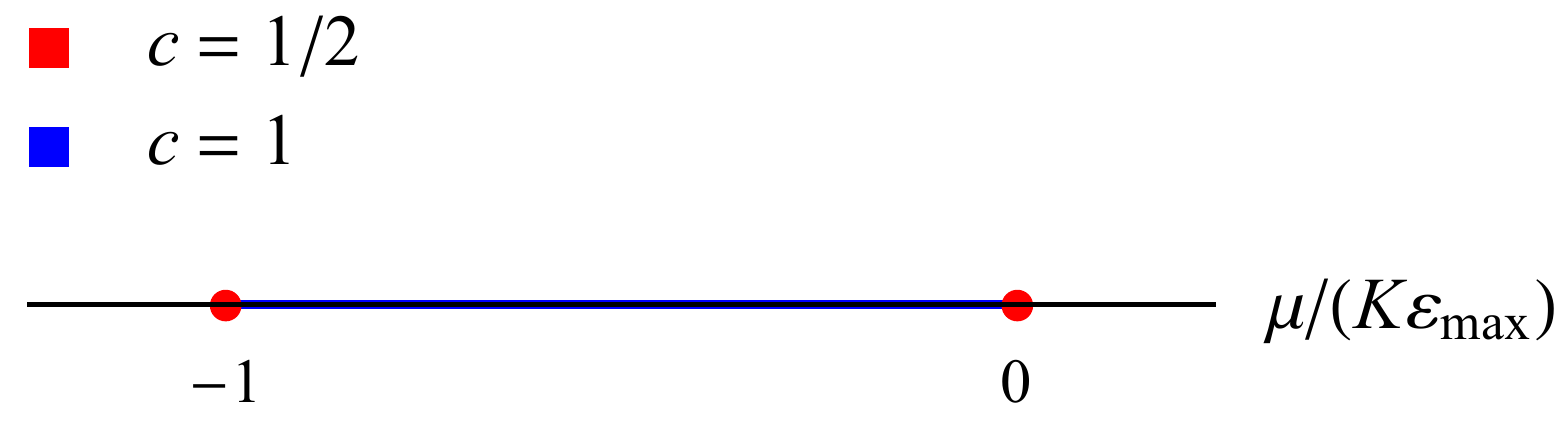}
  \caption{Phase diagram of the $\su(1|1)$ chains of HS type with $K>0$. The critical line and
    points are colored according to the value of the central charge $c$, as shown in the
    accompanying legend. The origin is not critical for the FI chain with~$\ga=0$, while the
    point~$-K\vepmax$ is not critical for the HS chain.}
  \label{fig.phase11}
\end{figure}

For $\mu=0$, a similar analysis shows that the HS, PF and FI (with~$\ga\ne0$) $\su(1|1)$
chains~\eqref{H0H1}-\eqref{FI} are again critical, but the central charge is now~$c=1/2$ (i.e.,
that of a free CFT with one fermionic field). On the other hand, the FI chain with~$\ga=\mu=0$ is
not critical, since following Ref.~\cite{CFGR17} it can be shown that in this case
\[
  f(T,0)=-\frac12\,\sqrt{\frac{\pi}K}\,\bigg(1-\frac1{\sqrt2}\bigg)\,\ze(3/2)\,T^{3/2}+\Or(T^2),
\]
where~$\ze(z)$ denotes Riemann's zeta function. Finally, for~$\mu=-K\vepmax$ the PF and FI chains
are critical with~$c=1/2$ (since the root~$x_0=1$ of $K\vep(x)-K\vepmax$ is simple in both cases),
while for the HS chain it was shown in Ref.~\cite{CFGRT16} that
\[
  f(T,-K/4)=\frac K6-\sqrt{\frac{\pi}K}\,\bigg(1-\frac1{\sqrt2}\bigg)\,\ze(3/2)\,T^{3/2}+\Or(T^2).
\]
In particular, the~$\su(1|1)$ HS chain with~$\mu=-K/4$ is not critical. In summary, the phase
diagram of the three $\su(1|1)$ chains of HS type is as represented schematically in
Fig.~\ref{fig.phase11}. For the $\su(1|1)$ HS chain, the above result follows from the general
ones in Ref.~\cite{CFGRT16} for a system of spinless free fermions, as well as the direct
calculation in Ref.~\cite{BBS08}. On the other hand, our result for the~$\su(1|1)$ PF chain
with~$\mu=0$ is in agreement with the heuristic analysis of Ref.~\cite{HB00}.

\subsection{Boson density at low temperatures}

\begin{figure}[h]
  \centering
  \includegraphics[width=8cm]{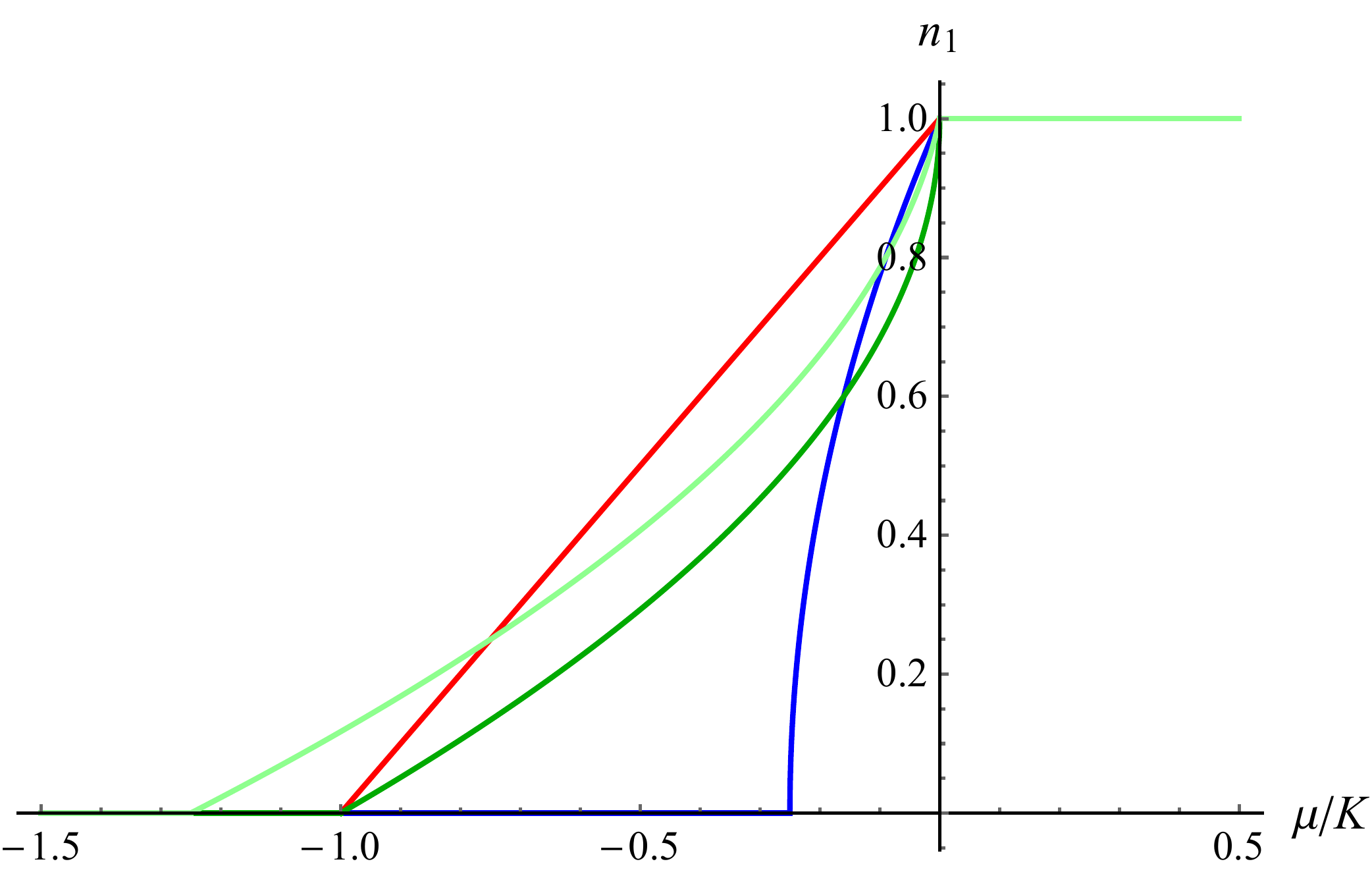}
  \caption{Zero temperature boson density~$n_1$ as a function of~$\mu/K$ for the HS (blue), PF
    (red) and FI (green for~$\ga=0$, light green for~$\ga=1/4$) chains.}
  \label{fig.n0}
\end{figure}

The low temperature behavior of the~$\su(1|1)$ boson density~\eqref{n11} can be analyzed using the
results of Ref.~\cite{CFGRT16}, which are valid for an arbitrary dispersion function~$\vep(x)$. To
this end, let us rewrite Eq.~\eqref{n11} as
\[
  n_1=\eta\int_0^{1/\eta}\frac{\diff x}{1+\e^{-\be(K\vep(x)+\mu)}},
\]
so that~$\vep(x)$ is monotonically increasing in the interval~$[0,1/\eta]$ for all three chains of
HS type. It readily follows from the latter expression that the value of the boson density
at~$T=0$ is given by
\[
  n_1(0,\mu)=\cases{0\,,& \hfill$\mu\le -K\vepmax$\hfill\\
    1-\eta x_0\,,& $-K\vepmax\le\mu\le0$\\
  1\,,& \hfill$\mu\ge0$,\hfill}
\]
where~$x_0$ is given by Eq.~\eqref{x0def}. Thus the $\su(1|1)$~boson density presents a
second-order (continuous) phase transition at zero temperature (cf.~Fig.~\ref{fig.n0}).

The low temperature behavior of the boson density for the~$\su(1|1)$ PF chain follows directly
from Eq.~\eqref{nB11PF}. For instance, in the critical region~$-K<\mu<0$ we have
\[
  n_1=1+\frac\mu{K}-\sgn(K+2\mu)\,\frac{T}K\,\e^{-\be\min(|\mu|,K+\mu)}+
  \Or\biggl(\frac{T}K\,\e^{-\ka\be}\biggr),
\]
with~$\ka=\min(2|\mu|,K+\mu)$ for~$-K/2\le \mu<0$ and~$\ka=\min(|\mu|,2K+2\mu)$
for~$-K< \mu\le-K/2$. For the HS and FI chains, an asymptotic approximation for $n_1$ at low
temperatures can be easily derived from the general formulas in Ref.~\cite{CFGRT16}. For instance,
in the critical region~$-K\vepmax<\mu<0$ we have
\begin{eqnarray*}
  \fl
  n_1&=1-\eta x_0+\frac{K\vep''(x_0)}{6\pi\eta^2v^3}\,T^2+\Or(T^3)\\\fl &=
                                                                          \cases{\sqrt{1+\tfrac{4\mu}K}-\tfrac{2\pi^2}{3K^2}\Big(1+\tfrac{4\mu}K\Big)^{-3/2}T^2+\Or(T^3)\,,&
                                                                                                                                                                             for the HS chain\vrule depth 12pt width0pt\\
  1+\tfrac12\Big(\ga-\sqrt{\ga^2-\tfrac{4\mu}K}\,\Big)
  +\tfrac{\pi^2}{3K^2}\Big(\ga^2-\tfrac{4\mu}K\Big)^{-3/2}T^2
  +\Or(T^3)\,,&
                for the FI chain.}
\end{eqnarray*}

\begin{figure}[h]
\includegraphics[width=.49\textwidth]{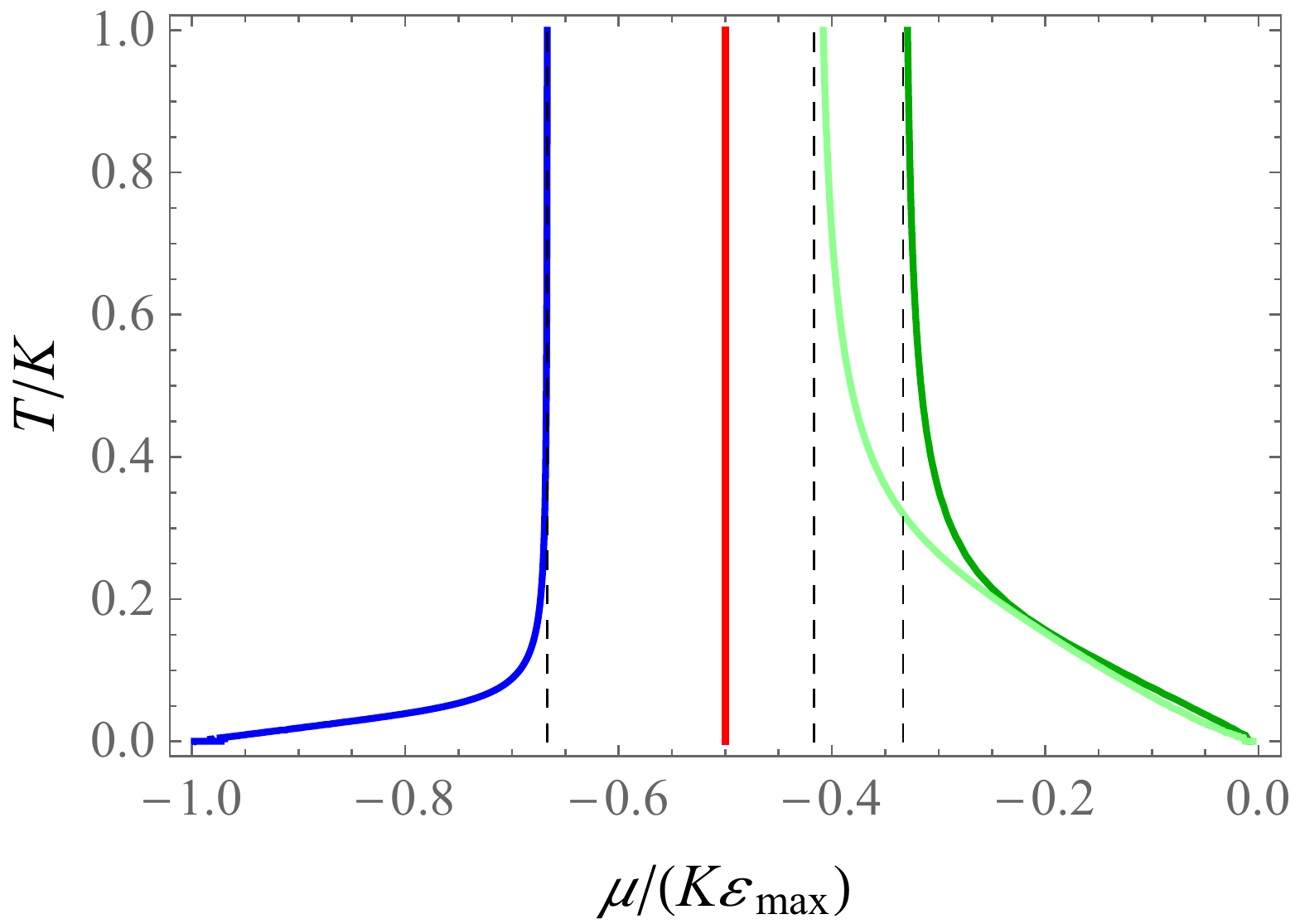}
  \hfill
\includegraphics[width=.49\textwidth]{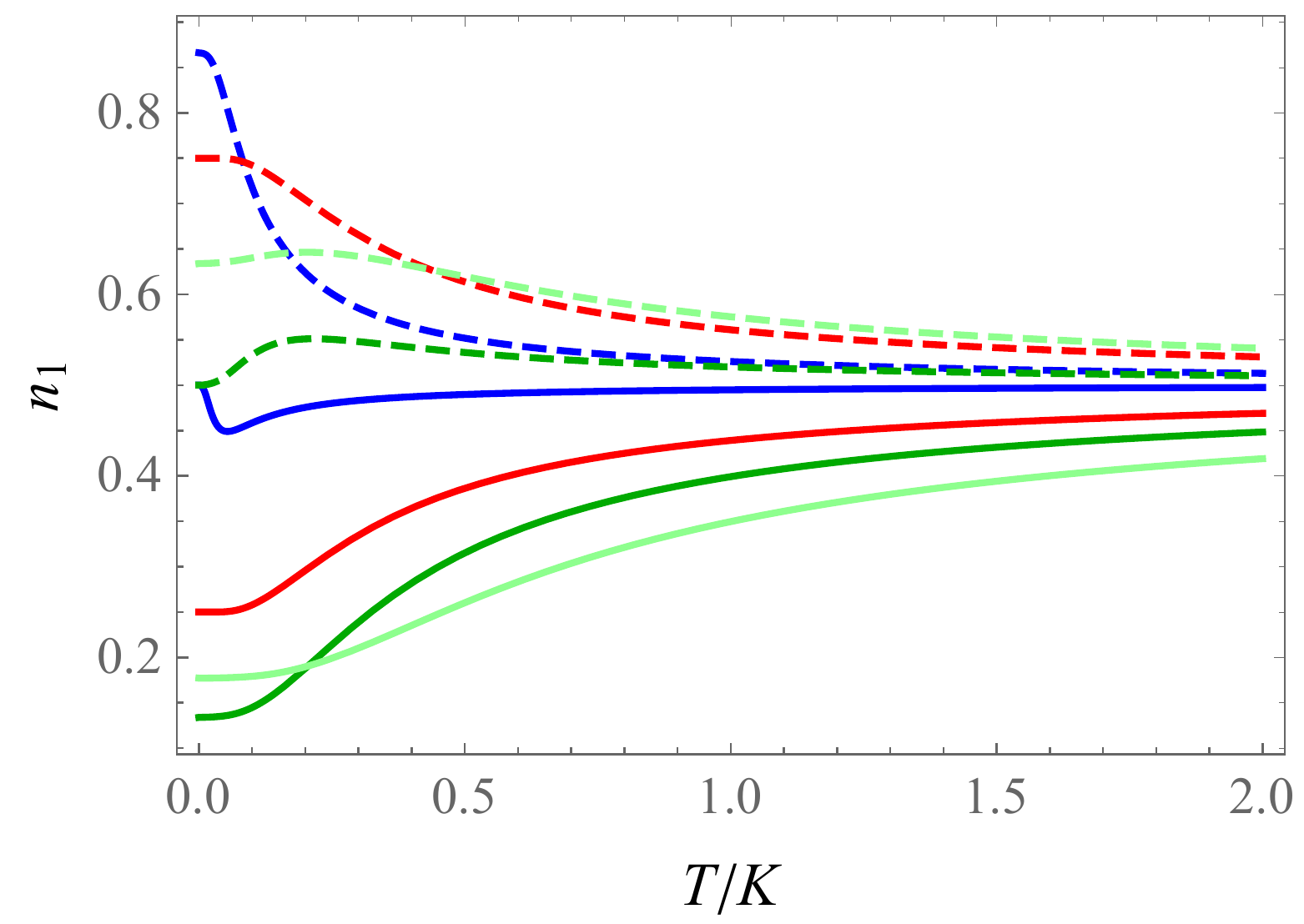}
\caption{Left: plot of the implicit curve~$\pdf{n_1}T=0$ for the~$\su(1|1)$ HS (blue), PF (red) and
  FI chains, with $\ga=0$ (green) and~$\ga=1$ (light green) in the latter case. The dashed black
  lines represent the vertical asymptotes $\mu/(K\vepmax)=-\vep_0/\vepmax=-2/3,-5/12,-1/3$. Right:
  boson density for~$\mu/(K\vepmax)=-3/4$ (solid lines) and~$\mu/(K\vepmax)=-1/4$ (dashed lines),
  with the same color code.}
  \label{fig.dnBnT}
\end{figure}

The qualitative behavior of the boson density for finite~$T$ can also be analyzed with the help of
the closed formula~\eqref{n11}. To begin with, since
\[
  \pdf{n_1}T=-\frac\eta{4T^2}\int_0^{1/\eta}
  \frac{K\vep(x)+\mu}{\cosh^2\bigl[\tfrac\be2\big(K\vep(x)+\mu\big)\bigr]}\,\diff x
\]
and we are taking~$K>0$, it is clear that for~$\mu\le-K\vepmax$ (resp.~$\mu\ge0$) the boson
density increases (resp.~decreases) monotonically to its $T\to\infty$ limit~$1/2$, as expected.
The qualitative behavior of the boson density is more subtle when~$\mu$ lies inside the critical
interval~$(-K\vepmax,0)$. To help analyze this behavior, in Fig.~\ref{fig.dnBnT} (left) we have
represented the implicit curve $\pdf{n_1}T=0$ for the three $\su(1|1)$ chains of HS type
and~$\mu/(K\vepmax)$ in the critical range~$(-1,0)$. From the latter plot it is clear that for the
HS and FI chains there is a range of values of~$\mu$ for which $n_1$ is not monotonic. More
precisely, for the~HS chain the boson density has a unique minimum at finite temperature
for~$-K\vepmax<\mu<\mu_c$ for a certain critical chemical potential~$\mu_c$, since~$n_1$ is
decreasing to the right of the curve~$\pdf{n_1}T=0$ and decreasing to its left (this is clear from
the behavior of~$n_1$ for $\mu>0$ and $\mu<-K\vepmax$). Similarly, the boson density of the~FI
chain presents a unique maximum at finite~$T$ in the range~$\mu_c<\mu<0$, where now the critical
value~$\mu_c$ of~$\mu$ depends on the chain parameter~$\ga$. The situation is totally different
for the PF chain, for which $n_1$ is monotonically increasing (resp.~decreasing) for $-K<\mu<-K/2$
(resp.~$-K/2<\mu<0$), since now~$n_1$ is constant for~$\mu=-K/2\equiv\mu_c$
(cf.~Fig.~\ref{fig.dnBnT}, left). The critical chemical potential~$\mu_c$ can be computed in all
cases from the condition
\[
  \lim_{T\to\infty}T^2\pdf{n_1}T(T,\mu_c)=-\frac14\,(K\vep_0+\mu_c)=0\,,
\]
which yields~$\mu_c=-K\vep_0$. The qualitative behavior of~$n_1$ just described is apparent in
Fig.~\ref{fig.dnBnT} (right), where we have plotted the boson density for the three chains of HS
type for $\mu=-3K\vepmax/4$ and~$\mu=-K\vepmax/4$.

\section{The $\su(2|1)$ chains}\label{sec.su21}

This case is of particular interest, since its dual~$\su(1|2)$ version with HS
interaction~\eqref{HS} can be mapped to the spin~$1/2$ Kuramoto--Yokoyama $t$-$J$ model in an
external magnetic field~\cite{Ka92,AS06} with a suitable choice of the chemical potentials. In
fact, the implications of our results for the latter model (including the discussion of its
critical behavior) will be presented in a forthcoming publication. We shall only mention in this
regard that, by contrast with the usual approaches, our method does not rely on any approximations
and is thus valid for arbitrary temperature.

\subsection{Free energy per site}

The transfer matrix~$A(x)$ is now given by
\[
  A(x)=\left(
  \begin{array}{ccc}
    q^{-\mu_1}& q^{-\frac12(\mu_1+\mu_2)} & -q^{-\frac{\mu_1}{2}}\\
    q^{K\vep(x)-\frac12(\mu_1+\mu_2)}& q^{-\mu_2} & q^{-\frac{\mu_2}2}\\
     q^{K\vep(x)-\frac{\mu_1}2}& q^{K\vep(x)-\frac{\mu_2}2} & q^{K\vep(x)}
  \end{array}
  \right),
\]
and its eigenvalues are zero and
\[
  \la_{\pm}(x)=a(x)\pm\sqrt{a(x)^2+q^{-(\mu_1+\mu_2)}(q^{K\vep(x)}-1)}\,,
\]
where
\[
  a(x)=\frac12\,\Big(q^{-\mu_1}+q^{-\mu_2}+q^{K\vep(x)}\Big)\,.
\]
Thus the Perron--Frobenius eigenvalue is~$\la_1(x)=\la_+(x)$. (Note that the term under the square
root is clearly positive, since it is strictly greater than $\frac12(q^{-\mu_1}-q^{-\mu_2})^2$.)
Moreover, the matrix $A(x)$ is again diagonalizable for~$0<x<1$, since for these values of~$x$ its
three eigenvalues are simple on account of the inequality~$q^{K\vep(x)}\ne1$. Hence condition~i)
of the previous section is again satisfied. Moreover, when $0<x\le1$ the matrix~$P(x)$ for the
$\su(2|1)$ PF and FI chains can be taken as\footnote{Equation~\eqref{P21} is valid for the HS
  chain only for $0<x<1$. However, as shown in the previous section, condition~ii) is always
  satisfied for this chain.}
\begin{equation}\label{P21}
  P(x)=\left(
    \begin{array}{ccc}
      q^{\frac12(\mu_2-\mu_1)}& 0& q^{\frac12(\mu_2-\mu_1)}\\[3pt]
      1+\frac{q^{-\mu_1}}{\la_+(x)}(q^{K\vep(x)}-1)&
        -q^{\frac{\mu_2}2}&1+\frac{q^{-\mu_1}}{\la_-(x)}(q^{K\vep(x)}-1)\\[3pt]
      q^{K\vep(x)+\frac{\mu_2}2}& 1& q^{K\vep(x)+\frac{\mu_2}2}
    \end{array}
  \right).
\end{equation}
Thus Eq.~\eqref{U11condalt} in this case reads
\[
  \fl
  \left|
\begin{array}{ccc}
  q^{-\frac{\mu_1}2}& 0& q^{\frac12(\mu_2-\mu_1)}\\[3pt]
  q^{-\frac{\mu_2}2}&
                      -q^{\frac{\mu_2}2}&1+\frac{q^{-\mu_1}}{\la_-(1)}(q^{K\vep(1)}-1)\\[3pt]
  1& 1& q^{K\vep(1)+\frac{\mu_2}2}
\end{array}
\right|=\frac{q^{-\frac{\mu_1}2+\mu_2}}{\la_-(1)}\,(1-q^{K\vep(1)})\Big(\la_-(1)+q^{-(\mu_1+\mu_2)}\Big)\ne0.
\]
For the~PF and FI chains~$\vep(1)=1$ and~$\vep(1)=1+\ga\ge1$, respectively, so that the second
factor never vanishes. The last one is positive, since it can be written
as~$\rho-\sqrt{\rho^2-\nu}$ with
\begin{eqnarray*}
  \rho&=\frac12\,(q^{-\mu_1}+q^{-\mu_2}+2q^{-(\mu_1+\mu_2)}+q^{K\vep(1)})>0,\\
  \nu&=q^{-2(\mu_1+\mu_2)}+q^{-(\mu_1+2\mu_2)}+q^{-(2\mu_1+\mu_2)}+q^{-(\mu_1+\mu_2)}>0.
\end{eqnarray*}
Thus condition~ii) is also satisfied in this case. Applying Eq.~\eqref{fTmu} we then obtain, after
a slight simplification,
\begin{equation}
  \label{f21}
  \fl
  f(T,\mu_1,\mu_2)=-\frac12(\mu_1+\mu_2)
  -T\int_0^1\log\Bigl(b(x)+\sqrt{b(x)^2+\e^{-K\be\vep(x)}-1}\,\Bigr)\diff x,
\end{equation}
with
\begin{equation}
  \label{b21}
  b(x)=\frac12\,\e^{-\be[K\vep(x)+\frac12(\mu_1+\mu_2)]}+\cosh\Bigl(\tfrac\be2(\mu_1-\mu_2)\Bigr).
\end{equation}
Comparing the expressions for the eigenvalue~$\la_1(x)=\la_+(x)$ from the~$\su(1|1)$ and
$\su(2|1)$ cases we conclude that the $\su(1|1)$ thermodynamic functions can be formally obtained
from the $\su(2|1)$ ones in the limit~$\mu_2\to-\infty$, as expected. The thermodynamic functions
of the~$\su(2|1)$ chains can be computed without difficulty from Eqs.~\eqref{f21}-\eqref{b21} and
the general equations~\eqref{ngen}, \eqref{varalpha} and~\eqref{thermo}, although the
corresponding expressions are rather cumbersome and shall therefore not be displayed here.

\subsection{Critical behavior}

With the help of the explicit formula~\eqref{f21}, we shall next briefly analyze the criticality
properties of the~$\su(2|1)$ chains of HS type as a function of the chemical potentials and the
interaction strength $K$. We shall see that these chains exhibit a richer critical behavior than
their $\su(1|1)$ counterparts, both in terms of the complexity of the critical region and the
possible values of the central charge.

\begin{figure}[h]
  \centering
  \includegraphics[width=.48\textwidth]{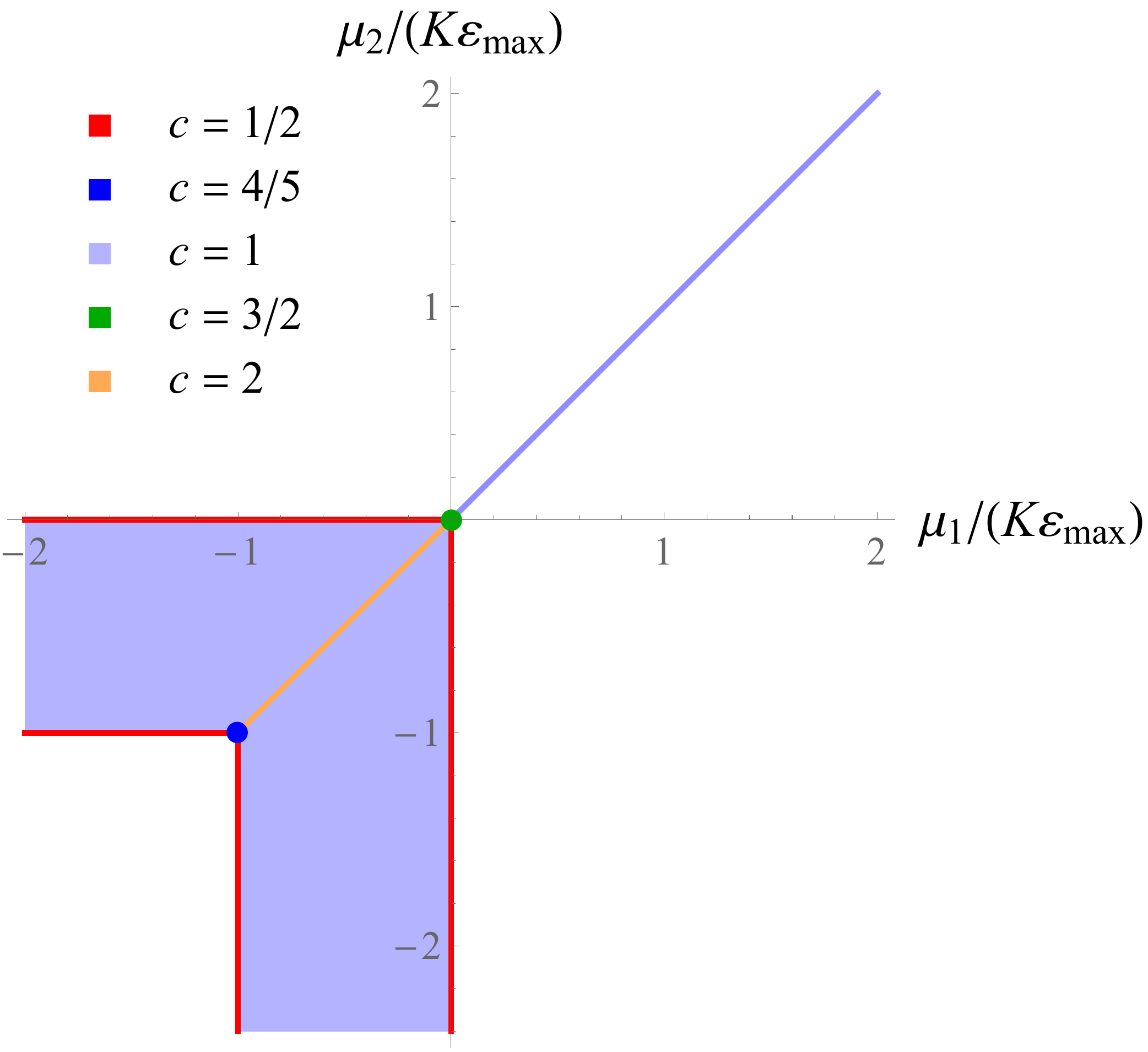}\hfill
  \includegraphics[width=.48\textwidth]{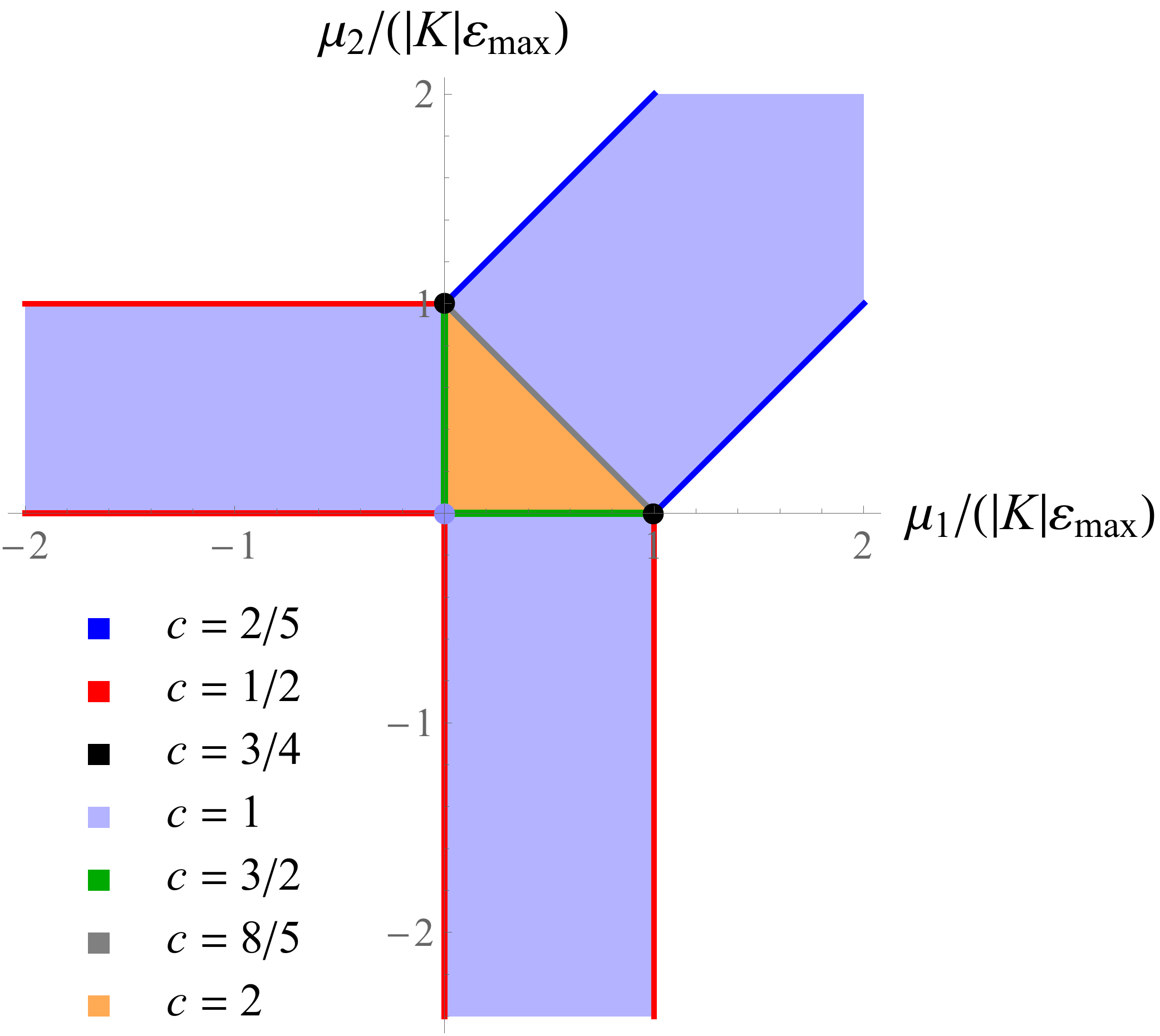}
  \caption{Phase diagram of the~$\su(2|1)$ chains of HS type with~$K>0$ (left) and $K<0$ (right).
    The critical regions, lines, and points are colored according to the central charge $c$ as
    shown in each plot's legend. For $K>0$, the origin and the half-lines~$\mu_1=0>\mu_2$,
    $\mu_2=0>\mu_1$, $\mu_1=\mu_2>0$ are not critical for the FI chain with~$\ga=0$, while the
    point~$(-K\vepmax,-K\vepmax)$ and the half-lines~$\mu_1=-K\vepmax>\mu_2$,
    $\mu_2=-K\vepmax>\mu_1$ are not critical for the HS chain. Similarly, for $K<0$ the origin and
    the half-lines~$\mu_1=0>\mu_2$, $\mu_2=0>\mu_1$, $\mu_1=\mu_2>0$ are not critical for the FI
    chain with~$\ga=0$, while the points~$(|K|\vepmax,0)$, $(0,|K|\vepmax)$, the
    segment~$\{\mu_1+\mu_2=|K|\vepmax,\ 0<\mu_1<|K|\vepmax\}$ and the
    half-lines~$\{\mu_1=|K|\vepmax,\ \mu_2<0\}$, $\{\mu_2=|K|\vepmax,\ \mu_1<0\}$,
    $\mu_1=\mu_2-|K|\vepmax>0$, $\mu_2=\mu_1-|K|\vepmax>0$ are not critical for the HS chain.}
  \label{fig.phase21}
\end{figure}

The phase diagram of the~$\su(2|1)$ chains is presented in Fig.~\ref{fig.phase21}, both for
positive and negative values of~$K$. For the sake of conciseness, we shall only present the
calculations for a few cases of special interest (the remaining ones can be analyzed in a similar
fashion). Note that, by Eq.~\eqref{falbe}, we can assume without loss of generality
that~$\mu_1\ge\mu_2$.

\subsubsection{$K>0$.}
Consider first the case~$K>0$. To begin with, it is clear that the open region~$\mu_1>0$,
$\mu_1>\mu_2$ is not critical. Indeed, in this region
\begin{equation}\label{bB}
  b(x)\underset{T\to0+}\simeq\frac12\Big(1+\e^{-\be(K\vep(x)+\mu_1)}\Big)\e^{\frac\be2(\mu_1-\mu_2)}\equiv
  B(x)\e^{\frac\be2(\mu_1-\mu_2)},
\end{equation}
where we have discarded the exponentially small term~$\e^{-\be(\mu_1-\mu_2)/2}/2$. It easily
follows that~$f(0)=-\frac12(\mu_1+\mu_2)-\frac12(\mu_1-\mu_2)=-\mu_1$, and hence when~$T\to0+$
\begin{eqnarray*}
  \fl
  |f(T)-f(0)|&\simeq\eta
  T\int_0^{1/\eta}\log\Bigl[B(x)+\sqrt{B^2(x)+\e^{-\be(\mu_1-\mu_2)}(\e^{-K\be\vep(x)}-1)}\,\Bigr]\diff
               x\\
             &\le\eta T\int_0^{1/\eta}\log\bigl[2B(x)\bigr]\,\diff x\le
               \eta T\int_0^{1/\eta}\e^{-\be(K\vep(x)+\mu_1)}\diff x\le T\e^{-\be\mu_1},
\end{eqnarray*}
since~$K>0$ by assumption. Similarly, if~$-K\vepmax>\mu_1>\mu_2$ we have
\[
  b(x)\simeq\frac12\Big(1+\e^{\be(K\vep(x)+\mu_1)}\Big)\e^{-\frac\be2(2K\vep(x)+\mu_1+\mu_2)}
  \equiv\widehat B(x)\e^{-\frac\be2(2K\vep(x)+\mu_1+\mu_2)},
\]
so that
\[
  f(0)=-\frac12\,(\mu_1+\mu_2)+\frac\eta2\int_0^{1/\eta}(2K\vep(x)+\mu_1+\mu_2)\diff x=K\vep_0
\]
and
\begin{eqnarray*}
  \fl
  |f(T)-f(0)|&\simeq\eta
               T\int_0^{1/\eta}\log\Bigl[\widehat B(x)+\sqrt{\widehat B^2(x)+\e^{\be(2K\vep(x)+\mu_1+\mu_2)}(\e^{-K\be\vep(x)}-1)}\,\Bigr]\diff
               x\\
             &\le\eta T\int_0^{1/\eta}\log\bigl[2\widehat B(x)\bigr]\,\diff x\le
               \eta T\int_0^{1/\eta}\e^{\be(K\vep(x)+\mu_1)}\diff x\le T\e^{\be(K\vepmax+\mu_1)}.
\end{eqnarray*}
Thus the triangular region $-K\vepmax>\mu_1>\mu_2$ is also noncritical.

Likewise, it can be shown that the open region~$\mu_1>\mu_2$, $-K\vepmax<\mu_1<0$ is critical,
with central charge~$c=1$. To this end, let us denote by~$x_0$ the unique root of the
equation~$K\vep(x)+\mu_1=0$ in the interval~$(0,1/\eta)$. We then have
\[
  f(T)\simeq-\mu_1-\eta
  T\int_0^{1/\eta}\log\Bigl[B(x)+\sqrt{B^2(x)+\e^{-\be(\mu_1-\mu_2)}(\e^{-K\be\vep(x)}-1)}\,\Bigr]\diff
  x,
\]
where the last term under the square root tends to~$0$ as $T\to0+$, and~$K\vep(x)+\mu_1$ is
positive for $x_0<x\le1/\eta$. It follows that
\[
  f(0)=-\mu_1+\eta\int_0^{x_0}(K\vep(x)+\mu_1)\,\diff x=\eta K\int_0^{x_0}\vep(x)\diff
  x-\mu_1(1-\eta x_0),
\]
and thus
\[
  f(T)-f(0)\simeq-T(I_1+I_2)\,,
\]
where
\begin{eqnarray}
  \label{I1}
  \fl
  I_1&=\eta\int_0^{x_0}\log\Bigl[\widehat B(x)+\sqrt{\widehat
       B^2(x)+\e^{\be(2K\vep(x)+\mu_1+\mu_2)}(\e^{-\be K\vep(x)}-1)}\,\Bigr]
       \diff x,\\\fl
  I_2&=\eta\int_{x_0}^{1/\eta}\log\Bigl[B(x)+\sqrt{B^2(x)
       +\e^{-\be(\mu_1-\mu_2)}(\e^{-K\be\vep(x)}-1)}\,\Bigr]
       \diff x.
        \label{I2}
\end{eqnarray}
As in Subsection~\ref{sub.crit11}, the main contribution to both integrals comes from an
arbitrarily small neighborhood of~$x_0$, where $|K\vep(x)+\mu_1|$ is small. In such a
neighborhood, the remaining terms under the square root are negligible, since their exponents are
strictly negative in the whole integration range. We thus have
\[
  I_1\simeq\eta\int_0^{x_0}\log\Bigl[2\widehat B(x)\Bigr]\diff x,\qquad
  I_2\simeq\eta\int_{x_0}^{1/\eta}\log\Bigl[2B(x)\Bigr]\diff x\,.
\]
We now perform in each of these integrals the change of variables~$y=\be|K\vep(x)+\mu_1|$. Taking
into account that in a small neighborhood of~$x_0$
\[
\frac\eta{K\vep'(x)}=\frac\eta{K\vep'(x_0)}+\Or(Ty)=\frac1{\pi v}+\Or(Ty)\,,
\]
where the effective speed of light~$v$ is given by Eq.~\eqref{vdef} with~$\mu=\mu_1$, we easily
obtain
\[
  I_1\simeq\frac T{\pi v}\int_0^{\be|\mu_1|}\log\Bigl(1+\e^{-y}\Bigr)\diff y\,,\qquad
  I_2\simeq\frac T{\pi v}\int_0^{\be|K\vepmax+\mu_1|}\log\Bigl(1+\e^{-y}\Bigr)\diff y
\]
up to a term of order~$\Or(T)$. Extending both integrals to~$+\infty$ (which produces an
exponentially small error in~$\be$, as shown
in Subsection~\ref{sub.crit11}) we finally obtain
\[
  I_{1,2}\simeq\frac T{\pi v}\int_0^{\infty}\log\Bigl(1+\e^{-y}\Bigr)\diff y=\frac{\pi T}{12v}
\]
and therefore
\[
  f(T)-f(0)\simeq -\frac{\pi T^2}{6v}\,.
\]
Comparing with Eq.~\eqref{fCFT} we conclude that the open set~$\mu_1>\mu_2,-K\vepmax<\mu_1<0$ is
indeed critical, with central charge~$c=1$.

The latter results, together with the symmetry of the free energy under exchange of the bosonic
chemical potentials, establish the validity of the phase diagram in Fig.~\ref{fig.phase21} (left)
in the ``generic'' subset~$\mu_1\ne\mu_2$ minus the half-lines~$\mu_1=0\ge\mu_2$,
$\mu_2=0\ge\mu_1$, $\mu_1=-K\vepmax\ge\mu_2$, $\mu_2=-K\vepmax\ge\mu_1$. To end the discussion for
$K>0$, we shall limit ourselves to analyzing the points $(0,0)$ and
$(-K\vepmax,-K\vepmax)$, which illustrate the general procedure.

First of all, at the origin we have
\[
  f(T)=-\eta T\int_0^{1/\eta}\log\Bigl[1+\tfrac12\,\e^{-K\be\vep(x)}+
  \sqrt{2\e^{-K\be\vep(x)}+\tfrac14\,\e^{-2K\be\vep(x)}}\,\Bigr]\diff x.
\]
Performing again the change of variables~$K\be\vep(x)=y$ and proceeding as above we obtain
\[
  f(T)\simeq-\frac{T^2}{\pi v}\int_0^\infty\log\Bigl[1+\tfrac12\,\e^{-y}+
  \sqrt{2\e^{-y}+\tfrac14\,\e^{-2y}}\,\Bigr]\diff y=-\frac{\pi T^2}{4v}\,,
\]
where now $v=K\vep'(0)/(\eta\pi)$. (Of course, the latter formula is clearly \emph{not} valid for
the FI chain with~$\ga=0$, as $\vep'(0)=0$ in this case. In fact, it is straightforward to show
that this chain is \emph{not} critical when~$\mu_1=\mu_2=0$, since~$f(T)-f(0)\propto T^{-3/2}$.)
Comparing with Eq.~\eqref{fCFT} we conclude that (except for the FI chain with~$\ga=0$) the model
is critical in this case with central charge $c=3/2$ (i.e., that of a free CFT with one boson and
one fermion). This is again in agreement with the general result of Ref.~\cite{HB00} for
the~$\su(m|n)$ PF chain with zero chemical potentials, according to which~$c=m-1+n/2$ for~$m\ge1$.
In fact, the same is true for the~$\su(1|2)$ chains with~$\mu_1=\mu_2=0$ (excluding again the FI
chain with~$\ga=0$). Indeed, using Eq.~\eqref{susysymmgen} we readily obtain
\[
  \fl f(T)-f(0)=f(T)-K\vep_0\simeq-\frac{T^2}{\pi v}\int_0^\infty\log\Bigl[\tfrac12+\e^{-y}+
  \sqrt{\tfrac14+2\e^{-y}}\,\Bigr]\diff y=-\frac{\pi T^2}{6v},
\]
so that~$c=1=m-1+n/2$ also in this case.

Consider, finally, the case~$\mu_1=\mu_2=-K\vepmax=-K\vep(1/\eta)\equiv\mu$, for which
\begin{eqnarray*}
  \fl
  f(T)=-\mu-\eta T\int_0^{1/\eta}\log\Bigl[1&+\tfrac12\,\e^{-\be(K\vep(x)+\mu)}\\&+
  \sqrt{\e^{-\be(K\vep(x)+\mu)}+\tfrac14\,\e^{-2\be(K\vep(x)+\mu)}+\e^{-K\be\vep(x)}}\,\Bigr]\diff x,
\end{eqnarray*}
and hence
\[
  f(0)=-\mu+\eta\int_0^{1/\eta}(K\vep(x)+\mu)\diff x=K\vep_0\,.
\]
We thus have
\begin{eqnarray*}
    \fl
  f(T)-f(0)&=-\eta T\int_0^{1/\eta}\log\Bigl[\tfrac12+\e^{\be(K\vep(x)+\mu)}+
  \sqrt{\tfrac14+\e^{\be(K\vep(x)+\mu)}+\e^{\be(K\vep(x)+2\mu)}}\,\Bigr]\diff
             x,\\
  &\simeq-\eta T\int_0^{1/\eta}\log\Bigl[\tfrac12+\e^{\be(K\vep(x)+\mu)}+
  \sqrt{\tfrac14+\e^{\be(K\vep(x)+\mu)}}\,\Bigr]\diff x\,,
\end{eqnarray*}
since~$\mu<0$. The PF and FI chains both satisfy the condition~$\vep'(1)\ne0$. In this case,
performing the usual change of variables~$\be(K\vep(x)+\mu)=-y$ and taking into account the
definition~\eqref{vdefgen} of the effective speed of light~$v$ we obtain
\[
  f(T)-f(0)\simeq-\frac{T^2}{\pi v}\int_0^\infty
  \log\Bigl[\tfrac12+\e^{-y}+
  \sqrt{\tfrac14+\e^{-y}}\,\Bigr]\diff y=-\frac{2\pi T^2}{15v}\,.
\]
Thus the~$\su(2|1)$ PF and FI chains with~$\mu_1=\mu_2=-K\vepmax$ are both critical with central
charge~$c=4/5$, as claimed. Remarkably, this value of~$c$ coincides with the central charge of the
three-state Potts model~\cite{Po52,Bax82} (or, indeed, of any unitary minimal model~\cite{FMS99}
with $p=5$, where~$c=1-6/[p(p+1)]$). Obviously, the latter conclusions do not hold for the HS
chain, since in this case we have~$\vep'(1/\eta)=\vep(1/2)=0$. In fact, it can be shown without
difficulty that this chain is \emph{not} critical when~$\mu_1=\mu_2=-K/4$.

\subsubsection{$K<0$.}
The phase diagram for~$K<0$ is more complex than its $K>0$ counterpart, as is apparent from
Fig.~\ref{fig.phase21}. We shall therefore limit ourselves to discussing the two most interesting
cases, namely the half-line $\mu_2=\mu_1-|K|\vepmax>0$ and the point $(|K|\vepmax,0)$. (Note that,
by Eq.~\eqref{falbe}, the results we shall obtain automatically apply to the half-line
$\mu_1=\mu_2-|K|\vepmax>0$ and the point $(0,|K|\vepmax)$.)

Consider, to begin with, the half-line $\mu_2=\mu_1-|K|\vepmax>0$, on which
\begin{equation}\label{bxKm0}
  b(x)=\frac12\Big(1+\e^{\be(|K|\vep(x)-\mu_1)}+\e^{-|K|\be\vepmax}\Big)\e^{\frac\be2|K|\vepmax}\equiv
  \hat b(x)\,\e^{\frac\be2|K|\vepmax}\,,
\end{equation}
and consequently
\begin{equation}\label{fKm0}
  \fl
  f(T)=-\mu_1-\eta T\int_0^{1/\eta}\log\Bigl[\hat b(x)+\sqrt{\hat
    b(x)^2+\e^{|K|\be(\vep(x)-\vepmax)}-\e^{-|K|\be\vepmax}}\,\Bigr]\diff x.
\end{equation}
Since~$\e^{\be(|K|\vep(x)-\mu_1)}\le\e^{-\be|K|\mu_2}$, the function~$\hat b(x)$ differs from
$1/2$ by terms that are exponentially small in~$\be$. Discarding these terms we obtain the
approximation
\[
  f(T)-f(0)\simeq-\eta
  T\int_0^{1/\eta}\log\Bigl[\tfrac12+\sqrt{\tfrac14+\e^{|K|\be(\vep(x)-\vepmax)}}\,\Bigr]\diff
  x\,,
\]
with~$f(0)=-\mu_1$. We now perform the usual change of variable
\begin{equation}\label{chvar}
  y=|K|\be\big(\vepmax-\vep(x)\big),
\end{equation}
which yields
\begin{equation}\label{fTf0pre}
  f(T)-f(0)\simeq-\frac{\eta
    T^2}{|K|}\int_0^{|K|\be\vepmax}\log\Bigl[\tfrac12+\sqrt{\tfrac14+\e^{-y}}\,\Bigr] \frac{\diff
    y}{\vep'(x)}\,,
\end{equation}
where $x$ should be expressed in terms of~$y$ inverting Eq.~\eqref{chvar}. As before, we can
replace the term~$1/\vep'(x)$ by its approximation near the lower endpoint of the integral (i.e.,
near~$x=1/\eta$), where the integrand is not exponentially small. For the PF and FI
chains~$\vep'(1/\eta)=\vep'(1)\ne0$, so that we can use Eqs.~\eqref{vepx} and~\eqref{vdefgen}
with~$x_0=\eta=1$. Extending the integral to~$+\infty$ (which, as we saw above, produces an
exponentially small error) we thus obtain
\[
  f(T)-f(0)\simeq-\frac{
    T^2}{\pi v}\int_0^{\infty}\log\Bigl[\tfrac12+\sqrt{\tfrac14+\e^{-y}}\,\Bigr]\diff y
  =-\frac{\pi T^2}{15v}\,.
\]
Comparing with Eq.~\eqref{fCFT} we conclude that in this case the PF and FI chains are critical,
with~$c=2/5$. Interestingly, this value of~$c$ does not coincide with the central charge of a
minimal unitary model (nor even, to the best of our knowledge, of a nonunitary minimal model).

The situation is quite different for the HS chain, since in this case~$\vep'(x)$ vanishes at the
endpoint~$x=1/\eta=1/2$. We now have
\[
  y=|K|\be\big(\tfrac12-x\big)^2\,,\qquad\vep'(x)=1-2x=2\,\sqrt{\frac{Ty}{|K|}}\,
\]
and hence
\[
  \fl
  f(T)-f(0)\simeq-\frac{\ka T^{3/2}}{\sqrt{|K|}}\,,\qquad
  \ka\equiv\frac14\int_0^{\infty}y^{-1/2}\log\Bigl[\tfrac12+\sqrt{\tfrac14+\e^{-y}}\,\Bigr]\diff y
  \simeq 0.254707\,.
\]
Thus the $\su(2|1)$ HS chain with $K<0$ is not critical along the half-line~$\mu_2=\mu_1-|K|\vepmax>0$.

Let us now turn to the point~$(|K|\vepmax,0)$, i.e., the endpoint of the half-line just considered.
Using Eq.~\eqref{fKm0} with
\[
  \hat b(x)=\frac12\Big(1+\e^{\be(|K|(\vep(x)-\vepmax)}+\e^{-|K|\be\vepmax}\Big)
\]
(cf.~Eq.~\eqref{bxKm0}) and discarding the exponentially small term $\e^{-|K|\be\vepmax}$ we
easily arrive at the asymptotic formula
\[
  f(T)-f(0)\simeq-\frac{\eta T^2}{|K|}\int_0^{|K|\be\vepmax}\log\Bigl[\tfrac12\big(1+\e^{-y}\big)
  +\sqrt{\tfrac14\,\big(1+\e^{-y}\big)^2+\e^{-y}}\,\Bigr]\,\frac{\diff y}{\vep'(x)}\,,
\]
where~$x$ and~$y$ are related by the change of variables~\eqref{chvar}. As explained above, for
the PF and FI chains we can replace~$\vep'(x)$ by~$\vep'(1/\eta)=\vep'(1)$ and extend the integral
to~$+\infty$, obtaining
\[
  \fl
  f(T)-f(0)\simeq-\frac{T^2}{\pi v}\int_0^{\infty}\log\Bigl[\tfrac12\big(1+\e^{-y}\big)
  +\sqrt{\tfrac14\,\big(1+\e^{-y}\big)^2+\e^{-y}}\,\Bigr]\,\diff y=-\frac{\pi T^2}{8v}\,,
\]
where~$v=|K|\,\vep'(1)/\pi$. Thus the~$\su(2|1)$ PF and FI chains are critical in this case, with
$c=3/4$ (cf.~Eq.~\eqref{fCFT}). Again, this value of~$c$ does not coincide with the central charge
of a unitary (or, to the best of our knowledge, nonunitary) minimal model. Finally, for the HS
chain proceeding as above we again obtain
\[
  f(T)-f(0)\simeq-\frac{\ka T^{3/2}}{\sqrt{|K|}}\,,
\]
where now
\[
  \ka=\frac14\int_0^{\infty}y^{-1/2}\log\Bigl[\tfrac12\big(1+\e^{-y}\big)
  +\sqrt{\tfrac14\,\big(1+\e^{-y}\big)^2+\e^{-y}}\,\Bigr]\,\diff y
  \simeq 0.471976\,.
\]
Thus the~$\su(2|1)$ HS chain with~$K<0$ is not critical at the point~$(|K|\vepmax,0)$.

\subsection{Zero-temperature densities}

From Eqs.~\eqref{ngen} and~\eqref{f21}-\eqref{b21} we obtain the following explicit expressions
for the particle densities of the~$\su(2|1)$ chains of HS type:
\begin{eqnarray}
  \label{n121}
  n_1(T,\mu_1,\mu_2)&=\frac12+\frac\eta2\,\int_0^{1/\eta}\frac{\sinh(\be\mu_-)-\frac12\e^{-\be(K\vep(x)+\mu_+)}}{\sqrt{b(x)^2+\e^{-K\be\vep(x)}-1}}\,\diff
                      x\,,\\ n_2(T,\mu_1,\mu_2)&=n_1(T,\mu_2,\mu_1)\,,\label{n221}\\
  n_3(T,\mu_1,\mu_2)&=
                      \frac\eta2\,\int_0^{1/\eta}\frac{\e^{-\be(K\vep(x)+\mu_+)}}{\sqrt{b(x)^2+\e^{-K\be\vep(x)}-1}}\,\diff
                      x\,,
                      \label{n321}
\end{eqnarray}
with
\[
  \mu_{\pm}\equiv\frac12\,(\mu_1\pm\mu_2)\,.
\]
We shall limit ourselves to analyzing the behavior of these densities at zero temperature. By
contrast with the $\su(2|0)$ and~$\su(1|1)$ cases, we shall show that in this case the bosonic
densities exhibit a first-order (discontinuous) phase transition across the
half-line~$\mu_1=\mu_2>-K\vepmax$ for $K>0$.

\begin{figure}[h]
  \includegraphics[width=.48\textwidth]{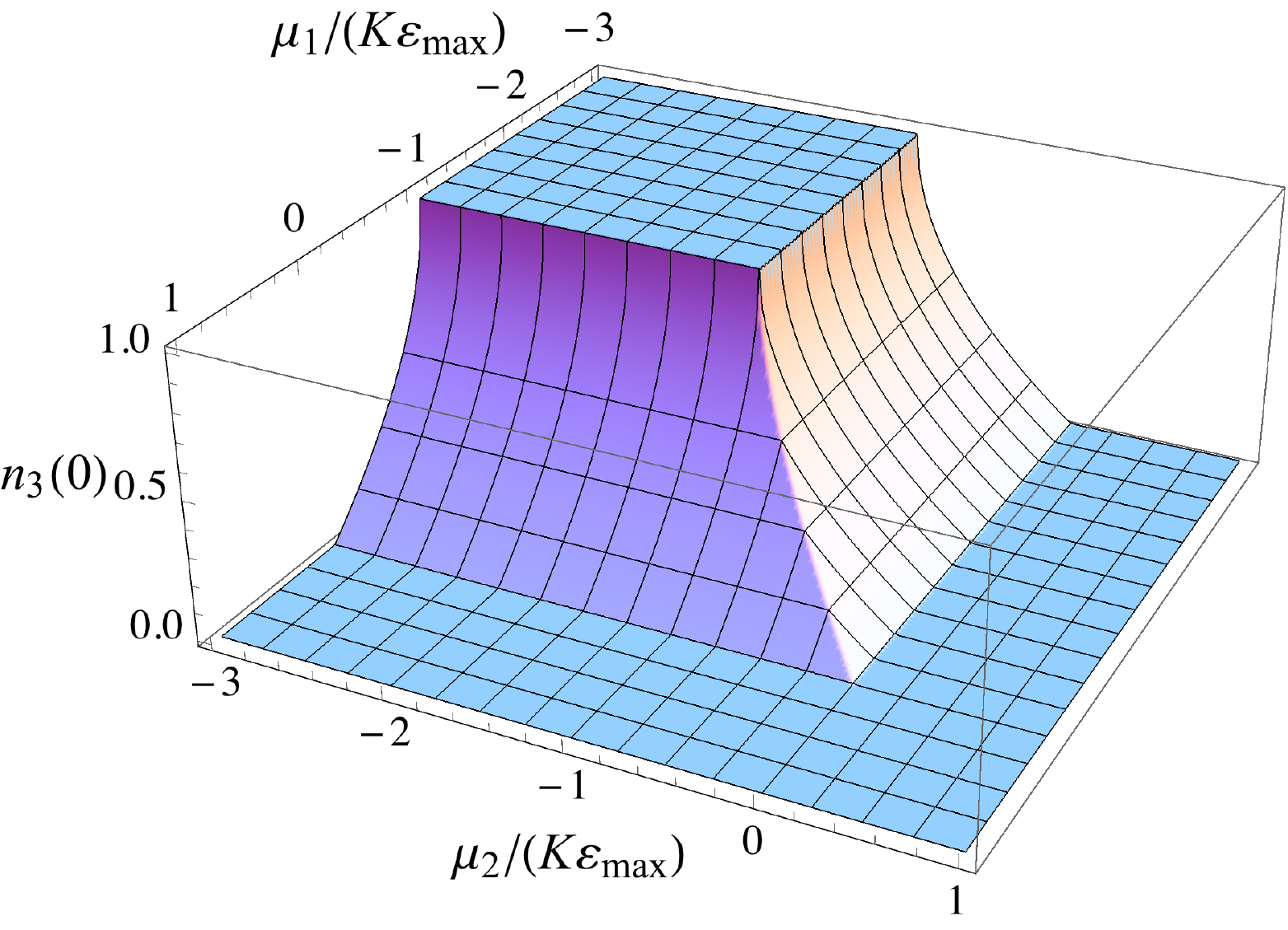}\hfill
  \includegraphics[width=.48\textwidth]{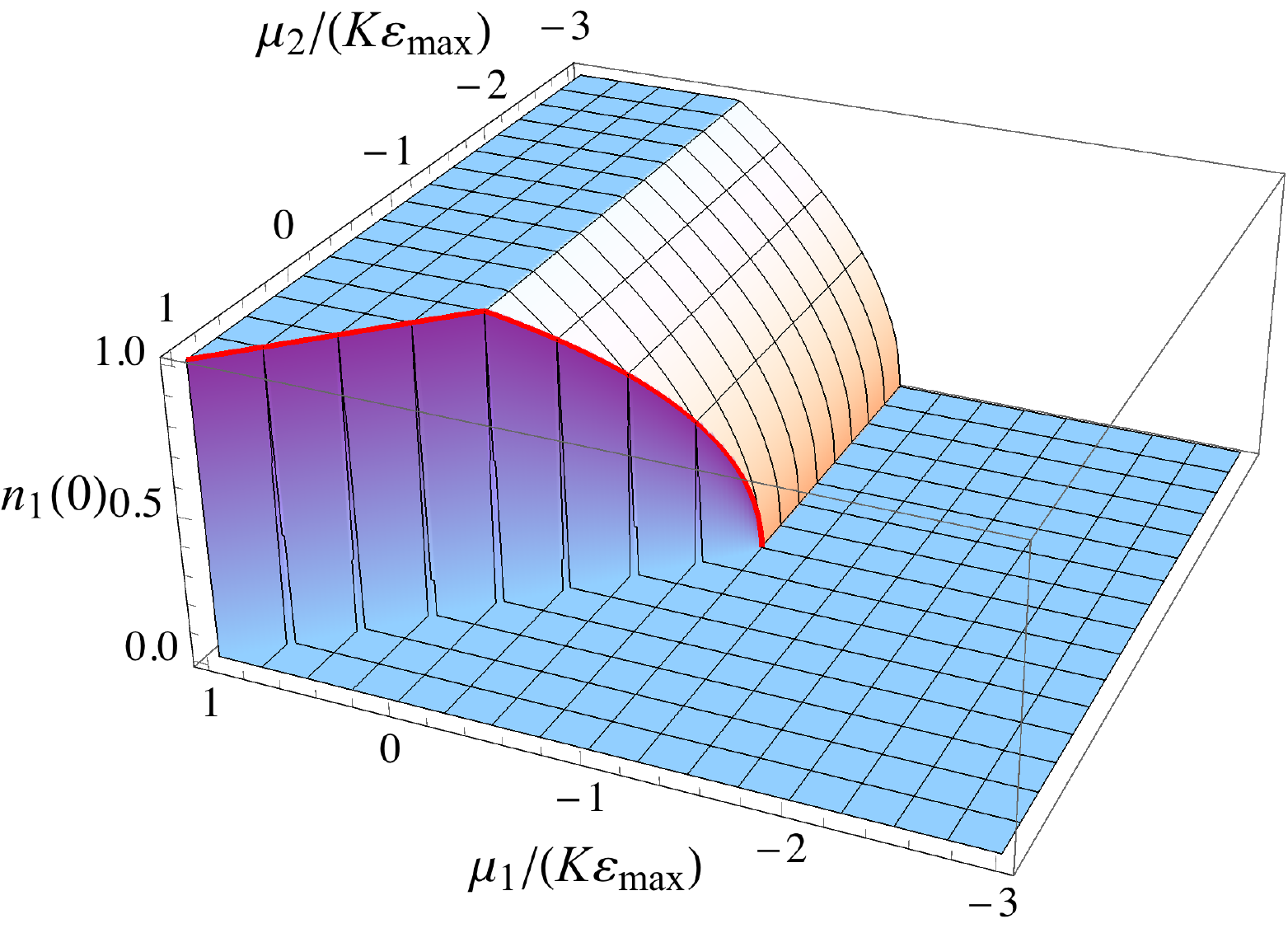}
  \caption{Left: fermion density at zero temperature for the~$\su(2|1)$ HS chain with $K>0$.
    Right: same plot for the bosonic density $n_1$, with a red line drawn to illustrate the
    discontinuity along the half-line $\mu_1=\mu_2\ge-K\vepmax$.}
  \label{fig.n021}
\end{figure}

Suppose, to begin with, that $K>0$, and consider first the fermionic density~$n_3$. Since this
density is clearly symmetric under exchange of the bosonic chemical potentials, we shall restrict
ourselves without loss of generality to the case~$\mu_1\ge\mu_2$. When $\mu_1>\mu_2$, using the
low temperature approximation~\eqref{bB} we have
\begin{equation}
  \label{n3D}
  n_3\simeq\eta\int_0^{1/\eta}\Big[(1+\e^{\be(K\vep(x)+\mu_1)})^2-4\e^{2\be(K\vep(x)+\mu_+)}\Big]^{-1/2}\diff x\,,
\end{equation}
where we have taken into account that $\e^{\be(K\vep(x)+2\mu_+)}\ll\e^{2\be(K\vep(x)+\mu_+)}$ as
$T\to0+$ for $0<x\le1/\eta$. When~$K\vep(x)+\mu_1>0$, the term~$\e^{2\be(K\vep(x)+\mu_1)}$
dominates over the remaining ones as~$T\to0+$, so that the integrand tends to zero in this region.
On the other hand, when $K\vep(x)+\mu_1<0$ the integrand clearly tends to~$1$ as~$T\to0+$. We thus
have
\[
  n_3(0)=\eta|I(\mu_1)|\,,
\]
where~$|I(\mu)|$ denotes the length of the (possibly empty)
interval
\[
  I(\mu)=\{x\in[0,1/\eta]:K\vep(x)+\mu<0\}.
\]
Denoting by~$x_{0}(\mu)$ the unique root of the equation~$K\vep(x)+\mu=0$ in the
interval~$[0,\eta]$ (cf.~Eq.~\eqref{x0def}), we conclude that for~$K>0$ and~$\mu_1>\mu_2$ the zero
temperature fermionic density is given by
\[
n_3(0)=\cases{0,& $\mu_1>0,$\\ \eta\,x_{0}(\mu_1),& $-K\vepmax\le\mu_1\le0,$\\1,&$\mu_1<-K\vepmax$\,.}
\]
Likewise, for~$K>0$ and~$\mu_1=\mu_2\equiv\mu$ we have
\[
  n_3=\eta\int_0^{1/\eta}\Big[1+4\e^{\be(K\vep(x)+\mu)}+4\e^{\be(K\vep(x)+2\mu)}\Big]^{-1/2}\diff x\,, 
\]
so that again~$n_3(0)=\eta|I(\mu)|$. From the previous formulas and the symmetry of~$n_3$ under
the exchange of~$\mu_1$ with~$\mu_2$ we obtain the following expression for~$n_3(0)$, valid in the
whole~$(\mu_1,\mu_2)$ plane when $K>0$:
\begin{equation}
  \label{n30final}
  n_3(0)=\cases{0,& $\max(\mu_1,\mu_2)>0$,\\
    \eta x_0(\max(\mu_1,\mu_2)),& $-K\vepmax\le\max(\mu_1,\mu_2)\le0$,\\
  1,& $\max(\mu_1,\mu_2)<-K\vepmax$\,.}
\end{equation}
It is apparent from the latter expression that~$n_3$ is continuous, but its gradient is
discontinuous along the segment~$-K\vepmax\le\mu_1=\mu_2\le0$ and the half-lines $\mu_1=0>\mu_2$,
$\mu_2=0>\mu_1$, $\mu_1=-K\vepmax>\mu_2$, $\mu_2=-K\vepmax>\mu_1$ (cf.~Fig.~\ref{fig.n021}, left).

\begin{figure}[t]
  \includegraphics[height=5cm]{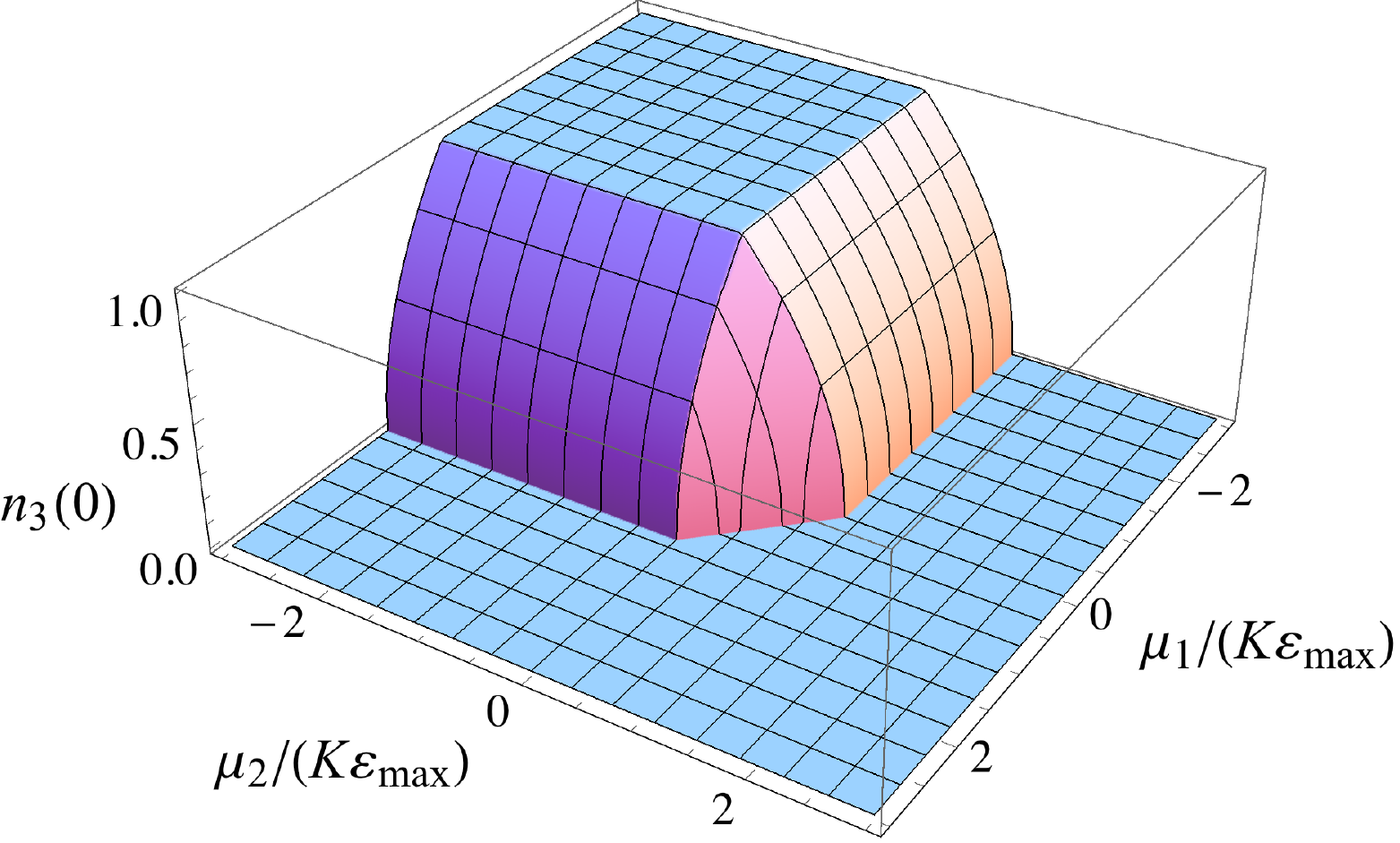}\hfill
  \includegraphics[height=5cm]{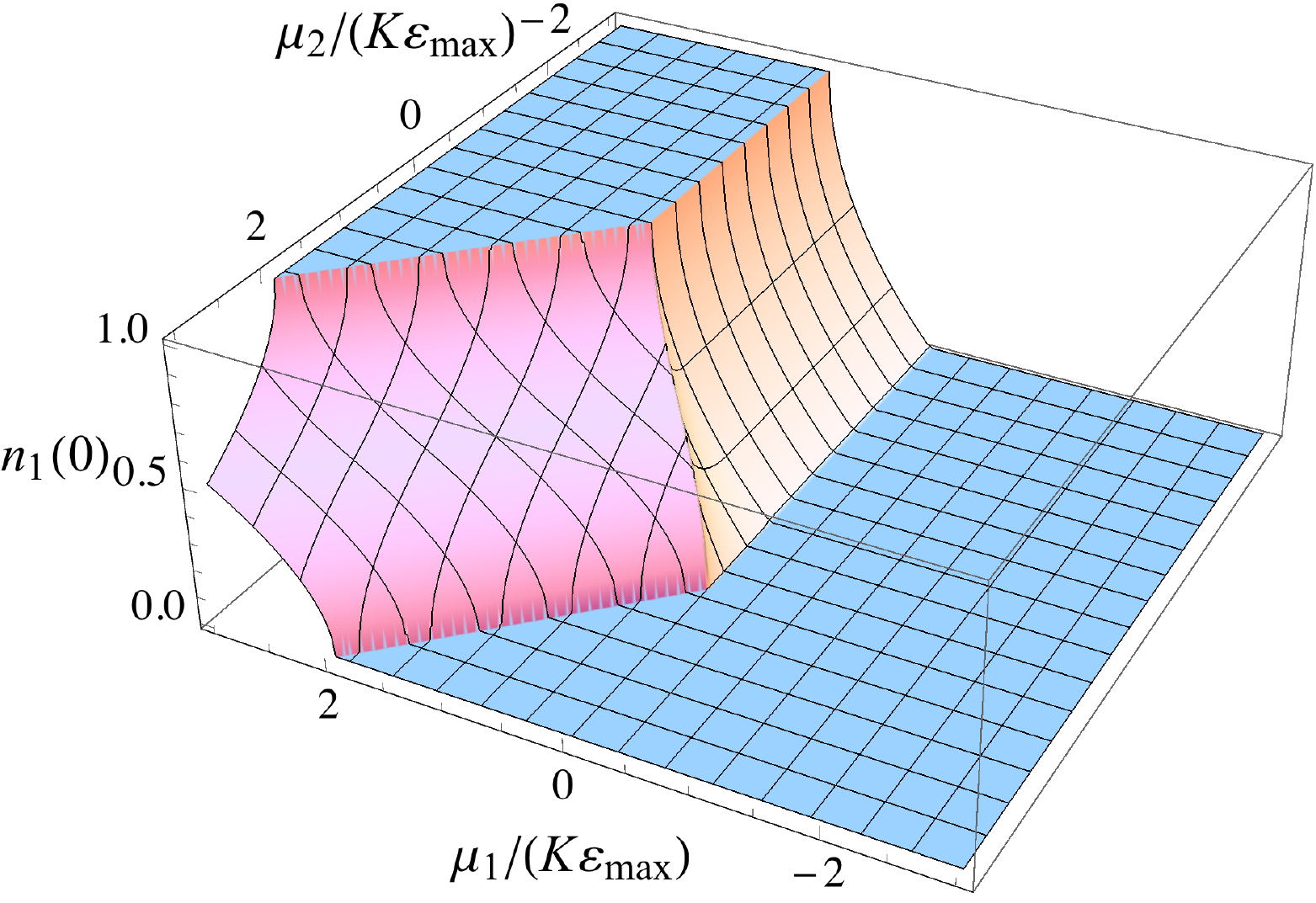}
  \caption{Left: fermion density at zero temperature for the~$\su(2|1)$ HS chain with $K<0$.
    Right: same plot for the bosonic density $n_1$.}
  \label{fig.n021m}
\end{figure}

Consider next the density~$n_1$ of the first species of bosons, which for~$\mu_1>\mu_2$ can be
expressed as
\begin{equation}\label{n1D}
  \fl
  n_1=\frac12\,(1-n_3)+\frac\eta2\,\int_0^{1/\eta}D(x)^{-1/2}\diff x\,,\qquad
  D(x)\equiv\frac{b(x)^2+\e^{-K\be\vep(x)}-1}{\sinh^2(\be\mu_-)}\,.
\end{equation}
At low temperatures we have
\[
  D(x)=\Big(1+\e^{-\be(K\vep(x)+\mu_1)}\Big)^2+\Or(\e^{-2\be\mu_-}),
\]
so that
\[
  \fl
  \lim_{T\to0+}\int_0^{1/\eta}D(x)^{-1/2}\diff x
  =\Big|\{x\in[0,1/\eta]:K\vep(x)+\mu_1>0\}\Big|
  =\frac1\eta-|I(\mu_1)|,
\]
and thus
\[
  n_1(0)=\frac12\Big(1-n_3(0)\Big)+\frac12\Big(1-\eta |I(\mu_1)|\Big)\,.
\]
Using the previous formula for~$n_3(0)$ we obtain the following expression for~$n_1(0)$ when~$K>0$
and~$\mu_1>\mu_2$:
\[
  n_1(0)=\cases{1\,,& $\mu_1>0,$\\ 1-\eta\,x_{0}(\mu_1),&
    $-K\vepmax\le\mu_1\le0,$\\0,&$\mu_1<-K\vepmax$\,.}
\]
On the other, when~$K>0$ and~$\mu_2>\mu_1$ we have
\[
  n_1=\frac12\,(1-n_3)-\frac\eta2\,\int_0^{1/\eta}D(x)^{-1/2}\diff x\,,
\]
so that proceeding as before we obtain
\[
  n_1(0)=\frac12\Big(1-n_3(0)\Big)-\frac12\Big(1-\eta |I(\mu_2)|\Big)=0\,.
\]
Finally, when~$\mu_1=\mu_2$ by Eq.~\eqref{n1D} we simply have~$n_1(0)=(1-n_3(0))/2$. Taking into
account the symmetry of~$n_1$ under exchange of~$\mu_1$ with~$\mu_2$ we obtain the following
general expression for the latter density (for $K>0$):
\begin{equation}
  \label{n1021}
  n_1(0)=\cases{0,& $\mu_1<\mu_2$ or $-K\vepmax\ge\mu_1>\mu_2$,\\
    \tfrac12,& $\mu_1=\mu_2\ge0,$\\
    \tfrac12(1-\eta x_0(\mu_1)),& $-K\vepmax\le\mu_1=\mu_2<0,$\\
    1-\eta x_0(\mu_1),& $-K\vepmax\le\mu_1\le0,\en \mu_1>\mu_2$,\\
  1,& $\mu_1>0,\en \mu_1>\mu_2$\,.}
\end{equation}
It follows from the previous expression that~$n_1$ is discontinuous along the
half-line~$\mu_1=\mu_2\ge-K\vepmax$, and has a discontinuous gradient along the
half-lines~$\mu_1=-K\vepmax\ge\mu_2$ and~$\mu_1=0\ge\mu_2$. Thus in this case the bosonic
density~$n_1$ (and hence $n_2$) presents both first- and second-order phase transitions for
appropriate values of the chemical potentials~$\mu_1$ and~$\mu_2$.
A very similar calculation, that we shall omit for the sake of conciseness, shows that when~$K<0$
the fermionic density is given by
\begin{equation*}
  \fl
  n_3(0)=\cases{0,&$\mu_1-\mu_2\ge|K|\vepmax$ or $\max(\mu_1,\mu_2)\ge|K|\vepmax$,\\
    1-\eta x_0(\max(\mu_1,\mu_2)),& $0\le\max(\mu_1,\mu_2)\le|K|\vepmax,\;\min(\mu_1,\mu_2)\le0$,\\
    1-\eta x_0(\mu_1+\mu_2),& $\mu_1\ge0,\;\mu_2\ge0,\;\mu_1+\mu_2\le|K|\vepmax$,\\
    1,& $\mu_1\le0,\;\mu_2\le0$,}
\end{equation*}
while the bosonic density $n_1(0)$ reads
\begin{eqnarray*}
  \fl
  &n_1(0)=\\\fl&
  \cases{0,& $\mu_1\le0$ or $\mu_2-\mu_1\ge|K|\vepmax$,\\
    \eta x_0(\mu_1),& $0\le\mu_1\le|K|\vepmax,\; \mu_2\le0$,\\
  \tfrac\eta2\Big[x_0(\mu_1+\mu_2)+\sgn(\mu_1-\mu_2)x_0(|\mu_1-\mu_2|)\Big],&
  $\mu_1\ge0,\;\mu_2\ge0,\;\mu_1+\mu_2\le|K|\vepmax,$\\
  \tfrac12\Big[1+\eta\sgn(\mu_1-\mu_2)x_0(|\mu_1-\mu_2|)\Big],&
  $|\mu_1-\mu_2|\le|K|\vepmax,\;\mu_1+\mu_2\ge|K|\vepmax,$\\
  1,& $\mu_1-\mu_2\ge|K|\vepmax,\;\mu_1\ge|K|\vepmax$\,.
}
\end{eqnarray*}
It can be easily checked that both densities (and hence the remaining one~$n_2(0)$) are
continuous, although their gradient is discontinuous along several segments and half-lines
(cf.~Fig.~\ref{fig.n021m}). Thus when $K<0$ the chains~\eqref{H0H1}--\eqref{FI} exhibit only
second-order phase transitions at zero temperature.

\section{The $\su(2|2)$ chains}\label{sec.su22}

The eigenvalues of the~$\su(2|2)$ transfer matrix
\[
  A(x)=\left(
    \begin{array}{cccc}
      q^{-\mu_1}& q^{-\frac12(\mu_1+\mu_2)} & q^{-\frac12(\mu_1+\mu_3)}& q^{-\frac{\mu_1}{2}}\\
       q^{K\vep(x)-\frac12(\mu_1+\mu_2)}& q^{-\mu_2} & q^{-\frac12(\mu_2+\mu_3)} & q^{-\frac{\mu_2}2}\\
      q^{K\vep(x)-\frac12(\mu_1+\mu_3)}& q^{K\vep(x)-\frac12(\mu_2+\mu_3)}& q^{K\vep(x)-\mu_3} & q^{-\frac{\mu_3}2}\\
      q^{K\vep(x)-\frac{\mu_1}2}& q^{K\vep(x)-\frac{\mu_2}2} & q^{K\vep(x)-\frac{\mu_3}2}& q^{K\vep(x)}
    \end{array}
  \right)
\]
are zero (double) and
\[
  \la_{\pm}(x)=a(x)\pm\sqrt{a(x)^2+(q^{K\vep(x)}-1)(q^{-(\mu_1+\mu_2)}-q^{K\vep(x)-\mu_3})}\,,
\]
where now
\[
  a(x)=\frac12\,\Big(q^{-\mu_1}+q^{-\mu_2}+q^{K\vep(x)-\mu_3}+q^{K\vep(x)}\Big)\,.
\]
Thus the Perron--Frobenius eigenvalue is again~$\la_1(x)=\la_+(x)$. However, in this case $A(x)$
is \emph{not} diagonalizable when~$x\in(0,1)$. More precisely, for~$0< x< 1$ its Jordan canonical
form can be taken as
\[
  J(x)=\left(
    \begin{array}{cccc}
      \la_+(x)& 0& 0& 0\\
      0& \la_-(x)& \de_{0,\la_-(x)}& 0\\
      0& 0& 0& 1\\
      0& 0& 0& 0       
    \end{array}
  \right)
\]
where~$\de_{0.\la_-(x)}$ denotes Kronecker's delta. Indeed, the eigenvalue~$\la_-(x)$ vanishes if
and only if $K\vep(x)=\mu_3-\mu_1-\mu_2$ (i.e., for at most two values of~$x$ for the HS chain and
one such value for the PF and FI chains), and when this happens it can be shown that the geometric
multiplicity of the zero eigenvalue is one\footnote{The eigenvalue $\la_-(x)$ also vanishes at
  $x=0$ and, in the case of the HS chain, at $x=1$. The matrix~$J_0$ (or $J_1$, in the latter
  case) is diagonal, although this has no influence on condition~i).}. It follows that the product
$J_1\cdots J_{N-1}$ is diagonal in either case provided that~$N\ge4$, so that the first condition
is again satisfied. As to the second condition, we shall not present the matrix~$P(x)$ in this
case, since it is too unwieldy to display. However, a long but elementary calculation with the
help of the symbolic package \emph{Mathematica\texttrademark} shows that the latter condition is
also satisfied in this case. Thus the free energy per spin is again given by Eq.~\eqref{fTmu}, or
equivalently
\begin{eqnarray}
  \fl
  f(&T,\mu_1,\mu_2,\mu_3)=-\frac12\,(\mu_1+\mu_2)\nonumber
  \\
  \fl
  &-T\int_0^1\log\Bigl[b(x)+\sqrt{b(x)^2-
    (1-\e^{-K\be\vep(x)})(1-\e^{-\be(K\vep(x)+\mu_1+\mu_2-\mu_3)})}\,\Bigr]
    \diff x,
  \label{f22}
\end{eqnarray}
where now
\begin{equation}\label{b22}
  b(x)=\e^{-\be[K\vep(x)+\frac12(\mu_1+\mu_2-\mu_3)]}\cosh\Bigl(\tfrac\be2\mu_3\Bigr)+
  \cosh\Bigl(\tfrac\be2(\mu_1-\mu_2)\Bigr).
\end{equation}
Comparing with Eqs.~\eqref{f21}-\eqref{b21} we deduce that the thermodynamic functions of the
$\su(2|1)$ chain can be formally obtained from those of its~$\su(2|2)$ counterpart in the
limit~$\mu_3\to-\infty$.

Although the thermodynamic functions can be computed without difficulty from
Eqs.~\eqref{f22}-\eqref{b22}, we shall not present here the corresponding expressions as they are
excessively long. An important exception occurs when all chemical potentials vanish, so that the
previous expression for the free energy per site simplifies to
\begin{equation}\label{f2220}
  f(T,0,0,0)=-2\eta T\int_0^{1/\eta}\log\Big(1+\e^{-\frac K2\be\vep(x)}\Big)\diff x=2f^{(2|0)}(T,0)\,.
\end{equation}
Thus the energy, specific heat and entropy of the~$\su(2|2)$ chains of HS type with~$\mu_\al=0$
for all~$\al$ are twice the corresponding values for their~$\su(2|0)$ counterparts with $\mu_1=0$
and the same interaction strength~$K$. Moreover, since the latter chains are all critical (except
for the FI chain with~$\ga=0$), with central charge~$c=1$, it follows that the $\su(2|2)$ chains
with zero chemical potentials are also critical with $c=2$ (again with the exception of the FI
chain with~$\ga=0$). This is once more in agreement with the general formula for the central
charge of the $\su(m|n)$ PF chain with zero chemical potentials in Ref.~\cite{HB00}.

\begin{figure}[t]
  \includegraphics[width=.48\textwidth]{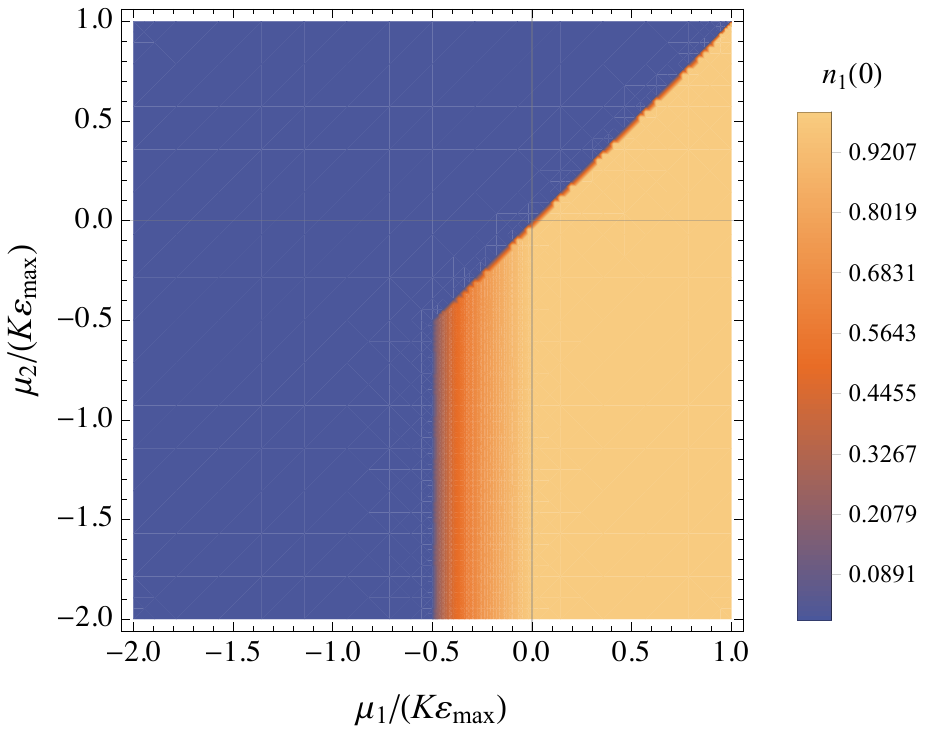}\hfill
  \includegraphics[width=.48\textwidth]{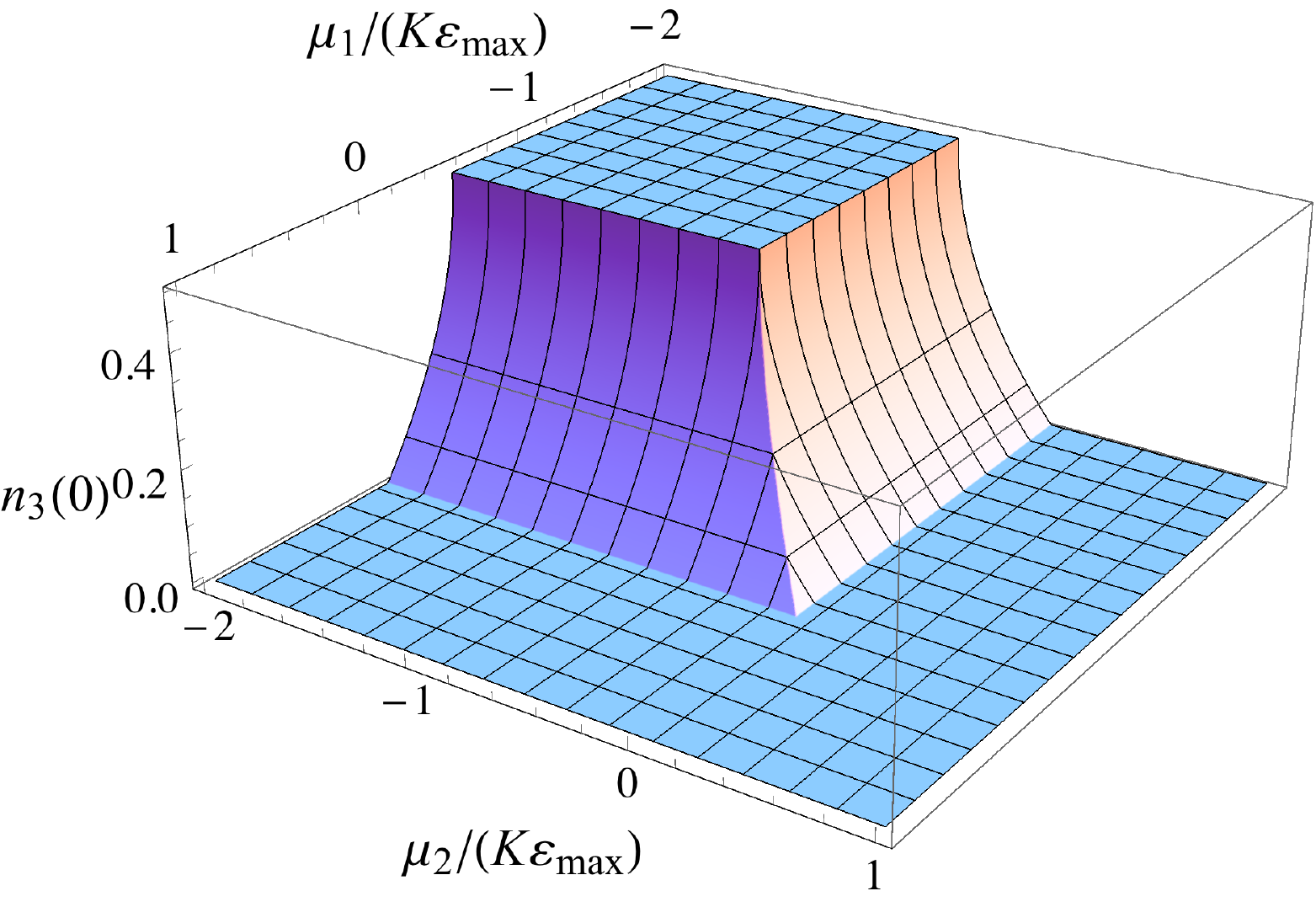}\\
  \includegraphics[width=.48\textwidth]{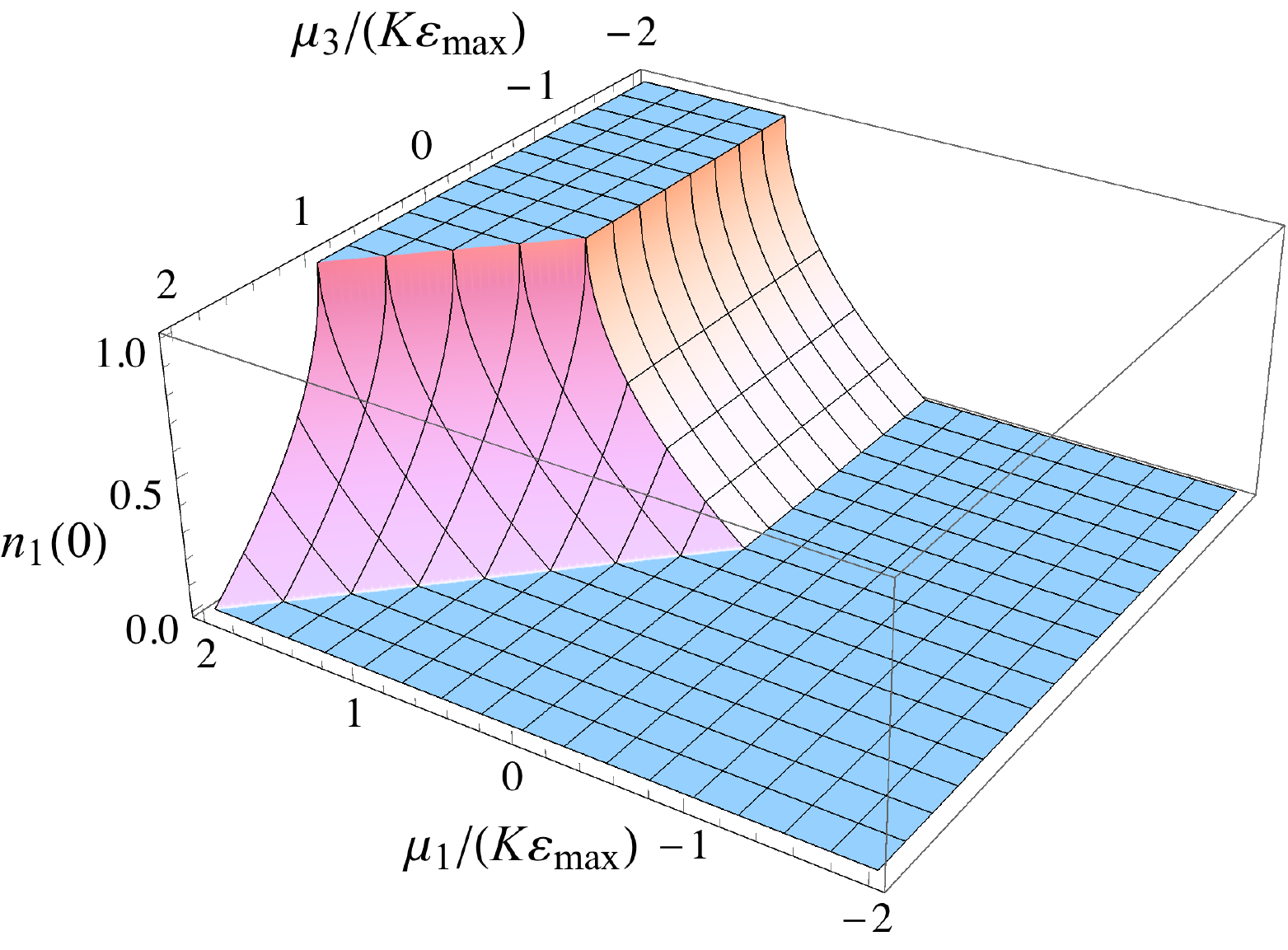}\hfill
  \includegraphics[width=.48\textwidth]{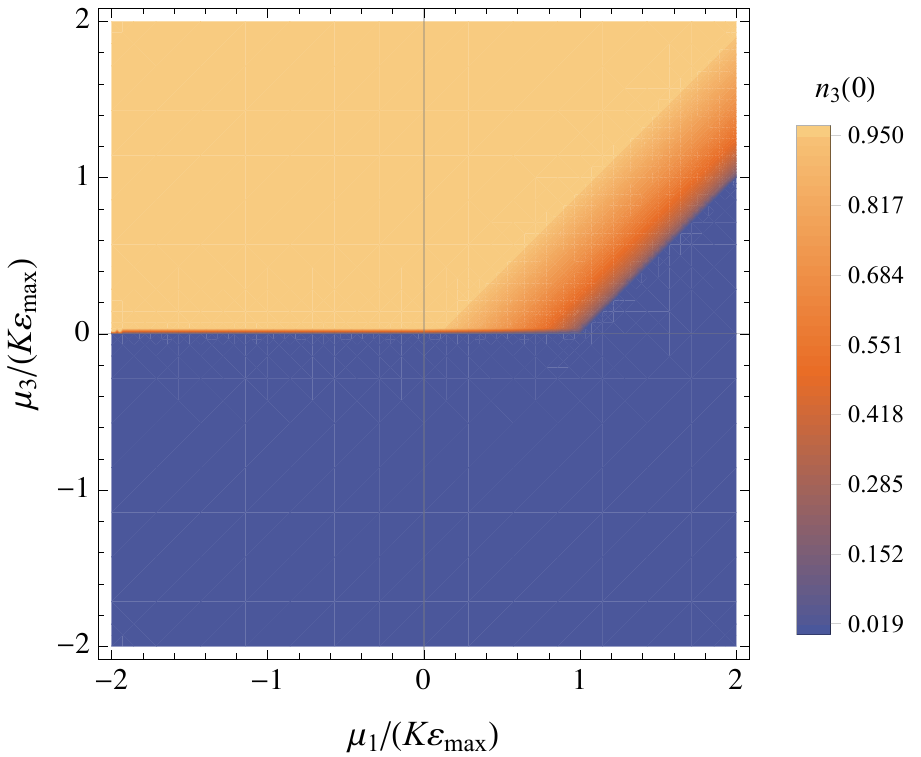}
  \caption{Contour and surface plots of the densities~$n_1(0)$ (left) and $n_3(0)$ (right) for the
    $\su(2|2)$ HS chain with~$\mu_3=0$, $K>0$ (top), and $\mu_2=0$, $K<0$ (bottom).}
  \label{fig.n022}
\end{figure}

We shall not exhaustively analyze the zero-temperature behavior of the particle densities, given
the relative complexity of their explicit expressions. However, our numerical calculations based
on the latter expressions clearly indicate that for~$K>0$ the fermionic densities~$n_{3,4}$
exhibit only second-order phase transitions at~$T=0$, while the bosonic ones~$n_{1,2}$ undergo
also a first-order phase transition across (a subset of) the plane~$\mu_1=\mu_2$ (see
Fig.~\ref{fig.n022}, top). On the other hand, from Eq.~\eqref{nalmnnm} we deduce that
\[
  n_\al(\mu_1,\mu_2,\mu_3;K)=n_{5-\al}(-\mu_1,\mu_3-\mu_1,\mu_2-\mu_1;-K).
\]
From this equation and the previous observation it follows that for $K<0$ the situation is
reversed, i.e., the fermionic densities feature only second-order phase transitions at zero
temperature while the bosonic ones present also a first-order phase transition across (a subset
of) the plane~$\mu_3=0$. Again, this statement is fully corroborated by our numerical calculations
(cf.~Fig.~\ref{fig.n022}, bottom).

\section{Conclusions}\label{sec.conc}

In this paper we study the thermodynamics and critical behavior of the three families of
$\su(m|n)$ supersymmetric spin chains of Haldane--Shastry type with an additional chemical
potential term. Our analysis is based on two main results, namely the computation in closed form
of the partition function for an arbitrary (finite) number of spins and the derivation of a simple
description of the spectrum in terms of supersymmetric motifs. By means of the transfer matrix
method, we obtain an analytic expression for the free energy per site, and hence the main
thermodynamic functions, in the thermodynamic limit. For the~$\su(1|1)$, $\su(2|1)$
(or~$\su(1|2)$) and $\su(2|2)$ chains, we identify the values of the chemical potentials for which
the models are critical (gapless) by studying the low-temperature behavior of the free energy per
site. In particular, we show that the central charge can take rational values that are not
integers or half-integers, thus excluding the equivalence to a CFT with free bosons and/or
fermions. Note, in this respect, that in order to establish the equivalence at low energies of a
critical quantum system (in the thermodynamic limit) to a CFT Eq.~\eqref{fCFT} is necessary but
not sufficient. For instance, the system's ground state should have finite degeneracy for this
equivalence to hold. Although we shall not go into specifics here, it is apparent that this
condition will hold for the Hamiltonian~\eqref{H0H1} for generic values of the chemical
potentials, since the term $H_1$ will break the degeneracy that the ground state of~$H_0$ may
possess (see Ref.~\cite{BBS08} for more details). We also analyze the existence of
zero-temperature phase transitions in the spin densities. More precisely, we show that in
the~$\su(1|1)$ case there are only second-order (continuous) phase transitions, while
for~$\su(2|1)$ and~$\su(2|2)$ first-order (discontinuous) phase transitions occur in the bosonic
densities when the interaction strength~$K$ is positive. Moreover, for~$\su(1|2)$ and $\su(2|2)$
the fermionic densities also undergo a first-order phase transition at $T=0$ for negative values
of~$K$.

The present work suggests several possible lines for future research. In the first place, the
previous results and those of Ref.~\cite{EFG12} seem to indicate that first-order transitions in
the spin densities at~$T=0$ will occur provided that~$m+n>2$. It would be of interest to ascertain
the validity of this conjecture, for instance by studying the behavior of these densities in
the~$\su(3)$ case. It would also be of interest to study the existence of a suitable recurrence
relation for the (generalized) partition function of the models under study for arbitrary values
of the chemical potentials~$\mu_\al$, similar to the one derived in Ref.~\cite{HB00}
for~$\mu_\al=0$. Such a relation could then be used, by the method explained in the latter
reference, to compute the central charge without explicit knowledge of the highest eigenvalue of
the transfer matrix. Finally, another open problem that comes to mind is the extension of the
above results to spin chains of HS type associated with root systems different than~$A_{N-1}$,
like~$BC_N$, $B_N$ or $D_N$. A key step in this endeavor would be the deduction of a description
of the spectrum in terms of suitable motifs. Note, in this respect, that the partition function of
the supersymmetric Polychronakos--Frahm spin chain of $BC_N$ type with~$\mu_\al=0$ is
known~\cite{BFGR09}, and the same is true for the ordinary (non-supersymmetric) PF chain of $D_N$
type~\cite{BFG09} and the~$BC_N$, $B_N$ and~$D_N$ Haldane--Shastry
chains~\cite{EFGR05,BFG13,BFG11}. However, for neither of these models an expression of the
energies in terms of motifs akin to Eq.~\eqref{specH} has been found so far.

\section*{Acknowledgments}

FF, AG-L and MAR were partially supported by Spain's MINECO under research grant no.\
FIS2015-63966-P.

\section*{Appendix}

In this Appendix we provide a justification of the different behavior of the free energy per site
at finite $N$ of the~$\su(1|1)$ chains for positive and negative values of~the chemical potential
$\mu$ when~$K>0$ (see, e.g.,~Fig.~\ref{fig.fNmu}). For simplicity, we shall restrict ourselves
to the PF and FI chains (the argument for the~HS chain is very similar). To begin with,
for~$m=n=1$ the value of~$f$ at zero temperature for the PF and FI chains is given by
\[
  f(0)=\cases{K\vep_0\,, &\hfill$\mu<-K\vepmax$,\hfill\cr
    K\int_0^{\mathrlap{x_0}}\vep(x)\,\diff x-\mu(1-x_0)\,,&\hfill$-K\vepmax\le\mu\le 0$,\hfill\cr
    -\mu\,,&\hfill$\mu>0$,\hfill}
\]
with $K\vep(x_0)+\mu=0$ and~$\vep_0$ defined in Eq.~\eqref{vep0} (cf.~Section~\ref{sec.su11}). On
the other hand, from Eq.~\eqref{specH} for $m=n=1$ it follows that the ground state of
the~$\su(1|1)$ PF and FI chains with $K>0$ is nondegenerate for~$\mu<-K\vepmax$ and~$\mu>0$, since
it is obtained from the unique values~$\bsi=(2,\dots,2)$ and~$\bsi=(1,\dots,1)$, respectively. The
ground state energy~$\Egs$ is thus given by
\[
  \Egs=\cases{J\sum_{i=1}^{N-1}\cE(i)=K\sum_{i=1}^{N-1}\vep(x_i)\underset{N\to\infty}\simeq
    NK\vep_0\,,& $\mu<-K\vepmax$\,,\cr -N\mu\,,&$\mu>0$.}
\]
For~$-K\vepmax\le\mu\le0$ and large $N$, the ground state (still nondegenerate, or with very
little degeneration) is instead obtained from a vector~$\bsi$ of the
form~$(2,\dots,2,1,\dots,1,2)$ with $Nx$ $2$'s and~$N(1-x)$ $1$'s (where~$0\le x\le 1$). The
parameter~$x$ is easily computed by minimizing the energy~$E(x)$ corresponding to such a
vector~$\bsi$, given by
\[
  E(x)=J\sum_{i=1}^{Nx-1}\cE(i)-\mu N(1-x)\underset{N\to\infty}\simeq
  N\left(K\int_0^x\vep(s)\,\diff s-\mu(1-x)\right).
\]
Differentiating with respect to~$x$ we easily obtain~$K\vep(x)+\mu=0$, so that~$x=x_0$.
Thus in this case we have
\begin{eqnarray*}
  E_{\mathrm{GS}}&\simeq
  J\,\sum_{i=1}^{\mathclap{Nx_0-1}}\cE(i)-N\mu(1-x_0)=K\,\sum_{i=1}^{\mathclap{Nx_0-1}}\vep(x_i)-N\mu(1-x_0)\\
  &\underset{N\to\infty}\simeq
  N\bigg[K\int_0^{\mathrlap{x_0}}\vep(x)\,\diff x-\mu(1-x_0)\bigg],\qquad -K\vepmax\le\mu\le0\,.
\end{eqnarray*}
In all cases, when~$T\to0$ we have~$\cZ\simeq\e^{-\be\Egs}$, and consequently $f_N(0)=\Egs/N$.
From the previous expressions for the ground state energy we indeed conclude that
\[
  \lim_{N\to\infty}f_N(0)=f(0)\,,
\]
as expected. However, for large though finite~$N$ the value of $f_N(0)$ is \emph{exactly} equal
to~$f(0)$ when~$\mu\ge0$, while for~$\mu<0$
\[
  f(0)-f_N(0)=K\bigg[\int_0^{\mathrlap{x_0}}\vep(x)\,\diff x-\frac1N\,\sum_{i=1}^{\mathclap{Nx_0-1}}\vep(x_i)\bigg]
\]
(where~$x_0$ should be interpreted as~$1$ for~$\mu<-K\vepmax$) is nonvanishing and~$O(N^{-1})$.

\section*{References}


\begin{thebibliography}{10}
\providecommand{\url}[1]{\texttt{#1}}
\providecommand{\urlprefix}{URL }
\providecommand{\eprint}[2][]{\url{#2}}

\bibitem{Ha88}
Haldane F~D~M, \emph{{E}xact {J}astrow--{G}utzwiller resonating-valence-bond
  ground state of the spin-$1/2$ antiferromagnetic {H}eisenberg chain with
  $1/r^2$ exchange,}  1988 \emph{Phys. Rev. Lett.} \textbf{60} 635

\bibitem{Sh88}
Shastry B~S, \emph{{E}xact solution of an ${S}=1/2$ {H}eisenberg
  antiferromagnetic chain with long-ranged interactions,}  1988 \emph{Phys.
  Rev. Lett.} \textbf{60} 639

\bibitem{FM93}
Fowler M and Minahan J~A, \emph{{I}nvariants of the {H}aldane--{S}hastry
  $\mathrm{SU}(n)$ chain,}  1993 \emph{Phys. Rev. Lett.} \textbf{70} 2325

\bibitem{BGHP93}
Bernard D, Gaudin M, Haldane F~D~M and Pasquier V, \emph{{Y}ang--{B}axter
  equation in long-range interacting systems,}  1993 \emph{J. Phys. A: Math.
  Gen.} \textbf{26} 5219

\bibitem{HHTBP92}
Haldane F~D~M, Ha Z~N~C, Talstra J~C, Bernard D and Pasquier V, \emph{{Y}angian
  symmetry of integrable quantum chains with long-range interactions and a new
  description of states in conformal field theory,}  1992 \emph{Phys. Rev.
  Lett.} \textbf{69} 2021

\bibitem{Ha91}
Haldane F~D~M, \emph{``{S}pinon gas'' description of the ${S}=\frac12$
  {H}eisenberg chain with inverse-square exchange: exact spectrum and
  thermodynamics,}  1991 \emph{Phys. Rev. Lett.} \textbf{66} 1529

\bibitem{BBS08}
Basu-Mallick B, Bondyopadhaya N and Sen D, \emph{{L}ow energy properties of the
  {$\mathrm{SU}(m|n)$} supersymmetric {H}aldane--{S}hastry spin chain,}  2008
  \emph{Nucl. Phys. B} \textbf{795} 596

\bibitem{CS10}
Cirac J~I and Sierra G, \emph{Infinite matrix product states, conformal field
  theory, and the {H}aldane--{S}hastry model,}  2010 \emph{Phys. Rev. B}
  \textbf{81} 104431(4)

\bibitem{GS05}
Greiter M and Schuricht D, \emph{No attraction between spinons in the
  {H}aldane--{S}hastry model,}  2005 \emph{Phys. Rev. B} \textbf{71} 224424(4)

\bibitem{Gr09}
Greiter M, \emph{{S}tatistical phases and momentum spacings for one-dimensional
  anyons,}  2009 \emph{Phys. Rev. B} \textbf{79} 064409(5)

\bibitem{FG05}
Finkel F and Gonz{\'a}lez-L{\'o}pez A, \emph{{G}lobal properties of the
  spectrum of the {H}aldane--{S}hastry spin chain,}  2005 \emph{Phys. Rev. B}
  \textbf{72} 174411(6)

\bibitem{BFGR08epl}
Barba J~C, Finkel F, Gonz\'alez-L\'opez A and Rodr{\'\i}guez M~A, \emph{{T}he
  {B}erry--{T}abor conjecture for spin chains of {H}aldane--{S}hastry type,}
  2008 \emph{Europhys. Lett.} \textbf{83} 27005(6)

\bibitem{BFGR10}
Barba J~C, Finkel F, Gonz\'alez-L\'opez A and Rodr{\'\i}guez M~A,
  \emph{Inozemtsev's hyperbolic spin model and its related spin chain,}  2010
  \emph{Nucl. Phys. B} \textbf{839} 499

\bibitem{EFG09}
Enciso A, Finkel F and Gonz{\'a}lez-L{\'o}pez A, \emph{Spin chains of
  {H}aldane--{S}hastry type and a generalized central limit theorem,}  2009
  \emph{Phys. Rev. E} \textbf{79} 060105(4)

\bibitem{EFG10}
Enciso A, Finkel F and Gonz{\'a}lez-L{\'o}pez A, \emph{Level density of spin
  chains of {H}aldane--{S}hastry type,}  2010 \emph{Phys. Rev. E} \textbf{82}
  051117(6)

\bibitem{GSFPA10}
Giuliano D, Sindona A, Falcone G, Plastina F and Amico L, \emph{Entanglement in
  a spin system with inverse square statistical interaction,}  2010 \emph{New
  J. Phys.} \textbf{12} 025022(15)

\bibitem{HGCK16}
Hung C~L, Gonz{\'a}lez-Tudela A, Cirac J~I and Kimble H~J, \emph{Quantum spin
  dynamics with pairwise-tunable, long-range interactions,}  2016 \emph{Proc.
  Natl. Acad. Sci. U. S. A.} \textbf{113} E4946

\bibitem{HH92}
Ha Z~N~C and Haldane F~D~M, \emph{{M}odels with inverse-square exchange,}  1992
  \emph{Phys. Rev. B} \textbf{46} 9359

\bibitem{Po94}
Polychronakos A~P, \emph{{E}xact spectrum of {$\mathrm{SU}(n)$} spin chain with
  inverse-square exchange,}  1994 \emph{Nucl. Phys. B} \textbf{419} 553

\bibitem{MP93}
Minahan J~A and Polychronakos A~P, \emph{{I}ntegrable systems for particles
  with internal degrees of freedom,}  1993 \emph{Phys. Lett. B} \textbf{302}
  265

\bibitem{In96}
Inozemtsev V~I, \emph{Exactly solvable model of interacting electrons confined
  by the {M}orse potential,}  1996 \emph{Phys. Scr.} \textbf{53} 516

\bibitem{Po93}
Polychronakos A~P, \emph{{L}attice integrable systems of {H}aldane--{S}hastry
  type,}  1993 \emph{Phys. Rev. Lett.} \textbf{70} 2329

\bibitem{Fr93}
Frahm H, \emph{{S}pectrum of a spin chain with inverse-square exchange,}  1993
  \emph{J. Phys. A: Math. Gen.} \textbf{26} L473

\bibitem{FI94}
Frahm H and Inozemtsev V~I, \emph{New family of solvable 1{D} {H}eisenberg
  models,}  1994 \emph{J. Phys. A: Math. Gen.} \textbf{27} L801

\bibitem{Ha93}
Haldane F~D~M, \emph{Physics of the ideal semion gas: spinons and quantum
  symmetries of the integrable {H}aldane--{S}hastry spin chain}, in A~Okiji and
  N~Kawakami, eds., \emph{Correlation Effects in Low-dimensional Electron
  Systems}, \emph{Springer Series in Solid-state Sciences}, volume 118, pp.
  3--20

\bibitem{BUW99}
Basu-Mallick B, Ujino H and Wadati M, \emph{{E}xact spectrum and partition
  function of {$\mathrm{SU}(m|n)$} supersymmetric {P}olychronakos model,}  1999
  \emph{J. Phys. Soc. Jpn.} \textbf{68} 3219

\bibitem{BB06}
Basu-Mallick B and Bondyopadhaya N, \emph{{E}xact partition functions of
  {$\mathrm{SU}(m|n)$} supersymmetric {H}aldane--{S}hastry spin chain,}  2006
  \emph{Nucl. Phys. B} \textbf{757} 280

\bibitem{BB09}
Basu-Mallick B and Bondyopadhaya N, \emph{{S}pectral properties of
  supersymmetric {P}olychronakos spin chain associated with {$A_{N-1}$} root
  system,}  2009 \emph{Phys. Lett. A} \textbf{373} 2831

\bibitem{HB00}
Hikami K and Basu-Mallick B, \emph{{S}upersymmetric {P}olychronakos spin chain:
  motif, distribution function, and character,}  2000 \emph{Nucl. Phys. B}
  \textbf{566} 511

\bibitem{BBHS07}
Basu-Mallick B, Bondyopadhaya N, Hikami K and Sen D, \emph{{B}oson-fermion
  duality in {$\mathrm{SU}(m|n)$} supersymmetric {H}aldane--{S}hastry spin
  chain,}  2007 \emph{Nucl. Phys. B} \textbf{782} 276

\bibitem{BBH10}
Basu-Mallick B, Bondyopadhaya N and Hikami K, \emph{One-dimensional vertex
  models associated with a class of {Y}angian invariant {H}aldane--{S}hastry
  like spin chains,}  2010 \emph{Symmetry Integr. Geom.} \textbf{6} 091(13)

\bibitem{KY91}
Kuramoto Y and Yokoyama H, \emph{Exactly soluble supersymmetric $t$-{$J$}-type
  model with long-range exchange and transfer,}  1991 \emph{Phys. Rev. Lett.}
  \textbf{67} 1338

\bibitem{EFG12}
Enciso A, Finkel F and Gonz{\'a}lez-L{\'o}pez A, \emph{Thermodynamics of spin
  chains of {H}aldane--{S}hastry type and one-dimensional vertex models,}  2012
  \emph{Ann. Phys.-New York} \textbf{327} 2627

\bibitem{CFGRT16}
Carrasco J~A, Finkel F, Gonz{\'a}lez-L{\'o}pez A, Rodr{\'\i}guez M~A and
  Tempesta P, \emph{Critical behavior of $\mathrm{su}(1|1)$ supersymmetric spin
  chains with long-range interactions,}  2016 \emph{Phys. Rev. E} \textbf{93}
  062103(12)

\bibitem{BPS94}
Bernard D, Pasquier V and Serban D, \emph{Spinons in conformal field theory,}
  1994 \emph{Nucl. Phys. B} \textbf{428} 612

\bibitem{BS96}
Bouwknegt P and Schoutens K, \emph{{The {$\widehat{SU(n)}_1$} WZW models.
  Spinon decomposition and yangian structure},}  1996 \emph{Nucl. Phys. B}
  \textbf{482} 345

\bibitem{BFGR09}
Barba J~C, Finkel F, Gonz\'alez-L\'opez A and Rodr{\'\i}guez M~A, \emph{{A}n
  exactly solvable supersymmetric spin chain of {$BC_N$} type,}  2009
  \emph{Nucl. Phys. B} \textbf{806} 684

\bibitem{AK96}
Ahn C and Koo W~M, \emph{{$\mathrm{gl}(n|m)$} color {C}alogero--{S}utherland
  models and super {Y}angian algebra,}  1996 \emph{Phys. Lett. B} \textbf{365}
  105

\bibitem{ABCOP79}
Ahmed S, Bruschi M, Calogero F, Olshanetsky M~A and Perelomov A~M,
  \emph{{P}roperties of the zeros of the classical polynomials and of the
  {B}essel functions,}  1979 \emph{Nuovo Cimento B} \textbf{49} 173

\bibitem{BaXX}
Basu-Mallick B, \emph{{\rm private communication}}

\bibitem{Mu10}
Mussardo G, \emph{Statistical Field Theory: an Introduction to Exactly Solved
  Models in Statistical Physics} (Oxford: Oxford University Press) 2010

\bibitem{Le81}
Lewin L, \emph{Polylogarithms and associated functions} (New York: North
  Holland) 1981

\bibitem{OLBC10}
Olver F~W~J, Lozier D~W, Boisvert R~F and Clark C~W, eds., \emph{{NIST}
  Handbook of Mathematical Functions} (Cambridge University Press) 2010

\bibitem{BCN86}
Bl{\"o}te H~W~J, Cardy J~L and Nightingale M~P, \emph{Conformal invariance, the
  central charge, and universal finite-size amplitudes at criticality,}  1986
  \emph{Phys. Rev. Lett.} \textbf{56} 742

\bibitem{Af86}
Affleck I, \emph{Universal term in the free energy at a critical point and the
  conformal anomaly,}  1986 \emph{Phys. Rev. Lett.} \textbf{56} 746

\bibitem{CFGR17}
Carrasco J~A, Finkel F, Gonz{\'a}lez-L{\'o}pez A and Rodr{\'\i}guez M~A,
  \emph{Supersymmetric spin chains with nonmonotonic dispersion relation:
  {C}riticality and entanglement entropy,}  2017 \emph{Phys. Rev. E}
  \textbf{95} 012129(15)

\bibitem{Ka92}
Kawakami N, \emph{{A}symptotic {B}ethe-ansatz solution of multicomponent
  quantum systems with $1/r^2$ long-range interaction,}  1992 \emph{Phys. Rev.
  B} \textbf{46} 1005

\bibitem{AS06}
Arikawa M and Saiga Y, \emph{{E}xact spin dynamics of the $1/r^2$
  supersymmetric {$t$-$J$} model in a magnetic field,}  2006 \emph{J. Phys. A:
  Math. Gen.} \textbf{39} 10603

\bibitem{Po52}
Potts R~B, \emph{Some generalized order-disorder transformations,}  1952
  \emph{Math. Proc. Cambridge} \textbf{48} 106

\bibitem{Bax82}
Baxter R~J, \emph{Exactly Solved Models in Statistical Mechanics} (London:
  Academic Press) 1982

\bibitem{FMS99}
di~Francesco P, Mathieu P and S{\'e}n{\'e}chal D, \emph{Conformal Field Theory}
  (New York: Springer), corrected edition 1999

\bibitem{BFG09}
Basu-Mallick B, Finkel F and Gonz{\'a}lez-L{\'o}pez A, \emph{Exactly solvable
  {$D_N$}-type quantum spin models with long-range interaction,}  2009
  \emph{Nucl. Phys. B} \textbf{812} 402

\bibitem{EFGR05}
Enciso A, Finkel F, Gonz{\'a}lez-L\'opez A and Rodr{{\'\i}}guez M~A,
  \emph{{H}aldane--{S}hastry spin chains of {$BC_N$} type,}  2005 \emph{Nucl.
  Phys. B} \textbf{707} 553

\bibitem{BFG13}
Basu-Mallick B, Finkel F and Gonz{\'a}lez-L{\'o}pez A, \emph{The exactly
  solvable spin {S}utherland model of {$B_N$} type and its related spin chain,}
   2013 \emph{Nucl. Phys. B} \textbf{866} 391

\bibitem{BFG11}
Basu-Mallick B, Finkel F and Gonz{\'a}lez-L{\'o}pez A, \emph{The spin
  {S}utherland model of {$D_N$} type and its associated spin chain,}  2011
  \emph{Nucl. Phys. B} \textbf{843} 505

\end{thebibliography}

\end{document}